\documentclass[journal=jacsat,manuscript=article]{achemso}

\usepackage[version=3]{mhchem} 

\usepackage{amsmath}
\usepackage{amssymb}
\usepackage{graphicx}
\usepackage{siunitx}
\usepackage{xcolor}
\usepackage{soul}
\usepackage{braket}
\usepackage{pdfpages}

\graphicspath{{img/}}

\newcommand{\degree}{^{\circ}}
\DeclareSIUnit\angstrom{\text {Å}}

\newcommand*{\SIfigmasdepna}{S2.1}
\newcommand*{\SIfigpolbdp}{S2.2}
\newcommand*{\SIfigrelCna}{S2.3-S2.4} 

\newcommand*{\SIfigspindiffvsT}{S2.5} 
\newcommand*{\SIfigspindiffna}{S2.6}
\newcommand*{\SIzplanalysis}{S2.7}
\newcommand*{\SIsecNVzpl}{S2.2}

\newcommand*{\SIfigavgnonspinning}{S5.2}
\newcommand*{\SIfigLACs}{S6.1} 
\newcommand*{\SIfigfirstshelldynamics}{S6.2}
\newcommand*{\SIsecEPR}{S1} 
\newcommand*{\SIsecnvhamiltoniandiag}{S3.1}

\newcommand*{\SIsecnvopticalpump}{S3.2} 

\newcommand*{\SIsecsysdescription}{S4}
\newcommand*{\SIsecfullhamiltonian}{S4.1}
\newcommand*{\SIsecsysparameters}{S4.2}
\newcommand*{\SIsechyperfine}{S4.3}
\newcommand*{\SIsecnonspinning}{S5}
\newcommand*{\SIsectheoryspinning}{S6} 
\newcommand*{\SIsecLZ}{S6.2}
\newcommand*{\SIsecextspinhp}{S7}

\title{Microwave-Free $^{13}$C Hyperpolarization of Diamond Particles Enabled by Magic Angle Spinning and  NV Centers}

\author{Rémi~Blinder}
\email{remi.blinder@uni-ulm.de}
\affiliation[IQO]
{Institut für Quantenoptik, Albert-Einstein Allee 11, Universität Ulm, 89081 Ulm, Germany}
\author{Guzel Musabirova}
\affiliation[Leipzig]
{Institut für Analytische Chemie, Universität Leipzig, Linnéstrasse 3, 04103 Leipzig, Germany}
\author{Daehee Kim}
\affiliation[OIST]
{Quantum Machines Unit, Okinawa Institute of Science and Technology Graduate University,
Onna, Okinawa 904-0495, Japan}
\author{Anshuman Nayak}
\affiliation[OIST]
{Quantum Machines Unit, Okinawa Institute of Science and Technology Graduate University,
Onna, Okinawa 904-0495, Japan}
\author{Jason Twamley}
\affiliation[OIST]
{Quantum Machines Unit, Okinawa Institute of Science and Technology Graduate University,
Onna, Okinawa 904-0495, Japan}

\author{Viateschlav N. Agafonov}
\affiliation[Tours]{GREMAN UMR 7347, University F. Rabelais, 37200 Tours, France}

\author{Raiker Witter}
\affiliation[IQO]
{Institut für Quantenoptik, Albert-Einstein Allee 11, Universität Ulm, 89081 Ulm, Germany}

\author{Jörg Matysik}
\affiliation[Leipzig]
{Institut für Analytische Chemie, Universität Leipzig, Linnéstrasse 3, 04103 Leipzig, Germany}
\author{Fedor Jelezko}
\affiliation[IQO]
{Institut für Quantenoptik, Albert-Einstein Allee 11, Universität Ulm, 89081 Ulm, Germany}
\alsoaffiliation[IQST]
{Centre for Integrated Quantum Science and Technology (IQST), Ulm 89081, Germany	}

\begin{document}


\begin{abstract}

Nuclear hyperpolarization from optically pumped color centers in solids offers an alternative to conventional microwave-driven dynamic nuclear polarization (DNP). 
Diamond can host the nitrogen vacancy (NV) center, whose ground spin state can be readily polarized by light  at room temperature, making diamond a candidate platform for nuclear hyperpolarization.
We report $^{13}$C nuclear hyperpolarization in  randomly oriented diamond particles with sizes ranging from 0.2 to \SI{2}{\micro\meter}, both at natural $^{13}$C abundance (1.1\%) and  at 20\% isotopic enrichment, at magnetic fields of \SI{7.1}{\tesla} and \SI{9.4}{\tesla}. The protocol combines optical  illumination with  magic angle spinning (MAS)   and does not require microwave irradiation. 
By investigating the nuclear polarization as a function of the MAS frequency between $0$ and $\SI{6}{\kilo\hertz}$ at the magnetic field of \SI{7.1}{\tesla}, we find maximum light-induced polarization  enhancements of  $280$--fold for the isotopically enriched sample   and $411$--fold for the  natural abundance sample. 
Under continuous illumination, steady-state absolute $^{13}$C polarization levels above $0.1\%$ are reached. A model involving optical pumping of NV centers and spin dynamics near level anticrossings (LACs) in  three-spin clusters formed by NV, a substitutional nitrogen (P1)  and  $^{13}$C  is used to describe these findings. The protocol strongly mitigates the effect of the anisotropy of the NV spin Hamiltonian, allowing more than $99.9\%$ of NV orientations to participate in the polarization transfer process. These results represent a first step toward transferring nuclear polarization from diamond particles to external nuclei, with potential  applications in  sensitive  and high-resolution NMR   at room temperature.

\end{abstract}

\section{Introduction}


Under typical experimental conditions, particularly at room temperature, the  spin polarization of nuclei induced by an external magnetic field is extremely low, often on the order of a few parts per million.
Methods that are capable to transiently enhance the nuclear polarization beyond thermal equilibrium, a process known as hyperpolarization, have considerably expanded since their initial discovery by Overhauser into an array of techniques. Hyperpolarization enables substantial improvement in the sensitivity of Nuclear Magnetic Resonance (NMR)\cite{eills2023} and provides as well opportunities in magnetic field sensing\cite{sahin2022,moon2026} and  in the investigation of quantum many-body effects.\cite{cai2013,vetter2026}%

Some hyperpolarization techniques are based on the application of light, which under certain conditions can drive nuclear spins out of equilibrium. In particular, light--induced chemical reactions leading to enhanced nuclear polarization have been observed in donor-acceptor molecules, both in the liquid state~\cite{cocivera1968, theiss2025} and in the solid state~\cite{zysmilich1994,matysik2021,matysik2023struc}. This effect, known as photo-chemically induced dynamic nuclear polarization (photo-CIDNP), results from the formation of spin-correlated radical pairs (SCRPs), whose magnetic evolution and spin-selective recombination, in the presence of hyperfine-coupled nuclei, create non-equilibrium nuclear spin populations upon return to the electronic ground state. In a theoretical description of photo-CIDNP, Ivanov and coworkers~\cite{sosnovsky2016} emphasize the importance of level anticrossings (LACs) involving the electron spins of the radical pair and nearby nuclear spins, both in the liquid- and solid-state.

Diamond, containing color centers such as the nitrogen-vacancy center (NV), constitutes an alternative solid-state system for hyperpolarization, where the favorable coherence time of NVs at room temperature enables implementation under ambient conditions. Several studies have shown evidence for light-induced polarization of $^{13}$C in single-crystalline diamond: while some report only a phenomenological description, \cite{scott2016} typically,  the results  are  successfully explained by  the physics of 
magnetic-field-induced LACs. The latter can occur in a two-spins system\cite{wang2013} (NV, $^{13}$C), or with three spins (NV, $^{13}$C, X) where X is a third nearby electron spin --- possibly a substitutional nitrogen (P1 center),\cite{wunderlich2017, pagliero2018, zangara2019, henshaw2019} or yet another NV.\cite{wunderlich2021, plotzki2025} 

Transferring these approaches to powdered diamond remains challenging 
due to the strong anisotropy of the NV  spin Hamiltonian, which leads to orientation-dependent spin transition energies and, consequently, severe inhomogeneous broadening. 
This broadening requires  hyperpolarization protocols that are robust to spin orientation dispersion. Hyperpolarization protocols reported so far for microdiamonds~\cite{ajoy2018,miyanishi2021} and nanodiamonds~\cite{blinder2025} have relied on applying specific microwave pulse sequences in synchronization with optical excitation. These hybrid approaches exploit not only optical spin polarization of NV,  but also robust  microwave driving on a subset of these centers, to drive polarization transfer to nearby nuclei. Nevertheless, a  demonstration of purely optically driven (i.e., microwave-free) nuclear hyperpolarization in diamond powder is missing.

In this work, we demonstrate the generation of $^{13}$C hyperpolarization \emph{via} optical excitation in diamond particles containing NV centers  under  magic angle spinning  (MAS), at magnetic fields of 7.1 and \SI{9.4}{\tesla},  without microwave irradiation. 
This is achieved by exploiting the physics of rotation-induced LACs in a system made of three spins: $^{13}$C, NV, and another electronic spin (such as P1).
This approach remains efficient in an arbitrarily oriented ensemble of color centers, thereby overcoming the previous limitations associated with the anisotropy of the NV spin Hamiltonian. Furthermore, the method does not rely on a specific geometry of the spin system, as  the electronic spins (NV, P1) are  distributed randomly in the diamond lattice. Our results open a new path of microwave-free hyperpolarization at ambient conditions, with potential for future applications in ultrasensitive NMR of substances in contact with the surface of nanoparticles.  

\section{Results}

\subsubsection{Experimental observation of light-induced $^{13}$C hyperpolarization in diamond particles under MAS}

Two batches of diamond particles differing by their $^{13}$C content were investigated and are referred to hereafter as Sample~A and Sample~B. Sample~A consists of  sub-micron particles, with a $^{13}$C content of $20\pm2$\%,  synthesized by  high-pressure high-temperature (HPHT) growth, as detailed in a previous report\cite{mindarava2020_13C}. For Sample~B, commercial diamond particles  with  natural $^{13}$C  abundance (1.1\%)   and particle  size in the range 1.5-\SI{2.5}{\micro\meter} were used as source material (Pureon AG, type Ib, HPHT, MSY1.5-2.5).   To create NV centers, the particles in Samples~A and B were subjected to simultaneous electron-irradiation and annealing,    following a  procedure described in an earlier work (\SI{10}{\mega\electronvolt}, \SI{800}{\celsius} annealing).\cite{mindarava2020_conv} An electron dose of  \SI{6e18}{\per\square\centi\meter} was applied to Sample~A, while Sample~B was subjected to a dose of \SI{3e18}{\per\square\centi\meter}.   NV formation was confirmed by quantitative electron paramagnetic resonance (EPR) analysis, providing  the concentrations given in Table~\ref{table:sample_description}.

\begin{table*}
	\centering
    		\begin{tabular}{c c  c  c  c  }
      & $^{13}$C content  & Particle size & dark spins   (P1+other)  & NV \\
       &    &  &  
          ppm
          & ppm \\
        \hline 
	Sample~A & 20(2)\%   & 0.2-\SI{1}{\micro\meter}   &  186(11)  &  14.0(8)   \\ \hline 	
	Sample~B & 1.1\%  & 1.5-\SI{2.5}{\micro\meter} &  83(5)  &  9.2(5)  \\ \hline 
	\end{tabular}
		\caption[]{\textbf{Summary of sample properties.}  $^{13}$C content, particle size and spin concentrations determined by  EPR. Corresponding EPR   spectra and fits are represented in SI, section \SIsecEPR.  }	\label{table:sample_description}
\end{table*}

In addition to NV, the samples  comprise different types of paramagnetic point defects that do not get spin-polarized by light,  hereafter referred to as ``dark spins''. Most of these centers provide an EPR spectrum near the  resonance field of the free electron ($g=2.0023$), which can be decomposed into three components. We now discuss this decomposition, while corresponding EPR spectra and fits are shown in  Supporting Information (SI), section \SIsecEPR. 
The first spectral component corresponds to the isolated substitutional nitrogen  (N$_{\mathrm{s}}^{0}$ or P1 center) and accounts for a fraction of  47\% and 60\%  of the observed $g\sim 2$ spectra, respectively, in Sample~A and B. Second, one can observe a  featureless contribution providing a broad  background to the  spectrum of isolated P1 centers. The presence of clusters of exchange-coupled P1, commonly encountered in diamond grown by the HPHT technique, \cite{nir-arad2024,bussandri2024} likely explains a fraction of this contribution. Nevertheless, we note that a residual fraction of this background could correspond, in the case of Sample~A (which has particles in the sub-micrometer range) to subsurface defects commonly observed in nanodiamonds\cite{yavkin2015, boele2020}. The third contribution consists of a relatively narrow line around $g\sim 2$ and could represent bulk defects (such as lattice vacancies) or again near-surface defects.\cite{yavkin2015} This last component accounts, however, for a lower fraction of the EPR signal (1.3\% and 1.1\% in Sample A and B, respectively).

\begin{figure}[h]
\centering
\includegraphics[width=0.5\linewidth]{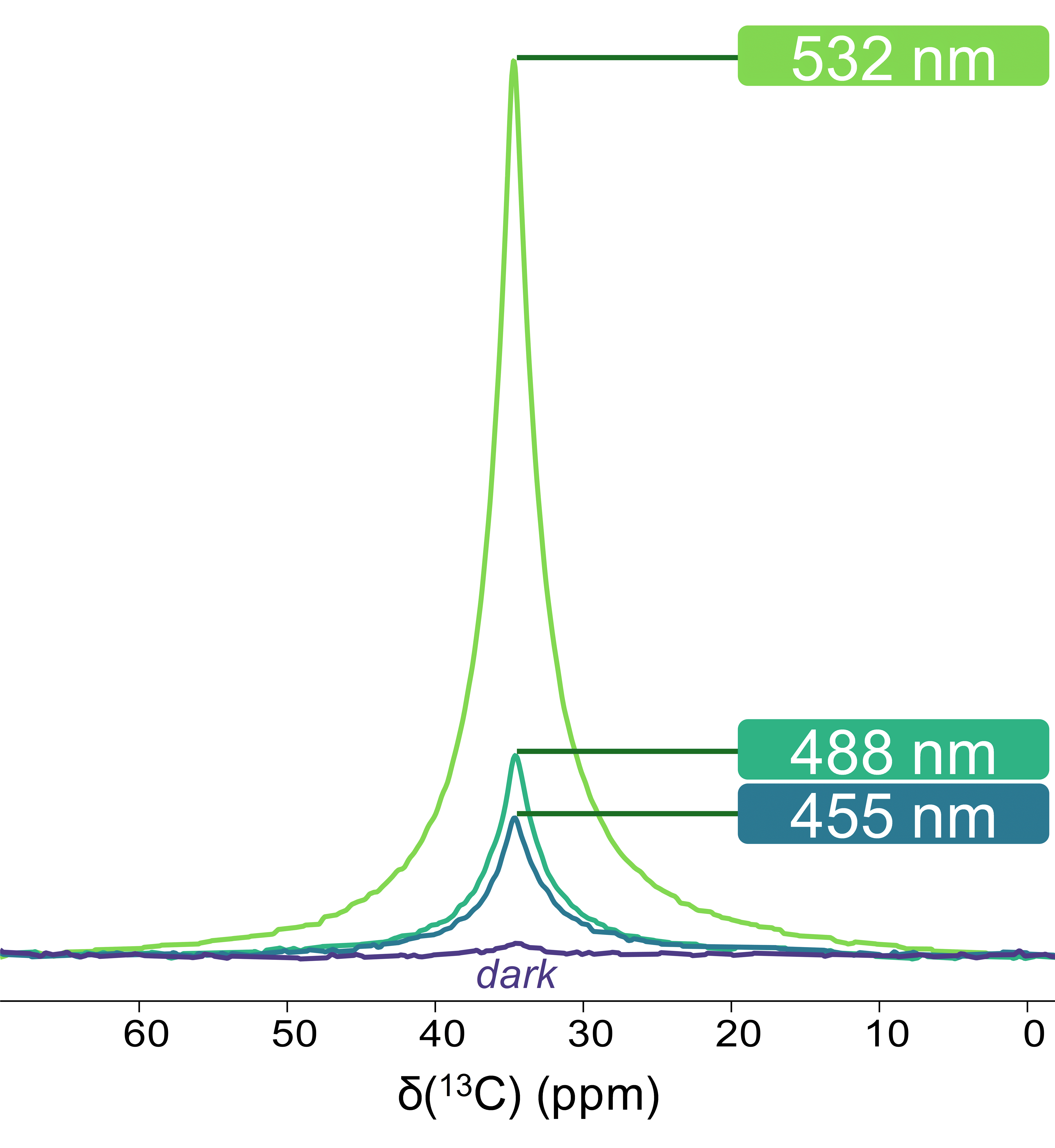}
\caption{\textbf{Light-induced  $^{13}$C hyperpolarization in  diamond particles under magic angle spinning at different excitation wavelengths.} The experiments were done in Sample~A (20\% $^{13}$C) at a magnetic field of $\SI{9.4}{\tesla}$ and a  MAS frequency of \SI{6}{\kilo\hertz}. Optical excitation was provided by lasers at 532, 488 and \SI{445}{\nano\meter} with equal power, \SI{650\pm5}{\milli\watt}. The signal was obtained by applying the sequence of Fig.~\ref{fig:setting_overview}, using as duration of illumination $d4=\SI{3}{\second}$. For comparison, a spectrum recorded without laser illumination (``dark'') is also shown.  }
\label{fig:lambda_variation}
\end{figure}

\begin{figure}[h]
\centering
\includegraphics[width=0.9\linewidth]{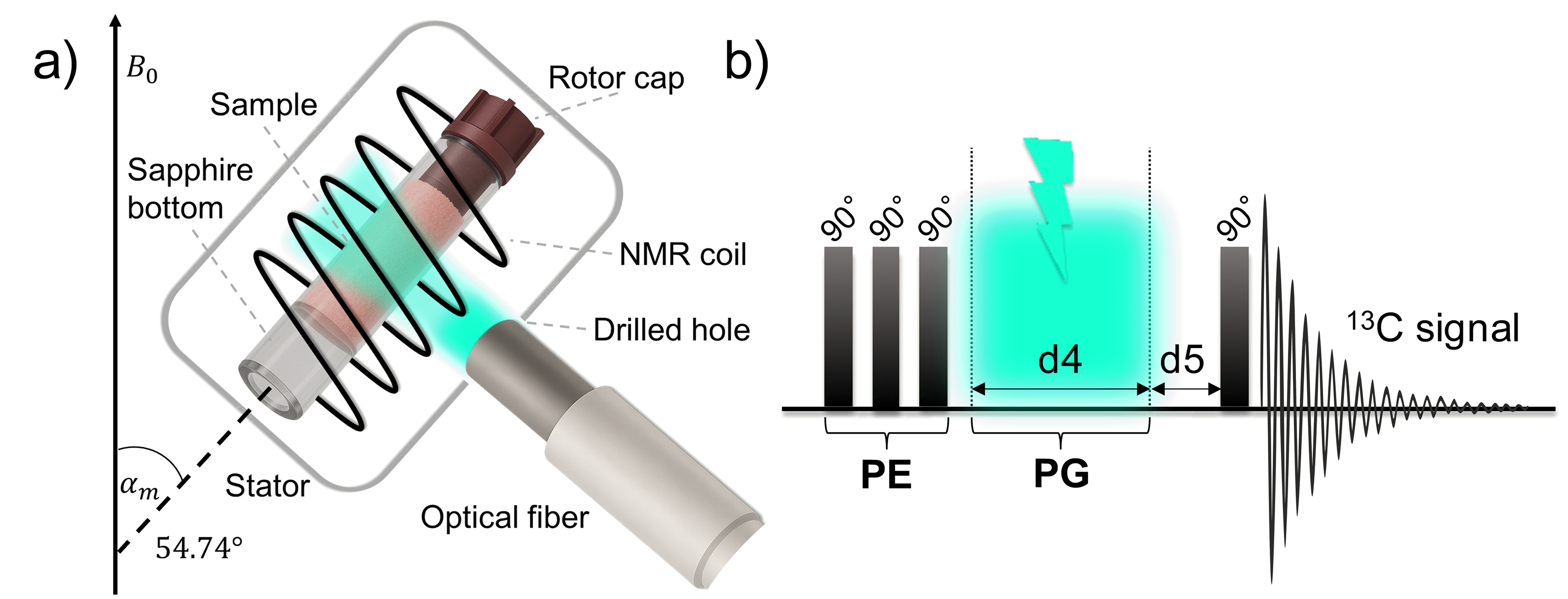}
\caption{\textbf{Overview of the  magic angle spinning experimental setup and pulse sequence.} 
(a) Schematic of the MAS NMR rotor oriented at the magic angle ($\alpha_{\rm m} = 54.74^\circ$) with respect to the external magnetic field, $B_0$. The diamond powder sample is packed into the rotor and illuminated with laser light delivered through an optical fiber. 
(b) Pulse sequence used in the experiments. The sequence consists of a polarization extinction block (PE), followed by a polarization generation period (PG) of length $d4$. During the delay $d4$, laser illumination induces nuclear spin polarization. After an optional   delay $d5$, the $^{13}$C signal is detected with a $90^\circ$ pulse.
}
\label{fig:setting_overview}
\end{figure}

\begin{figure}[h!]
\centering
\includegraphics[width=1\linewidth]{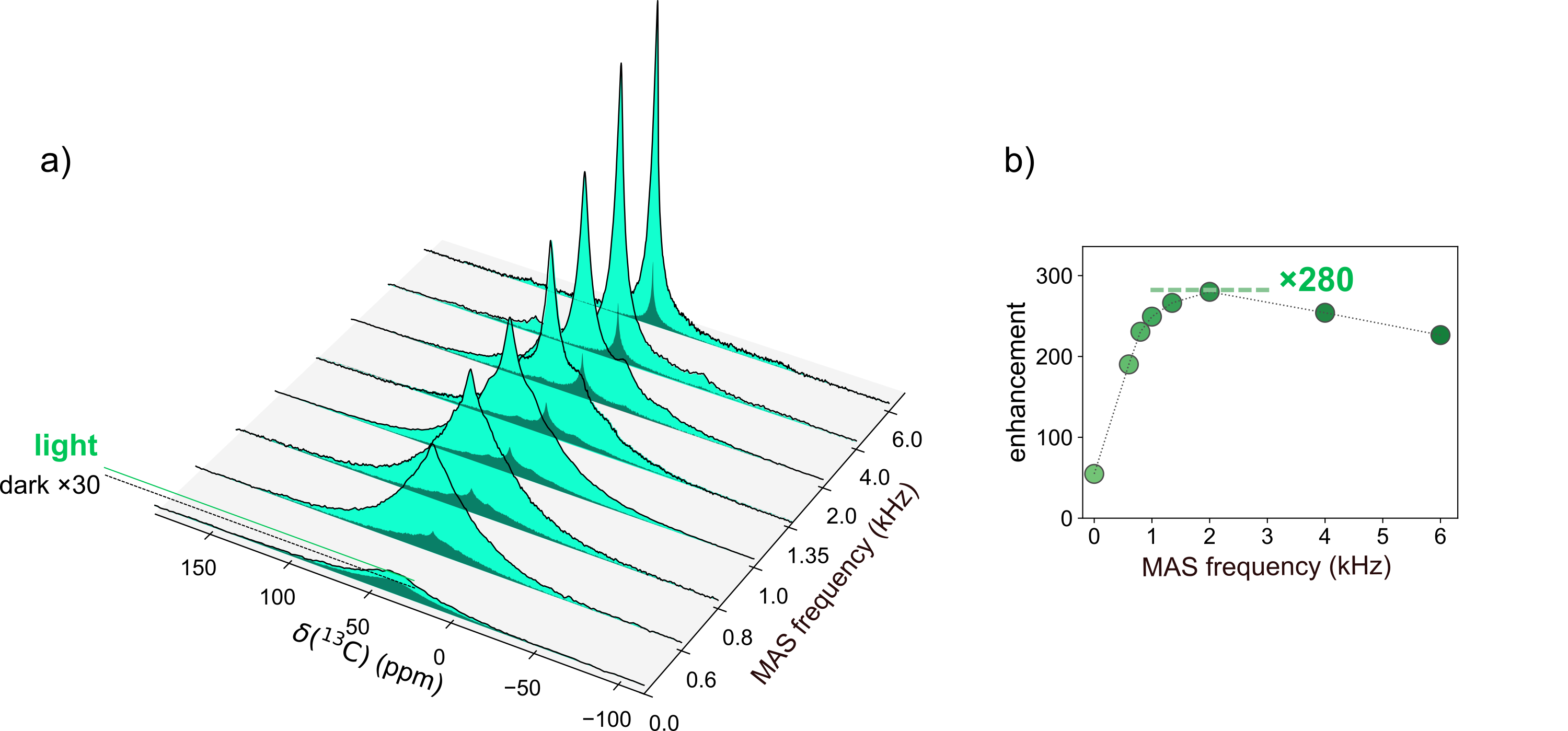}
\caption{\textbf{Light-induced hyperpolarized  $^{13}$C signal in Sample~A  as a function of the MAS frequency. } (a) Hyperpolarized $^{13}$C NMR spectra obtained by varying the MAS frequency between $0$ and $\SI{6}{\kilo\hertz}$, after an illumination time $d4=\SI{3}{\second}$ (green shaded areas). For comparison, the signal obtained without illumination,   magnified $\times$30, is also represented (dark shaded areas). (b) Corresponding light-induced signal enhancement relative to the dark acquisition, as a function of the MAS frequency. The experiments were performed using   green-light (\SI{532}{\nano\meter}) illumination, with a laser power of \SI{2}{\watt}, at a magnetic field of \SI{7.1}{\tesla}. }
\label{fig:mas_dependency_20pc}
\end{figure}

\begin{figure}[h]
\centering
\includegraphics[width=0.7\linewidth]{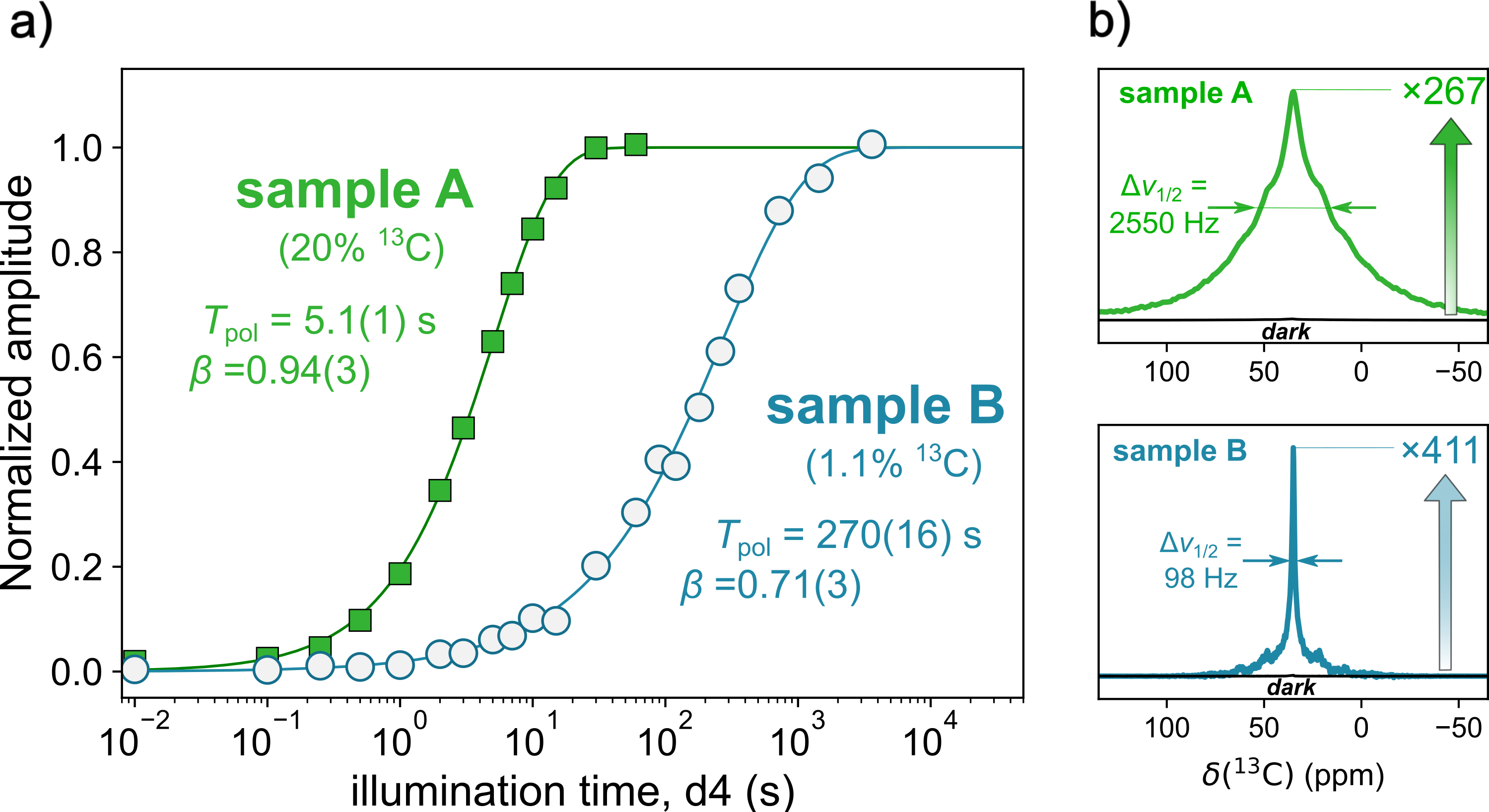}
\caption{\textbf{Effect of $^{13}$C isotopic enrichment on the polarization buildup  and NMR lineshape at a fixed MAS frequency of 1\,kHz.} a) Polarization buildup curves for Sample~A and B. The curves were obtained by integrating the full spectral line for different values of $d4$ in the pulse sequence shown in Fig.~\ref{fig:setting_overview}b), with the delay $d5$ set to 0.
The experimental data points were fitted with a stretched exponential function $M(t) = M_{\infty} \times \left\{ 1 - \mathrm{exp}\left(-t/T_{\mathrm{pol}})^{\beta} \right) \right\}$; both experimental points and fitted curves (continuous lines) were then  normalized to the saturation value $M_{\infty}$ for representation. b) $^{13}$C spectra 
acquired after illumination time $d4=\SI{3}{\second}$ in Sample A and $d4=\SI{120}{\second}$ in Sample~B, showing the impact of  the isotopic content on the spectral width (FWHM), $\Delta \nu_{1/2}$. Light/dark enhancement factors are indicated on the right in each plot. 
The experiments were performed using   green-light  illumination at \SI{532}{\nano\meter}, with a laser power of \SI{2}{\watt}, at a magnetic field of \SI{7.1}{\tesla}. }
\label{fig:buildup_dynamics}
\end{figure}

In the present work, we leverage the technique of  MAS in NMR at  $\SI{7.1}{\tesla}$ and $\SI{9.4}{\tesla}$ magnetic fields. In addition to being the central tool for resolution improvement in solid-state NMR,\cite{polenova2015}  MAS  has previously been used in the context of the hyperpolarization of amorphous samples by dynamical nuclear polarization (DNP).~\cite{thurber2012, akbey2013} 
Our MAS NMR systems    allow   the sample to,  simultaneously,  be spun at frequencies up to $\sim \SI{8}{\kilo\hertz}$ and  illuminated.\cite{bode2012} In Figure~\ref{fig:lambda_variation} is shown the enhanced $^{13}$C NMR signal resulting from the application of light at different wavelengths (405, 488 and \SI{532}{\nano\meter}) in Sample~A,   at a magnetic field of $\SI{9.4}{\tesla}$ and a MAS frequency of $\SI{6}{\kilo\hertz}$. The measurement setup and the NMR sequence  are  described in Fig.~\ref{fig:setting_overview}.
The comparison of the light-induced signals to the dark signal demonstrates that illumination at all these wavelengths  gives rise to hyperpolarization. Furthermore, the strongest enhancement is observed at the wavelength of \SI{532}{\nano\meter}, which hints at NV being involved.\cite{han2012,beha2012}
The NMR spectral shape and intensity exhibit a dependence on the MAS frequency, both in Sample~A (Fig.~\ref{fig:mas_dependency_20pc}a) and Sample~B (SI, Fig.~\SIfigmasdepna). Increasing the MAS rate narrows the $^{13}$C line and modulates the signal enhancement, measured as the  ratio of the signal intensity (integral over the full spectrum) under light and dark conditions. 
In Sample~A, although a 55-fold enhancement is already present without spinning the sample, the maximum, 280-fold, is  observed upon spinning at $\SI{2}{\kilo\hertz}$ frequency.  
In Sample~B, with natural $^{13}$C abundance,  the enhancement is 40--fold in static conditions and reaches a maximum of 411-fold at  $\SI{1}{\kilo\hertz}$ MAS frequency    (see Fig.~\ref{fig:buildup_dynamics}b). 
 The systematic reduction in the $^{13}$C spectral width upon faster spinning is an expected consequence of, either, magnetic dipole-dipole  interactions among $^{13}$C, or the hyperfine couplings of electronic defects (such as NV, P1 and other dark spins) to distant $^{13}$C, being gradually averaged out. 
 From the comparison of the spectra of Sample~A and B at the MAS frequency of \SI{1}{\kilo\hertz} (Fig.~\ref{fig:buildup_dynamics}b), one can conclude that  dipole-dipole interaction among $^{13}$C dominantly determines the NMR lineshape acquired in these conditions.

\vfill
\vfill 

\clearpage 
\pagebreak

Although  the strongest signal enhancement is found in the 1-2\,kHz MAS frequency regime for both samples, their polarization buildup dynamics  differ substantially. 
In Fig.~\ref{fig:buildup_dynamics}a
  the $^{13}$C polarization  buildup curves  are compared, for a MAS frequency of \SI{1}{\kilo\hertz}.
  In both samples, and in agreement with previous reports,\cite{reynhardt1998_dnp_1} the polarization  buildup time $T_{\mathrm{pol}}$ approaches the nuclear $T_1$  with $T_{\mathrm{pol}} \lesssim T_1$ (see in SI, Figures~\SIfigrelCna~for the $T_1$ at different MAS frequencies),  the buildup time is  substantially shorter in the $^{13}$C-enriched case (Sample~A). 
Here, the isotopic concentration appears, as well, to be the predominant factor. The   stronger magnetic interaction among $^{13}$C in Sample~A should lead to   faster nuclear spin diffusion towards paramagnetic impurities, favoring a shorter and more homogeneous $T_1$.\cite{seymour1985}  
To some extent, the higher paramagnetic defect concentration in the Sample~A (Table~\ref{table:sample_description}) might as well favor  nuclear spin diffusion, as the latter can be mediated by electronic spins. \cite{wittmann2018} 
We note that these mechanisms do affect the speed of buildup but not necessarily the light/dark enhancement: as spin diffusion should be involved in the growth of,  both, the dark and light signals in the PG block of Fig.~\ref{fig:setting_overview}b, its impact on the actual light/dark  enhancement  should be substantially reduced. Consistently, enhancements of similar orders of magnitude are seen in Sample~A and B.

\subsubsection*{Experimental evidence for the mediation of polarization by spin diffusion}

\begin{figure}[htp]
\centering
\includegraphics[width=0.85\linewidth]{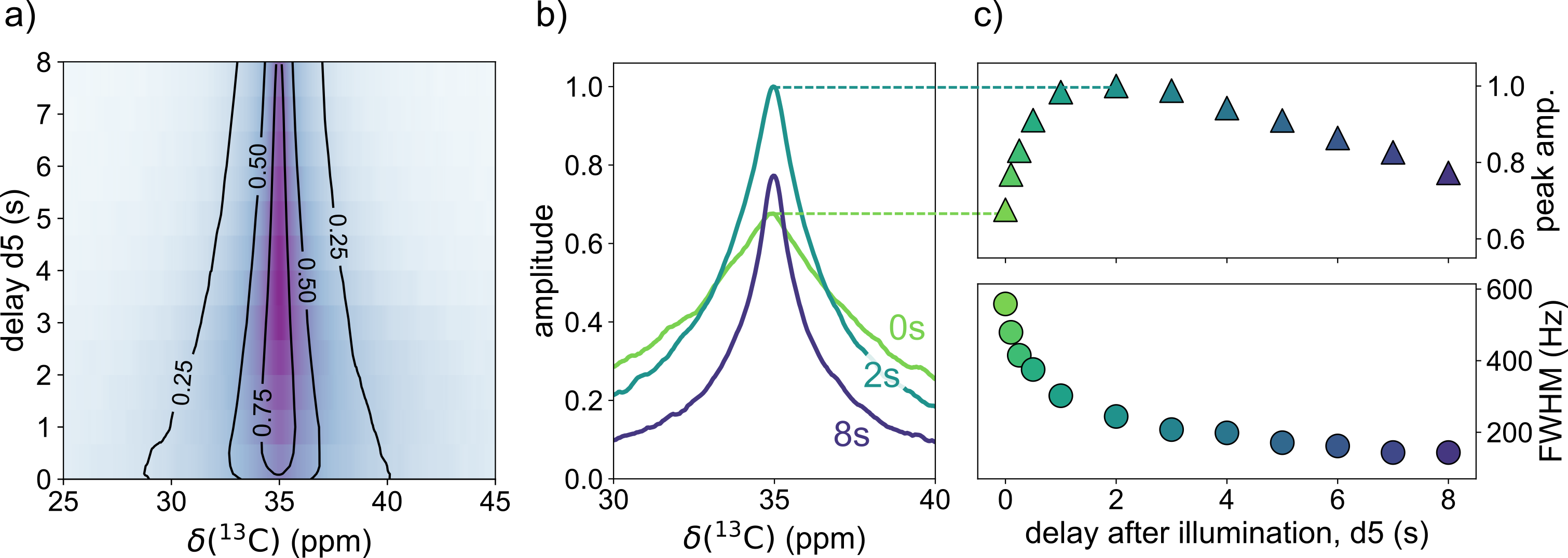}
\caption{\textbf{Observation of spin diffusion by varying the delay after illumination}. a) 2D plot of the hyperpolarized NMR spectrum measured as a function of the post-illumination delay ($d5$), following an illumination time $d4=\SI{0.5}{\second}$ in the sequence of Fig.~\ref{fig:setting_overview}b. The experiment was performed on Sample~A (20\% $^{13}$C) at a MAS rate of \SI{6}{\kilo\hertz} and a magnetic field of \SI{7.1}{\tesla}, using green-light illumination (\SI{532}{\nano\meter}). b) Corresponding spectra at $d5=0, 2$ and \SI{8}{\second}. c)  Peak amplitude and full width at half maximum of the NMR signal as a function of $d5$ (the peak amplitude is normalized to the value at $d5=\SI{2}{\second}$).}
\label{fig:spin_diffusion}
\end{figure}

Although the shape of the spectra in  Fig.~\ref{fig:buildup_dynamics}b is, as previously discussed, dominantly determined by the magnetic interaction among $^{13}$C, signatures of different interactions can appear at short illumination times.  
To investigate these features, we  acquire the NMR  signal after  a short illumination pulse of fixed duration ($d4$) followed by a variable post-illumination delay ($d5$), first in Sample~A. The outcome of this experiment (using as illumination time,  $d4=\SI{0.5}{\second}$) is shown in Fig.~\ref{fig:spin_diffusion}.  Upon increase of the post-illumination delay $d5$, two features are observed:  first,  the peak amplitude of the NMR signal rises up to $d5=\SI{2}{\second}$ and then decreases;  second,  as $d5$ increases, continuous reduction in the full width at half maximum occurs (Fig.~\ref{fig:spin_diffusion}c). Looking for the driving mechanisms, we remark that the evolution of the peak amplitude cannot be fully explained by nuclear  relaxation ($T_1$) behavior, as that mechanism should lead to signal decay in all regions of the spectrum. 
To complement the description, we observe that immediately after the short illumination pulse (that is, for $d5=0$), the $^{13}$C nuclear spins contributing to the NMR signal should locate in the vicinity of the paramagnetic centers that are involved in polarization transfer. A strong hyperfine coupling of these $^{13}$C to one paramagnetic center --or several, depending on the polarization mechanism--  is expected. For these $^{13}$C, line broadening naturally occurs  through, either, the distribution of hyperfine shifts or paramagnetic relaxation enhancement.\cite{beatrez2023} 
However, on the few seconds timescale, due to spin diffusion, the spin polarization can be transferred to more distant $^{13}$C.  Since these remote nuclei experience weaker hyperfine couplings, the NMR signal should sharpen while the area below the NMR spectrum should stay  constant (since, as long as $d5<T_1$,  the total polarization is essentially conserved). The combination of these two factors explains  the initial rise in the peak amplitude. Due to its slower dynamics, nuclear $T_1$ relaxation is only visible through the subsequent decay, at $d5>\SI{2}{\second}$. 
The presence of these two independent timescales is confirmed by variable temperature measurements, where the ($T_1$-driven)  decay is accelerated upon an increase in temperature while the initial (spin-diffusion-driven) rise remains unaffected (SI, Fig \SIfigspindiffvsT).  
The present method allows observing spin diffusion through the  evolution of the NMR lineshape, and therefore complements earlier protocols based on the observation of nuclear spin coherence under periodic RF driving.~\cite{beatrez2023}
As a high $^{13}$C content leads to a fast polarization rate and high NMR signal, this evolution is better resolved in Sample~A. However, the mechanism is also observable in sample~B, the principal difference being a slower dynamics (see SI, Fig.~\SIfigspindiffna).

\subsubsection{Level anticrossings in the NV-P1-$^{13}$C system and NV photophysics as the basis for polarization transfer}

\begin{figure}[htp]
\centering
\includegraphics[width=0.8\linewidth]{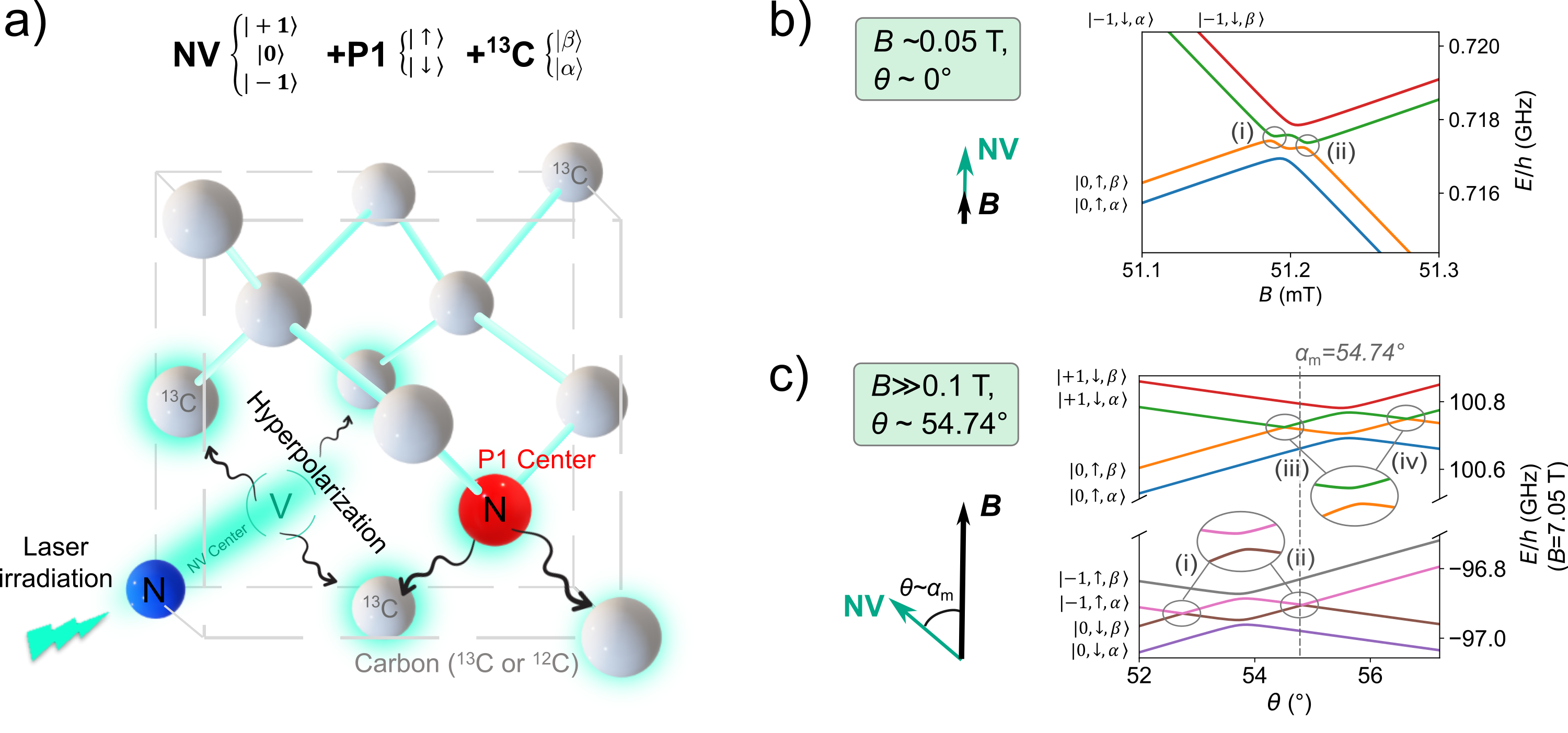}
\caption{\textbf{Level anticrossings involving NV, $^{13}$C and a dark spin as the basis for microwave-free hyperpolarization protocols.} a) Schematic representation of the defects in the diamond lattice: nitrogen-vacancy center (NV), substitutional nitrogen (P1) --which acts as a dark electron spin-- and surrounding $^{13}$C that can become hyperpolarized. For clarity, the NV and P1 centers are shown within the same unit cell; however   in the actual samples, their typical separation is expected to be larger. b) Description of the  field-induced LACs occuring in the NV-P1-$^{13}$C system under the setting of previously reported hyperpolarization protocols:~\cite{wunderlich2017, pagliero2018,henshaw2019} for a NV aligned to the magnetic field, level anticrossings occur at $\sim$\SI{51.2}{\milli\tesla},  LACs at which flip of the $^{13}$C nuclear spin can occur  are denoted  (i) and (ii). Energies were simulated assuming a strength of the NV-P1 flip-flop interaction  $a_{\mathrm{ee}}=\SI{0.5}{\mega\hertz}$  and $A_{zx} =\SI{0.1}{\mega\hertz},A_{zy} =0$ as the off-diagonal components of the $^{13}$C-NV hyperfine coupling, the state $m_I=0$ was assumed for the $^{14}$N nuclear spin within the P1 and NV  centers.  c) Description of  the alternative, rotation-induced level anticrossings, used in the present work, for a fixed magnetic field $B\gg \SI{0.1}{\tesla}$ (here, $B=\SI{7.1}{\tesla}$), that is, well above the ground state level anticrossing of NV. The anticrossings are centered at $\theta = \alpha_m = 54.74$°, as shown in the figure, or similarly  at $\theta = 180^{\circ}-\alpha_m$ (not represented). As one more electron spin state manifold is involved compared to (b), the $^{13}$C polarization can now occur at four distinct LACs, labelled (i-iv). In this simulation, a NV-P1 interaction strength  $a_{\mathrm{ee}}= \SI{5}{\mega\hertz}$ is taken, and the $^{13}$C-NV hyperfine tensor  corresponds to the coupling labeled `Weak 1'  in  Table~\ref{table:13Cgroups}.
}
\label{fig:mw_free_protocols}
\end{figure}

Next, we examine the mechanism underlying the observed $^{13}$C hyperpolarization.
 Rather than treating the NV as an isolated source of polarization, we introduce a minimal three-spin system composed of a NV electron spin, a $^{13}$C nuclear spin, and one dark electron spin. This framework provides a natural explanation for the experimental results in terms of level anticrossings that enable polarization transfer to $^{13}$C by a cross-effect-type mechanism\cite{thurber2012} (analogous to three-spin-mixing, as  described in the context of SCRPs\cite{jeschke1998,  sosnovsky2016}).  
In the following, we assume the dark spin to be a P1 center, since these centers likely represent the dominant contribution to the electron spin bath -- considering that P1 can be present either as isolated centers or in cluster form.  However, the dark electron spin does not necessarily have to be assigned to a P1 center. In principle, the proposed model is also valid if the dark spin is associated with any other EPR-active defect with resonance near $g\sim 2$, provided that the defect has  suitable coherence properties. 
The main difference of the current investigation to other protocols employing the physics of level anticrossings for performing nuclear hyperpolarization in diamond  (using \emph{e.g.}, the NV-P1-$^{13}$C system)\cite{wunderlich2017, pagliero2018, zangara2019,henshaw2019,  wunderlich2021, plotzki2025} 
consists in the way the LAC condition is  achieved. While previous studies have relied on LACs induced by the magnetic field in a sample kept static, our technique relies on  LACs induced \emph{by sample rotation}. A comparison of  the two approaches, considering the NV-P1-$^{13}$C system, is shown in  Figure~\ref{fig:mw_free_protocols}. 
The present protocol is thus  based on the orientation dependence (rather than the field dependence) of the NV energies. Given that our experiments are performed at a magnetic field $B \gg \SI{0.1}{\tesla}$, that is, well above the ground state level anticrossing of NV, the energies follow a simple angular dependence. To a very good approximation, the  transition frequencies are given by $\nu_{0,\pm 1}(\theta) = |\bar{\gamma}_{\mathrm{NV}}|B \pm \frac{1}{2}(3 \cos^2\theta - 1) D$ 
 for the $\ket{0}\leftrightarrow \ket{+1}$ (high-frequency)
 and $\ket{0}\leftrightarrow \ket{-1}$ (low-frequency)  transitions, where $|\bar{\gamma}_{\mathrm{NV}}| = \SI{28.032}{\giga\hertz\per\tesla}$ is the gyromagnetic ratio, $D=\SI{2.867}{\giga\hertz}$ is the zero-field splitting and  $\theta$ is the tilt of the NV axis with respect to the magnetic field (see SI, section \SIsecnvhamiltoniandiag, for the derivation). Here we label $\ket{m_s} = \ket{0}, \ket{\pm 1}$ the on-field eigenstates of NV, where $m_s$ denotes the spin projection number along the magnetic field axis, which  essentially sets the spin quantization axis. 
 Numerical diagonalization of the the Hamiltonian of the NV-P1-$^{13}$C system  shows that LACs locate in the vicinity of the tilt angles $\theta = \alpha_m$ (see Fig.~\ref{fig:mw_free_protocols}c) and $\theta =  \beta_m$, where $\alpha_m = \mathrm{acos}(1/\sqrt{3}) = 54.74$° and $\beta_m =  180^{\circ}-\alpha_m$. In these regions, the factor $\propto (3\cos^2\theta -1)$  is nearly canceled, which means the transition frequencies of NV and P1 are close enough to possibly  allow for a flip-flop of NV and P1 and a simultaneous flip of the $^{13}$C nuclear spin, satisfying energy conservation. A simulation of the  NV-P1-$^{13}$C energies at a magnetic field of $B=\SI{7.1}{\tesla}$ (Fig.~\ref{fig:mw_free_protocols}c) reveals that four LACs involving a nuclear spin flip ($\ket{\alpha}\leftrightarrow \ket{\beta}$), labeled (i-iv), are met in a first LAC region $|\theta-\alpha_m|<2.1\degree$. Four similar  LACs are present in a second  region $|\theta-\beta_m|<2.1\degree$
 (this region is not represented in Fig.~\ref{fig:mw_free_protocols}c, as the angle dependences of energies around angles $\alpha_m$ and $\beta_m$ are essentially the same).

\subsubsection{Simulations: static and spinning case}

\begin{figure}[htp]
\centering
\includegraphics[width=0.95\linewidth]{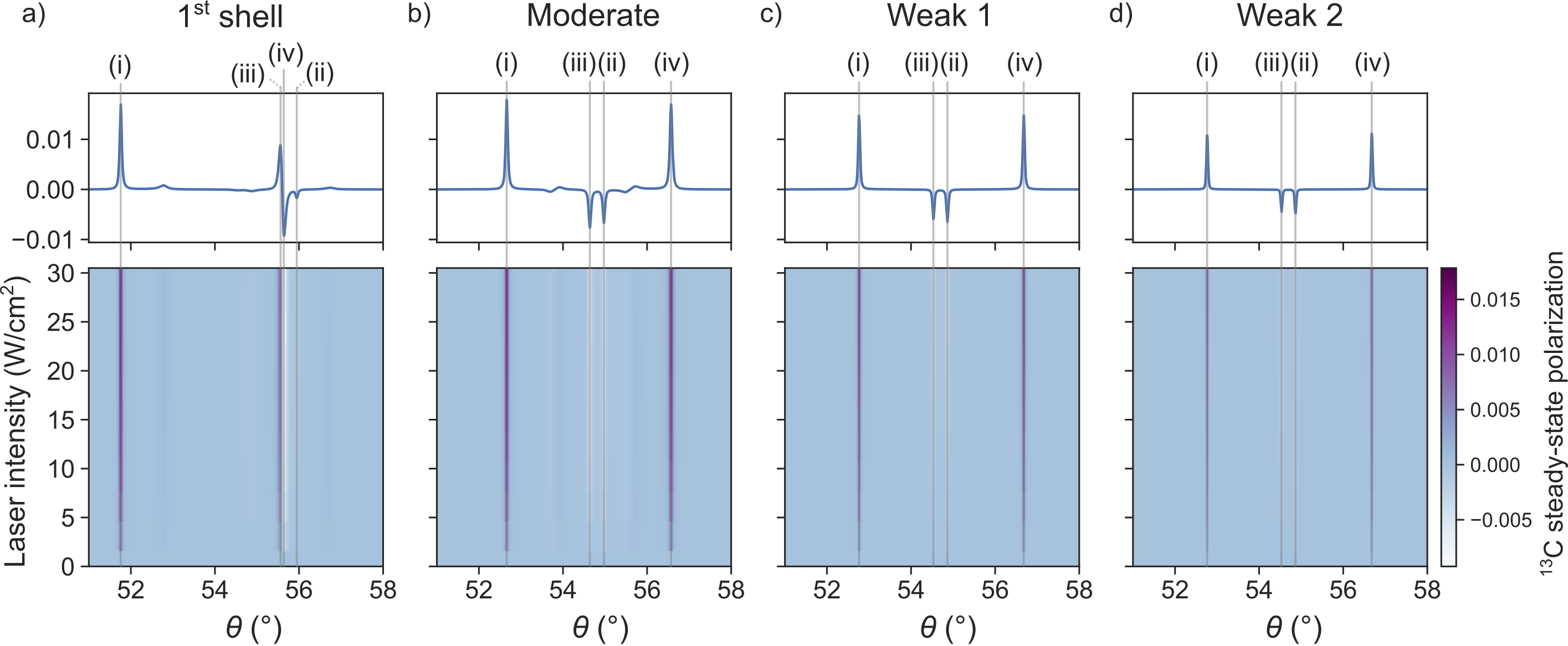}
\caption{ \textbf{Simulated steady-state $^{13}$C polarization in the static (non-spinning) case}. The graphs represent the steady-state $^{13}$C polarization ($P_{\rm 13C} = 2 \braket{I_z}$) obtained by simulating the non-unitary dynamics of the NV-P1-$^{13}$C system, as a function of the tilt of the NV axis to the magnetic field $\theta$ and the laser intensity $I$. The polarization is represented in the $\theta\sim \alpha_m$ region, where polarization transfer occurs at the electron-nuclear LACs labeled (i-iv). The maximum intensity $I_{\rm max}=\SI{30.5}{\watt\per\square\centi\meter}$ corresponds to that applied in the experiments performed at the magnetic field of $\SI{7.1}{\tesla}$. The various subplots (a-d) correspond to simulations considering different NV-$^{13}$C hyperfine coupling strengths, where the labels ($1^{\rm st}$ shell/Moderate/Weak 1/Weak 2) refer to the parameters given in  Table~\ref{table:13Cgroups}.  For a list of  simulation parameters and a detailed description of the  model, see SI, sect.~\SIsecsysparameters~and   \SIsectheoryspinning. }%
\label{fig:pol_static_case}
\end{figure}

We first describe hyperpolarization in the static (\emph{i.e.} non-spinning) regime, corresponding to the first dataset in Fig.~\ref{fig:mas_dependency_20pc}a. For this, we consider the dynamics of a NV-P1-$^{13}$C cluster at the different LACs, (i-iv), represented in Fig.~\ref{fig:mw_free_protocols}c. In a powder ensemble, due to the random NV orientations, a fraction of NV will have tilt angles close to each of these LACs. 
For these NVs, the combination of optical pumping and state mixing in the NV-P1-$^{13}$C system can lead to  polarization transfer to $^{13}$C nuclei. 
To model these dynamics, we first take into account  
the photophysics of NV centers  by using the seven-level model.\cite{robledo2011,tetienne2012} 
This model, detailed in SI, section~\SIsecnvopticalpump, describes the rates of population changes within the ground and excited triplets, as well as the intermediate singlet of NV.   
In the present work, considering the moderate illumination intensities, the model can be further  reduced to the ground spin state --that is, to three levels-- by adiabatic elimination of the fast relaxing states (see as well SI, sec.  \SIsecnvopticalpump). 
For a magnetic field aligned with the NV axis (or at zero magnetic field),  optical excitation  leads to preferential population of the $m_s=0$  spin state, as a consequence of spin-dependent decay through intersystem crossings (ISC).\cite{robledo2011}
If the magnetic field is tilted from the NV axis, mixing of the spin states occurs, leading to a change in the ISC rates. 
Thus, for an arbitrary orientation of the NV axis, all three levels in the ground spin state can get populated, with orientation-dependent pumping rates.\cite{tetienne2012,drake2015} 
We then simulate the non-unitary evolution of the NV-P1-$^{13}$C system under illumination using a Lindblad master equation. In the simulation,  different strengths of hyperfine interaction are considered, corresponding to the tensor parameters listed in Table~\ref{table:13Cgroups}, for a  $^{13}$C coupled to NV.   Further, a scalar interaction term of strength $a_{\rm ee} = \SI{5}{\mega\hertz}$ is taken to describe the coupling between NV and P1 (see SI, sect \SIsecsysdescription~and \SIsecnonspinning~for the full system's Hamiltonian  and the detailed model description). The presence of these two spin-spin interaction terms induces state mixing that allows polarization transfer to $^{13}$C.
The simulation leads to the steady-state $^{13}$C polarization  as a function of the laser intensity and the tilt of the NV axis to the magnetic field, represented in Fig.~\ref{fig:pol_static_case}. The simulation outcome demonstrates that illumination indeed leads to the creation of non-thermal $^{13}$C polarization,  for all considered hyperfine couplings,  in the vicinity of LACs. Besides, while  both positively and negatively enhanced $^{13}$C polarization can be found, the polarization is more efficiently generated in the positive case  (see \emph{e.g.}, in  Fig.~\ref{fig:pol_static_case}b,c,d: LACs \{i,iv\} corresp. to the positive case, while LACs \{ii,iii\} provide negative polarization of weaker amplitude).
As a consequence, we predict the powder-averaged  $^{13}$C polarization  to be  positively enhanced, which is consistent  with our experimental finding (see also SI, Fig. \SIfigavgnonspinning~for plots of the simulated powder-averaged polarization).  This observation provides a first agreement between  the experiment and  the model based on NV-P1-$^{13}$C dynamics.

\begin{table}[htp]
	\centering
		\begin{tabular}{ c  c  c  c c}
       $^{13}$C hyperfine coupling  & $A_{ZZ}$, MHz & $A_{XX} (=A_{YY})$, MHz   & source \\
        \hline 
	$1^{\mathrm{st}}$ shell  &  198.2 &  120.8  & Felton \emph{et al.}\cite{felton2009}    \\ \hline 
    Moderate  & 18.49  & 13.26   &  Felton \emph{et al.}\cite{felton2009}     \\ \hline 
    Weak 1 & 1.0  & -0.5    &   -- \\ \hline 
    Weak 2 & 0.5  & -0.25    &   -- \\ \hline 
	\end{tabular}
		\caption[]{\textbf{$^{13}$C hyperfine coupling parameters used in simulations.} In this Table are given the principal values of the hyperfine coupling tensors.
         The $^{13}$C is considered in all cases to be coupled to NV, and the tensor frame orientation is defined in see SI, section \SIsecfullhamiltonian. 
         While the First-shell and Moderate  cases correspond to  actual  positions in the diamond lattice, the Weak 1\&2 values were defined to  match typical cases of strong and intermediate hyperfine coupling at natural $^{13}$C abundance (1.1\%). Defining the hyperfine strength as $a_h =  \sqrt{\braket{ A_{zx}^2 + A_{zy}^2 }}$ (where $\braket{}$ refers to powder averaging and the elements $A_{zx}, A_{zy}$ are taken in the laboratory frame) and using the  parameters derived by DFT in Tak\'acs  \emph{et al.}\cite{takacs2024}, the probabilities of having one carbon with stronger hyperfine strength  than `Weak 1' and `Weak 2' considering a macroscopic 1.1\% $^{13}$C ensemble are  30\% and 50\%, respectively   (see SI, sect.~\SIsechyperfine~for the derivation).
         
         }
	\label{table:13Cgroups}
\end{table}

\begin{figure}[htp]
\centering
\includegraphics[width=0.6\linewidth]{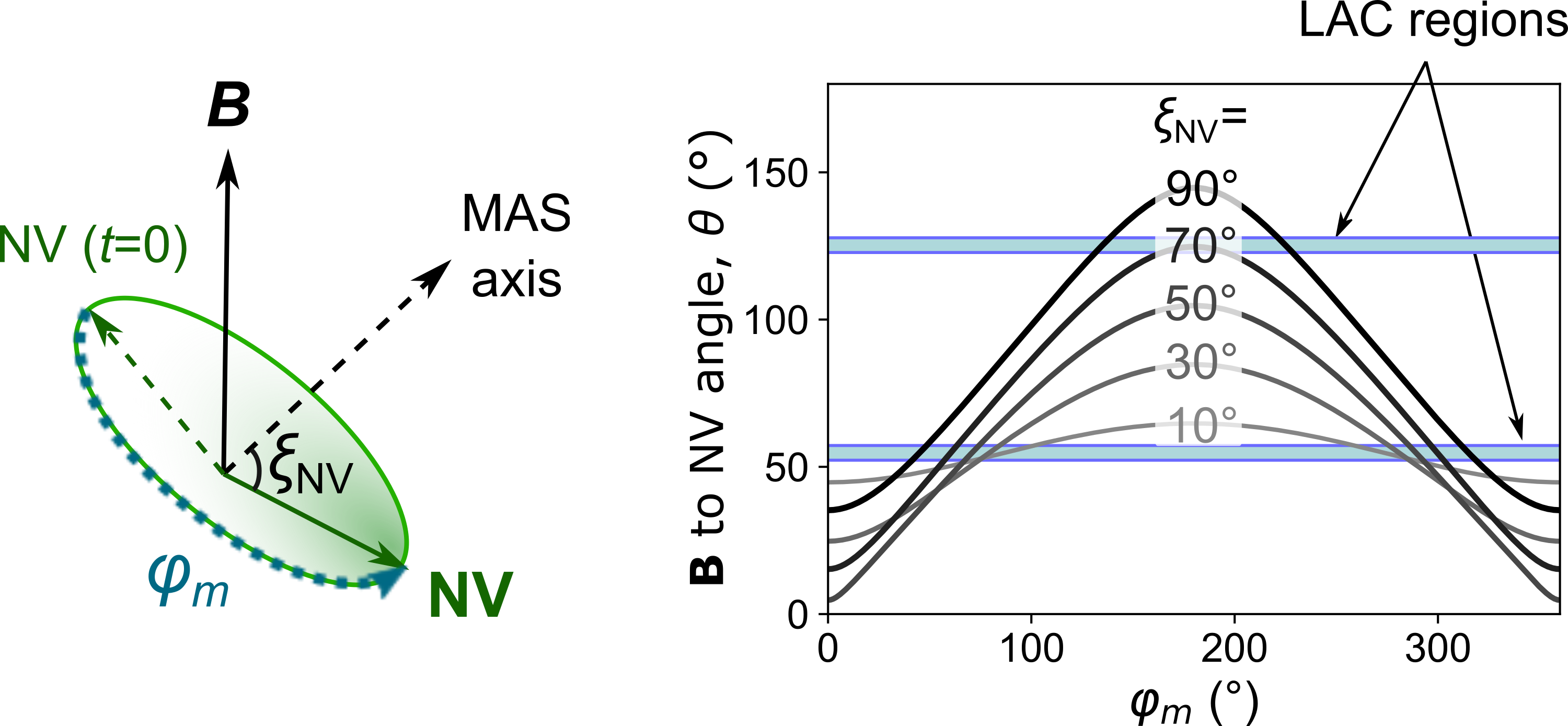}
\caption{\textbf{NV trajectories and level anticrossings under MAS}.  The plot shows the possible trajectories of the NV tilt angle relative to the magnetic field, $\theta$, upon sample rotation about the MAS axis, parameterized by the rotor phase $\varphi_{\mathrm{m}}$. Trajectories are characterized by the parameter $\xi_{\mathrm{NV}}$, defined as the angle between the NV axis and the MAS axis. The blue shaded areas indicate the angular regions where LACs occurs for the NV-P1-$^{13}$C system, corresponding to those shown in Fig.~\ref{fig:mw_free_protocols}c. The first LAC region, centered on $\theta= \alpha_m = 54.74^{\circ}$ is crossed twice per rotor period for NVs satisfying $\xi_{NV} > 2.1^{\circ}$. 
For $\xi_{NV} > 72.6^{\circ}$, the trajectory also crosses the second LAC region, centered at  $\theta= \beta_m = 180^{\circ}- \alpha_m$, twice per rotor period. The simulation parameters are the same as in Fig.~\ref{fig:mw_free_protocols}c.   }
\label{fig:lac_regions}
\end{figure}

The polarization process taking place in a static sample must therefore be restricted to a small subset of NV-P1-$^{13}$C clusters --  those in which the tilt of the NV axis brings the system at an avoided crossing. 
In contrast, when the sample is spun at the magic angle, the vast majority of NV-P1-$^{13}$C clusters will \emph{sweep} through at least one of the LAC regions. During MAS, individual NVs follow periodic  trajectories, such that their tilt angle  relative to the external magnetic field varies as  $\theta(t)$ (Fig.~\ref{fig:lac_regions}). These trajectories are characterized by the angle between the NV and the spinning axis, that we label $\xi_{\mathrm{NV}}$ (in that sense, all NVs having the same angle $\xi_{\mathrm{NV}}$ will describe the same trajectory).  Owing to the periodicity of the angular dependence of the NV energy levels, it is sufficient to consider 
 $\xi_{\mathrm{NV}}$ values in the range 
$0-90.0 \degree$. 
For $\xi_{\mathrm{NV}} >  2.1\degree$, the NV trajectory passes through the first LAC region twice per rotor period. For $\xi_{\mathrm{NV}}> 72.6 \degree$, in addition, the second LAC region will also be crossed twice per period. 
Although sweeps through the LACs  can be achieved by magnetic field sweeps around $\SI{51.2}{\mT}$ as  discussed in Henshaw \emph{et al.}\cite{henshaw2019}, the present protocol based on rotation-induced LACs   significantly enhances the robustness with respect to NV orientation, 
as all the color centers satisfying $\xi_{\mathrm{NV}} >  2.1\degree$, which corresponds to a fraction of $99.93\%$ of all NV centers, geometrically traverse at least one LAC region. 
We remark that the ability of such a large fraction of NVs to fulfill this crossing requirement is a specific consequence of the geometry of the MAS NMR experiment --where the tilt of the spinning axis with respect to the magnetic field is by $\alpha_m=54.74\degree$-- and the particular angular dependence of NV energies for  $B\gg \SI{0.1}{\tesla}$. For a different configuration of  the spinning axis, or for $B\lesssim \SI{0.1}{\tesla}$, a lower fraction of NVs would reach the tilt angles required for the LACs during rotation.

The $^{13}$C polarization dynamics under MAS are modeled using the Landau-Zener (LZ) formalism,  which allows determining the  transition probabilities at each  LAC. 
This approach follows the general description of MAS-DNP driven by the cross-effect\cite{thurber2012}. One must consider, in addition to the 
electron-nuclear LACs labeled in Fig.~\ref{fig:mw_free_protocols}c, also the \emph{electron-electron}  LACs -- which involve a change in the NV and P1 states while leaving the nuclear spin state identical (see SI, Fig.~\SIfigLACs~for a representation of the energy levels as a function of the rotor coordinate, including these LACs).  
While the LZ formula gives the probability for discrete population jumps at LACs, the 
 evolution aside avoided crossings is described using a standard rate equation model including relaxation and, for NV, the rates of optical pumping obtained  through the seven-level model. In contrast with the static scenario discussed earlier,  the optical pumping rates are now time-dependent, as the tilt of the NV axis to the magnetic field evolves with time. The full simulation model is described in SI, section~\SIsectheoryspinning.

\begin{figure}[htp]
\centering
\includegraphics[width=0.9\linewidth]{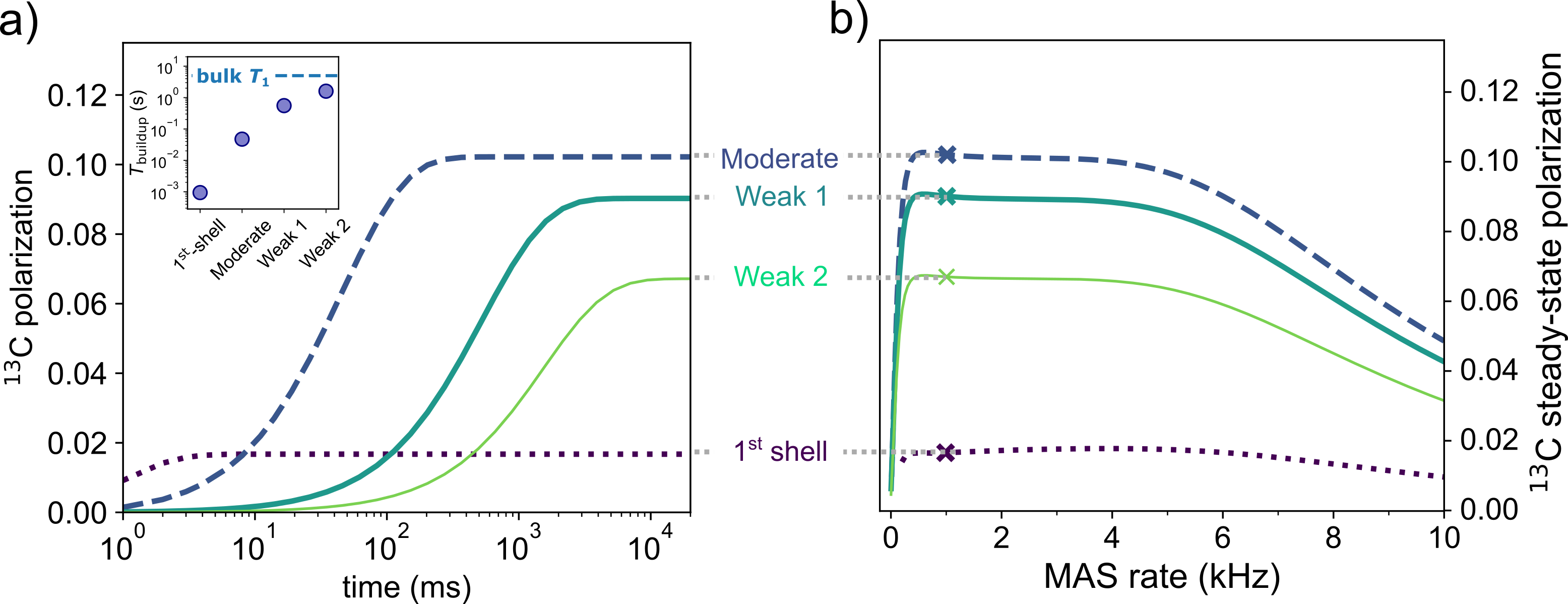}
\caption{\textbf{Simulated $^{13}$C polarization under MAS}. a) Polarization buildup   for a $^{13}$C in a NV-P1-$^{13}$C cluster using the $^{13}$C-NV hyperfine coupling parameters given in Table~\ref{table:13Cgroups}. The simulations were performed for a MAS frequency of \SI{1}{\kilo\hertz},  and a magnetic field of  \SI{7.1}{\tesla}.  The model includes the dynamics at LACs and the orientation dependence of the optical pumping on NV. The represented curves correspond to the $^{13}$C polarization $P_{\mathrm{13C}} = 2\braket{\hat{I}_z}$, where $\hat{I}_z$ is the nuclear spin operator along the direction of the external magnetic field.  
Powder averaging is performed to include all NV orientations ($0 \leq\xi_{\mathrm{NV}} \leq 90^{\circ}$). The polarization is     
 sampled at the initial point of each rotor period for representation.  Inset: corresponding buildup times $T_{\mathrm{buildup}}$ obtained by fitting the curves  with an exponential function. The dashed line indicates the  background polarization decay process, represented in the simulation by the lifetime $T_{1,\rm 13C \,bulk}=\SI{5}{\second}$,  which sets an upper limit to $T_{\mathrm{buildup}}$. Further description of the model and its implementation is given in SI, sect.~\SIsectheoryspinning. b) Corresponding  steady-state $^{13}$C polarization as a function of the MAS frequency.  For each $^{13}$C group, curves are shown only within the range of MAS frequencies for which our model is valid: as discussed in SI, sect. \SIsecLZ, this requires the  MAS frequency to be above a certain threshold (200, 17, 4 and \SI{2}{\hertz}  for the $^{13}$C hyperfine couplings: 1$^{\rm st}$-shell, Moderate,  Weak 1 and Weak 2, respectively). Crosses indicate the steady-state polarization at \SI{1}{\kilo\hertz} MAS rate.    }
\label{fig:sim_13C_buildup_and_steady_pol}
\end{figure}

The simulation results are shown in Fig.~\ref{fig:sim_13C_buildup_and_steady_pol}.  Fig.~\ref{fig:sim_13C_buildup_and_steady_pol}a   shows the simulated polarization buildups for the different $^{13}$C   
at the MAS frequency of \SI{1}{\kilo\hertz} while Fig ~\ref{fig:sim_13C_buildup_and_steady_pol}b shows the simulated steady-state $^{13}$C polarization as a function of the MAS frequency. 
Considering Fig.~\ref{fig:sim_13C_buildup_and_steady_pol}a, it is visible  that nuclei with stronger hyperfine couplings experience faster polarization buildup. Upon decreasing  the hyperfine coupling strength, the buildup time increases and approaches the bulk $^{13}$C $T_1$, set in the simulation to $T_{1,\rm 13C \,bulk}=\SI{5}{\second}$ (inset in Fig.~\ref{fig:sim_13C_buildup_and_steady_pol}a). 
Besides, if we exclude the first-shell $^{13}$C,    stronger hyperfine couplings tend to correlate with a higher steady-state polarization at various MAS frequencies, as visible in  Fig.~\ref{fig:sim_13C_buildup_and_steady_pol}b. 
The first-shell $^{13}$C  exhibits a steady-state polarization substantially below that of the other considered $^{13}$C and therefore escapes the second trend. As we determine, this  specific behavior originates in the strong hyperfine coupling ($A_{XX}, A_{ZZ} > \nu_{\mathrm{13C}}$, where $\nu_{\mathrm{13C}}$ is the bare Larmor frequency), which leads to partial inversion of the energies of the nuclear $\alpha$ and $\beta$ states (in the $m_s = +1$ manifold of NV) and thereby affects the order of the various LACs. This inversion results in negative polarization transfer at certain LACs and ultimately produces  a lower polarization efficiency (see SI, Fig.~\SIfigfirstshelldynamics, for details on the buildup dynamics for the $1^{\mathrm{st}}$-shell $^{13}$C). Last, although the simulations
were made considering a $^{13}$C in the vicinity of NV, we note that  nuclear spins with dominant hyperfine coupling to P1 --rather than NV-- are expected to get polarized with close efficiency.

\section{Discussion}

From the simulated dependence of the steady-state $^{13}$C polarization on the MAS frequency (Fig.~\ref{fig:sim_13C_buildup_and_steady_pol}b), one can distinguish three regions: a first domain in the $0-\SI{0.5}{\kilo\hertz}$ range where a sharp rise of the $^{13}$C polarization occurs for increasing MAS frequency,  a plateau  in the $0.5-\SI{5}{\kilo\hertz}$ range, and a continuous decrease  above $\SI{5}{\kilo\hertz}$.
Several differences between simulation and experiment can be observed, as we now discuss. 
 As a first difference,  the decay  seen in the experiment begins immediately after the polarization  maximum, occuring at $\SI{2}{\kilo\hertz}$ in Sample~A and   $\SI{1}{\kilo\hertz}$ in Sample~B (see  Fig.~\ref{fig:mas_dependency_20pc}b and SI, Fig.~\SIfigmasdepna). That is, no plateau of  $^{13}$C polarization is observed. To explain this behavior, we consider the parameters in the model that determine the $^{13}$C polarization at high MAS frequencies. 
 As pointed out by Thurber and Tycko in the context of conventional MAS-DNP,\cite{thurber2012} the reason for the  polarization efficiency  to 
 decrease at high MAS frequencies is that the sweep rate through the electron–electron LACs becomes too high  for the passages to remain fully adiabatic. The onset of the decay is determined uniquely by the strength of the electron-electron interaction, $a_{\rm ee}$ in our simulations. The value $a_{\rm ee}= \SI{5}{\mega\hertz}$  corresponds to the strength of the   dipolar interaction for an electron pair (\emph{e.g.}, NV-P1) separated by a distance $r \approx \SI{1.7}{\nano\meter}$ ($a_{\mathrm{ee}} \approx \frac{\mu_0  \hbar }{4 r^3} \bar{\gamma}_{\mathrm{NV} }\bar{\gamma}_{\mathrm{P1}}$ -- see SI, section \SIsecfullhamiltonian~for the derivation). Assuming a random distribution of paramagnetic centers,  taking into account the  concentrations given in Table~\ref{table:sample_description}, this separation matches the expected distance between NV and the nearest dark spin in Sample~A, that is  $r_{\mathrm{NN}} = \SI{1.75}{\nano\meter}$ ($r_{\mathrm{NN}} = 0.55 n^{-1/3}$, where $n$ is the volume concentration). Under this assumption that paramagnetic centers are randomly distributed, a possible explanation for the discrepancy to the simulation could be the presence of a distance distribution (leading to a distribution of $a_{\rm ee}$), or the angular dependence of the dipolar interaction  (SI, section  \SIsecfullhamiltonian). As another factor,    specific  charge state dynamics of nearby defects, which can occur for NV-P1 pairs,\cite{manson2018} might lead to an underestimation of $r_{\mathrm{NN}}$ and correspondingly an overestimation of $a_{\rm ee}$ -- possibly explaining as well the discrepancy between experiment and simulation. A second difference  can be noted, with regard to the signal dependence at low MAS frequencies: in the experiments on both Sample A and B, the signal continues rising till the MAS frequency of \SI{1}{\kilo\hertz}, while in the simulations, the plateau is reached at $\sim$\SI{0.5}{\kilo\hertz}. 
As a possible explanation, we note that the polarization transfer dynamics might involve the interaction of NV with further  electron spins beyond the nearest neighbor, and therefore might not be fully captured in our minimal three-spin model.  To address this limitation, further simulations involving realistic spin bath configurations would be required\cite{mentinkvigier2017}  (which are however beyond the scope of this manuscript). 
With the aim of further refining the description, one could monitor as well  the polarization of the \emph{electron spins} (such as NV, P1). Measurements of P1 centers by high-field EPR\cite{nirarad2024} and of NV  by  photoluminescence\cite{wood2018} have been achieved under MAS.
Analyzing simultaneously the EPR and NMR signals  could help investigating more precisely the interaction between  the various paramagnetic species and their role in the polarization transfer to  nuclear spins.

To evaluate the overall efficiency of our technique in comparison to previous reports, we consider the  absolute $^{13}$C polarization upon saturation of the buildup process (at illumination times $d4 \gg T_{\mathrm{pol}}$). At the magnetic field of $\SI{7.1}{\tesla}$ and at  a MAS frequency of \SI{1}{\kilo\hertz}, the enhancement at saturation is $191-$fold for Sample~A and
$368-$fold for Sample~B (see Fig.~\SIfigpolbdp~in SI for comparison). 
We note that these enhancements apply to the polarization in the dark which, under magic angle spinning, can be lower than that predicted by the Boltzmann statistics  owing to the known effect of spinning-induced nuclear depolarization.\cite{thurber2014} 
However, by examining the spectra acquired in the dark, we find that no substantial depolarization occurs at the MAS frequency of $\SI{1}{\kilo\hertz}$. Therefore, we assume $P_{\rm dark}\approx P_{\rm Boltzmann} \approx \frac{h \nu_{\rm 13C}}{2 k_{\rm B} T}$ which gives (at $T=\SI{293}{\kelvin}$) $P_{\rm dark}=\num{6.18e-6}$.  Correspondingly, the absolute light-induced $^{13}$C polarization evaluates to $0.12\%$ and $0.23\%$ for Sample A and B, respectively, that is almost an order of magnitude above the levels reported previously in micro- and nanodiamonds.\cite{blinder2025}

The signal enhancement observed in Sample~B with natural isotopic abundance   exceeds that seen in Sample~A with 20\%  isotopic content. A possible explanation  is the high probability 
of finding at least one $^{13}$C nucleus in the first shell of a given NV center in the 20\% $^{13}$C-enrichment  case. Assuming a random isotopic distribution and taking into account the presence of three equivalent first-shell carbon sites,\cite{felton2009} this probability is $1-(1-0.2)^3 = 48\%$. As we noted from our simulations, the  polarization of these strongly coupled carbons is inefficient. Besides and most importantly, even when a first-shell $^{13}$C is polarized, efficient diffusion of the spin polarization to other $^{13}$C is hindered  by the spin Hamiltonian. 
Indeed, for the polarization to propagate from the first shell to the reservoir constituted by other $^{13}$C, the difference in their Larmor frequency  must be weaker -- or of same order of magnitude -- than the  interaction driving spin diffusion.  
For NV in the $m_S = \pm 1$ state, the first-order effect of the hyperfine interaction leads to a  resonance frequency detuning   between the first-shell  $^{13}$C  and the reservoir in the $100-\SI{200}{\mega\hertz}$ range ($\sim A_{XX}, A_{ZZ}$ of the first-shell).
For NV in the $m_S = 0$ state, although the first-order shift disappears, numerical diagonalization (at the magnetic field of  \SI{7.1}{\tesla})  shows that  the second-order effect of the hyperfine interaction leads to detunings exceeding $\SI{100}{\kilo\hertz}$. This difference in resonance frequencies cannot be overcome by the weak ($<\SI{5}{\kilo\hertz}$) dipolar interaction driving spin diffusion.
Such behavior, where a nuclear spin acts as polarization `sink' for NV, already  observed with  the  $^{14}$N nucleus\cite{blinder2025}, can therefore also occur with a  $^{13}$C in the first shell.  
However, in the present experiments, one can not exclude part of the lower enhancement seen in Sample~A to originate from a weaker efficiency of 
 NV initialization, indeed, the sample consists of particles in the \SI{0.2}{}-\SI{1}{\micro\meter} size range, which act as stronger  scatterers of  light compared to the bigger particles ($1.5-\SI{2.5}{\micro\meter}$)  in Sample~B, which leads potentially to less homogeneous illumination. Owing to the multiple factors determining the enhancement, a precise determination of the impact of isotopic enrichment is left for further study.

We now discuss potential applications of our technique. Since NVs are known to be stable in diamond down to $\sim \SI{5}{\nano\meter}$-sized particles,\cite{terada2019} our protocol is amenable to powder ensembles with a higher surface-to-volume ratio. 
The present technique might thus be beneficial for enhancing the NMR signal from nuclear spins in external analytes. 
So far, substantial enhancement factors for nuclear spins located outside nanodiamonds have been demonstrated only using microwave-driven DNP at cryogenic temperatures \cite{kato2023} -- while, at room temperature, only modest enhancement factors were obtained.\cite{rej2016}

Assuming $^{13}$C inside the diamond lattice are polarized through the dynamics involving NV-P1-$^{13}$C clusters, a fraction of the $^{13}$C polarization is then expected to diffuse towards the surface. 
Once the polarization reaches near-surface $^{13}$C spins, it could in principle be transferred to external nuclear spins by a cross-polarization (CP) -based mechanism, for example first to nearby $^1$H nuclei and subsequently to other heteronuclei through an additional CP step. The important size ($>\SI{200}{\nano\meter}$) of the particles used in the present work is, however, not favorable to this application. To estimate a more favorable particle size, we note that    the nuclear polarization can spread around the electron centers over a  characteristic spin-diffusion length $ l_{\mathrm{d}} \sim \sqrt{D T_{\mathrm{pol}}} $, where $D$ is the spin-diffusion coefficient, $T_{\mathrm{pol}}$ the buildup time. When spin diffusion is driven by homonuclear dipole-dipole interaction, $D$ can be estimated as  $D=\frac{\Delta \nu_{\mathrm{dd}} a^2}{30}$, where $\Delta \nu_{\mathrm{dd}}$ is the dipolar linewidth,\cite{parker2019}  and $a$  the average spacing between $^{13}$C, which can be deduced from the concentration as $a=n^{-1/3}$. Considering the parameters of Sample~A at $\SI{1}{\kilo\hertz}$ MAS frequency, $T_{\mathrm{pol}} \approx \SI{5}{\second}$, $\Delta \nu_{\mathrm{dd}} \approx \SI{2.5}{\kilo\hertz}$ (we take the parameters from Fig.~\ref{fig:buildup_dynamics}, as remarked earlier, the full linewidth is mostly determined by the $^{13}$C-$^{13}$C interaction) and $a=\SI{3.05}{\angstrom}$, one obtains $l_{\mathrm{d}} \approx \SI{6}{\nano\meter}$. This value suggests that particles of size $\lesssim \SI{30}{\nano\meter}$, where the majority of  $^{13}$C nuclei would be located within the distance $l_{\mathrm{d}}$ from the surface, would be favorable for this protocol. Alternatively, rather than using the diamond $^{13}$C spins as intermediate polarization reservoir, one could attempt to directly polarize $^1$H spins in  the surface moeities, by using   subsurface paramagnetic species with strong (up to \SI{10}{\mega\hertz}) hyperfine coupling as dark spins.\cite{yavkin2019,sushkov2014} For discussion on the parameters determining the efficiency of this method, see SI, sect.~\SIsecextspinhp. By operating in the standard setting of solid-state NMR (using MAS), such protocols --if feasible--  would combine polarization enhancement with  improved chemical resolution, a feature that would be beneficial for the analysis of external compounds.

Alternatively, the improved resolution provided by MAS might also be exploited in the development of magnetic field sensors based on hyperpolarized nuclear spins. In this context, optimizing the magnetic field sensitivity requires in general extending the nuclear spin coherence, which so far was achieved through the use of RF homonuclear decoupling.\cite{sahin2022} In our experiments, the use of MAS provides not only the ability to perform \emph{in-situ} hyperpolarization but also achieves  decoupling of the homonuclear interactions (which, as we noted explains the observed behavior of the NMR linewidth upon increase of the MAS frequency). It can therefore be expected that replacing our simple $90^{\circ}$ readout pulse by methods of RF decoupling adapted to  MAS --known as Combined Rotation and Multiple Pulse Spectroscopy, or CRAMPS-- could lead to an improved sensitivity in both DC and AC magnetic field sensing.

Last, we note that  our hyperpolarization protocol can also be extended to solids hosting defects with properties comparable to those of NV. 
It is known that molecules with a photoexcited triplet -- such as pentacene -- can be introduced either in  molecular crystals (such as naphthalene or \emph{para}-Terphenyl)\cite{henstra1990,tateishi2014} or in glassy matrices,\cite{miyanishi2021} 
and can provide an analogue of NV in these systems. Similarly, one can generally introduce stable $S=1/2$ radicals (e.g. trityl, nitroxide), which would provide an analogue to P1. In fact, the co-introduction of pentacene derivatives and $S=1/2$ radicals has already been demonstrated  in various glassy matrices in tethered form,\cite{avalos2020}
 which suggests that their introduction as separate species   --  providing a counterpart to NV and  P1 in diamond -- would also be possible. 
 Compared to the approach based on `all-in-one' molecular structures (such as  
some flavoproteins\cite{ding2020})
and synthetic molecular structures 
 (such as Photopol\cite{debiasi2024jacs}), using separate species could be advantageous due to the degree of tunability it offers. 
 For instance, one could think of engineering nanoparticles   where one of the electron spin carriers is designed to be located preferentially near the surface, in order to achieve targetted   hyperpolarization of nuclear spins in that region.

\section{Conclusion}

We have demonstrated microwave-free $^{13}$C hyperpolarization in randomly oriented diamond particles under high-field MAS conditions. Light-induced polarization enhancements of up to $280-$fold in the  sample with isotopic enrichment and $411-$fold in the natural-abundance sample were observed, with steady-state absolute $^{13}$C polarization levels exceeding 0.1\% under continuous illumination.
Our protocol leverages  light-induced spin-polarization of NV  combined with periodic polarization transfer at level anticrossings. 
A minimal three-spin model considering NV, P1 and $^{13}$C,  where P1 denotes a substitutional nitrogen in diamond,  captures qualitatively the dependence of the $^{13}$C polarization as a function of the MAS frequency. 
Operating in the standard geometry of solid-state NMR experiments, with the spinning axis set at the magic angle relative to the magnetic field, mitigates the anisotropy of the NV spin Hamiltonian 
as more than  $99.9\%$ of NV orientations meet the geometrical requirement for polarization transfer.
Further developments of the protocol are expected to enable the hyperpolarization of nuclear spins external to the diamond particles, opening opportunities, for instance,  for high-sensitivity NMR at room temperature.

\clearpage

\section{Experimental}

The NMR measurements were performed using \SI{7.1}{\tesla} and \SI{9.4}{\tesla} Avance NMR spectrometers (Bruker Biospin GmbH, Ettlingen, Germany) equipped with 4 mm custom MAS probes operating at the respective $^{13}$C Larmor frequencies of \SI{75.51}{\mega\Hz} and \SI{100.62}{\mega\Hz}. To enable optically pumped nuclear hyperpolarization in this  setup, stable and efficient sample illumination had to be implemented. For this purpose, a fiber  is used to transfer the radiation from an external laser arrangement into the stator of the MAS probe. 
The fiber enters the probe from the bottom, passing through the body of the  probe until it reaches the stator where it is mounted. In Figure \ref{fig:setting_overview}a, the arrangement with the optical fiber mounted to the stator and the sample located inside the NMR coil is shown. To allow light to reach the sample, the coil windings are spaced at the illumination site, reducing light absorption and minimizing scattering \cite{pompe2020thesis,daviso2008book}. 

To prepare the samples, for the measurements under illumination, a mixture of diamond powder (in quantity \SI{0.29}{\milli\gram} for Sample~A and  \SI{0.39}{\milli\gram} for the Sample~B) and  UV-curing polymer (NOA65, Thorlabs) was deposited on the inner wall of a  quartz tube of \SI{3}{\milli\meter} outer diameter. The quartz tube with the diamond powder was inserted into a \SI{4}{\milli\meter} sapphire rotor (Bruker Biospin GmbH). To stabilize the diamond powder and polymer mixture, UV illumination (\SI{365}{\nano\meter}, \SI{1}{\watt}) was applied during approximately one minute.  To improve the efficiency of illumination in the NMR experiment, a small quantity of index-matching gel (Thorlabs G608N3) was inserted in the gap between the inner walls of the rotor and the outer walls of the \SI{3}{\milli\meter} insert. Remaining empty spaces within the rotor were filled with SiO$_2$ particles (\SI{150}{\nano\meter}, non-porous, Sigma-Aldrich) enabling  rotor balance, required for spinning at $\SI{}{\kilo\hertz}$ frequency. To enable acquisitions in the absence of illumination despite the weak signals, dark acquisitions were generally performed on a higher mass of the respective samples ($m=\SI{12}{\milli\gram}$ and $\SI{39}{\milli\gram}$ for Sample~A and B, respectively). In this case, the powders were packed into a \SI{3}{\milli\meter} Kel-F insert fitting a \SI{4}{\milli\meter} ZrO$_2$ rotor. The signal was subsequently normalized to the mass, to allow comparison to the data taken under illumination. This normalization is made possible by the fact that in all measurements, the powder is located in a region of high $B_1$-field homogeneity of the NMR coil, therefore, the signal acquired under dark conditions scales only with the mass. 
In all experiments, the rotor is spun along an axis tilted by $\alpha_m = 54.74^\circ$ to the magnetic field, as shown in Figure~\ref{fig:setting_overview} (\emph{i.e.}, following the standard geometry of MAS experiments).   The spinning frequency was varied between 
0 (static) and 6 kHz.  
The spinning rate is regulated by adjusting manually the bearing and drive gas pressure. While stable, deviations of ±30 Hz may occur.

The NMR sequence  consists of the polarization extinction (PE), polarization generation (PG) and readout parts (Fig.  \ref{fig:setting_overview}b). The optimized 90$^\circ$ hard pulse for $^{13}$C was \SI{10}{\micro\second}. 
The polarization extinction (PE) block~\cite{daviso2008}, consisting  of a train of at least three 90$^\circ$ pulses, is implemented before the PG block to eliminate any residual coherence and polarization from the previous scan, thereby ensuring a well-defined initial spin state with no constraint on the repetition rate of the experiments (Fig.  \ref{fig:setting_overview}b).

For the polarisation generation (PG) with light, different laser systems were used. In the experiments performed at $\SI{9.4}{\tesla}$ magnetic field (Fig.~\ref{fig:lambda_variation}), optical excitation was provided by three continuous-wave lasers: a Genesis MX488-1000 STM OPSL (Coherent) emitting at 488~nm, and two diode lasers from Ultralasers (Newmarket, Canada), MGL-FN-532 operating at 532~nm and MDL-III-445-1000mW emitting at 445~nm. The laser power at the optical fiber output was adjusted to approximately $\sim650$~mW for all wavelengths. For the experiments  
performed at $\SI{7.1}{\tesla}$ magnetic field, $\SI{532}{\nano\meter}$ illumination, a   diode-pumped solid-state laser (LRS-0532-PFN-03000-01, Laserglow Technologies, Toronto, Canada). The laser power was adjusted to a value of \SI{2}{\watt} at the light fiber output.

 In the experiments performed at \SI{9.4}{\tesla} (Fig.~\ref{fig:lambda_variation}), the temperature was regulated to \SI{293}{K} using a  SmartCooler$^{\mathrm{TM}}$ BCU II (Bruker).
  The experiments at the magnetic field of \SI{7.1}{\tesla} were performed without active temperature regulation. An upper bound to the laser-induced sample heating could be determined by analysis of the NV zero-phonon-line (ZPL) in the photoluminescence (PL) spectrum. PL spectra of the (spinning) sample inside the NMR probe  were acquired using an Ocean Optics HR4 UV-Vis spectrometer. The ZPL from the spectra was fitted, allowing to extract the wavelength of the   maximum $\lambda_{\rm ZPL}$ at two laser powers (see SI, Fig.~\SIzplanalysis). Knowing the temperature variation coefficient of the ZPL energy of \SI{46}{\micro\electronvolt\per\kelvin} \cite{li2017}, the heating coefficient could be determined to be below $\SI{10}{\kelvin\per\watt}$ yielding a maximum laser-induced temperature gain  under continuous \SI{2}{\watt} illumination of $\SI{20}{\kelvin}$ (see SI, sect. \SIsecNVzpl). We note that the strong air flows (bearing and drive) required for spinning the sample  limit sample heating. The non-spinning experiments were performed without air flow, consequently, the heating under steady illumination might be stronger. 
To reduce heating effects, the repetition delay in all the experiments performed at \SI{7.1}{\tesla} was adjusted to ensure an illumination duty cycle below 50\% (over one acquisition scan), with the exception of  the polarization buildup  measurements (Fig.~\ref{fig:buildup_dynamics}) where light was present all the time.

NMR data acquisition and processing were carried out using TopSpin 4.5.0 (Bruker), while spectral analysis and visualization were done with MestReNova 14.1.0 (Mestrelab Research S.L.) or Python.

For characterization with EPR, we used  a X-band EPR system (Bruker ELEXSYS II E580) with an ER4122 SHQE cavity, and the software xEPR for acquisition. The measurements were made at a microwave frequency $ \sim \SI{9.85}{\giga\hertz}$, with a resonance quality factor $Q \sim 8000$. Determination of the spin concentrations was performed using the built-in spin-counting feature from the acquisition software. 

Numerical simulations of the energies, spin dynamics of NV-P1-$^{13}$C clusters and the optical pumping of NV  were performed with Python, using the package Qutip.

\begin{acknowledgement}

J. M. acknowledges the Deutsche Forschungsgemeinschaft (DFG, German   Research Foundation) – TRR-386 (HYP*MOL) – A3 (project number   514664767). 
F. J acknowledges the German Federal Ministry of Research, Technology and Space (BMFTR) via future cluster QSENS   and projects EXTRASENS (13N16935 ) DIAQNOS (13N16463), quNV2.0 (13N16707),   the DFG via projects   387073854, 445243414, ,  491245864, 546850640, 560722984  and joint DFG/JST ASPIRE program via project  554644981 and JPMJAP24C1, European Union's HORIZON Europe program via projects CQuENS (101135359), QCIRCLE (101059999) and FLORIN (101086142), QuantERA ERA-NET Cofund in Quantum Technologies via AQuSeND (532771161) and MultiQS (583792014) projects, European Research Council (ERC) via Synergy grant HyperQ (856432),   and Carl-Zeiss-Stiftung  via Research Center QPhoton and project Ultrasens-vir.
J. T acknowledges  funding from the Okinawa Institute of Science and Technology (OIST);   support from the Scientific Computing and Data Analysis (SCDA) section  at OIST and the use of the DEIGO supercomputing cluster; and funding from Japan Science and Technology Agency (JST) as part of Adopting Sustainable Partnerships for Innovative Research Ecosystem (ASPIRE), Grant No. JPMJAP24C1.
  R. W. acknowledges the DFG  under Germany’s Excellence Strategy—EXC 2154—Project number 390874152, as well as ETAG (Estonian Research Agency) and Tallinn University of Technology (TalTech) under projects PRG1832 and STP57.

Personal acknowledgements: R. B. and F. J. acknowledge Yuliya Mindarava for assistance with sample characterization by EPR and for help with   the NV photoluminescence experiment, as well as Wolfgang Knolle and Christian Laube  for assistance with electron irradiation and sample surface treatment. R. W. acknowledges Mai-Liis Org and Ago Samoson for technical support with the NMR detector and engineering purchases. 

\end{acknowledgement}

\section*{Author contributions}

R.B. and G. M. performed the experiments at the magnetic fields of \SI{7.1}{\tesla} and \SI{9.4}{\tesla}, respectively. A. N., D. K. and J. T. performed the simulations in the non-spinning case and wrote the description of the optical pumping model. R. B. performed the simulations in the spinning case. V. A. assisted with the synthesis of the isotopically enriched sample. R. W.  provided the NMR probe used for the experiments performed at the magnetic field of \SI{7.1}{\tesla}. G. M., J. M,  R. B. and F. J. coordinated the project.  F. J., J. M., J. T. and R. W. secured funding.  All authors contributed to the manuscript. 



\bibliography{bib_spinNV}



\section*{Supplementary material}

# Supporting Information, to: Microwave-Free $^{13}C$ Hyperpolarization of Diamond Particles Enabled by Magic Angle Spinning and NV Centers

Rémi Blinder,[1,*], Guzel Musabirova,[2] Daehee Kim,[3] Anshuman Nayak,[3] Jason Twamley, [3] Viateschlav N. Agafonov,[4] Raiker Witter,[1] Jörg Matysik,[2], Fedor Jelezko,[1,5]
[1]Institut für Quantenoptik, Albert-Einstein Allee 11, Universität Ulm, 89081 Ulm, Germany, [2]Institut für Analytische Chemie, Universität Leipzig, Linnéstrasse 3, 04103 Leipzig, Germany, [3]Quantum Machines Unit, Okinawa Institute of Science and Technology Graduate University, Onna, Okinawa 904-0495, Japan, [4]GREMAN UMR 7347, University F. Rabelais, 37200 Tours, France, [5]Centre for Integrated Quantum Science and Technology (IQST), Ulm 89081, Germany

*To whom correspondence should be addressed; E-mail: remi.blinder@uni-ulm.de.

# Contents

## S1 EPR

The continuous-wave EPR acquisitions for the 20% $^{13}$C and 1.1% $^{13}$C samples are represented in Figure S1.1. Spectra taken in the $g \sim 2$ region (left panels) show contributions coming from P1 centers and other dark spins, while spectra taken in the half-field region (right panels) show the NV signal [1, 2].

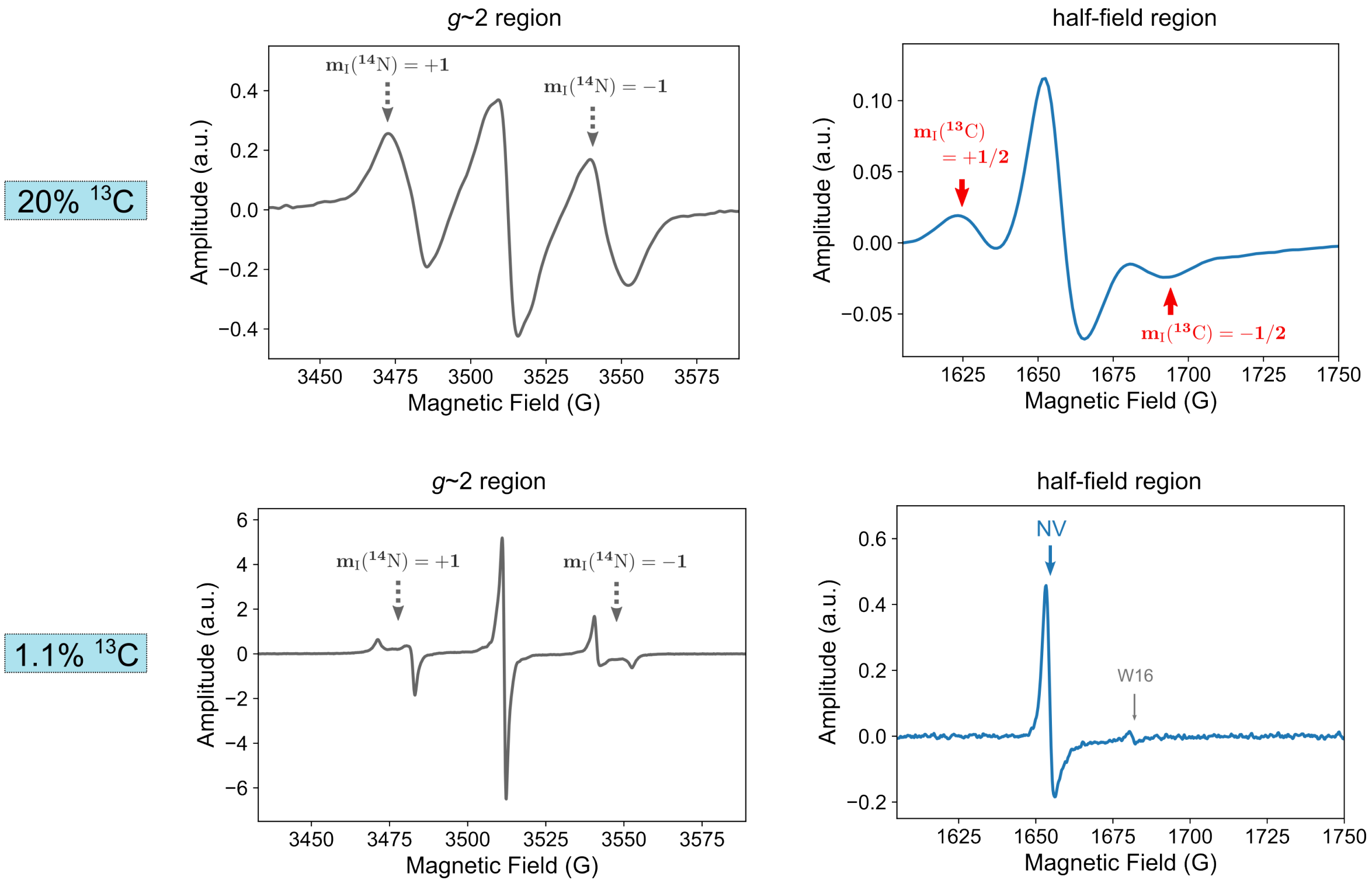


Fig. S1.1: **CW-EPR spectra of the 20% $^{13}$C and 1.1% $^{13}$C sample.** Left panels: spectra in the $g \sim 2$ region corresponding to P1 centers and other dark spins. The hyperfine lines of P1 with the respective state of the $^{14}$N nuclear spin are indicated by arrows. Right panels: spectra of NV in the half-field region. For the 20% $^{13}$C sample, arrows indicate hyperfine satellites corresponding to one $^{13}$C present in the first shell around NV. For the 1.1% $^{13}$C sample, an additional small contribution is observed, which is attributed to the W16 defect [2]. The microwave frequency in the measurements was 9.847 GHz, the parameters { microwave power attenuation, modulation amplitude} were: { 54 dB , 1 G} ($g \sim 2$ reg., 20% $^{13}$C), { 60 dB, 1 G} ($g \sim 2$ reg., 1.1% $^{13}$C), { 39 dB, 5 G} (half-field, 20% $^{13}$C), { 45 dB, 2 G} (half-field, 1.1% $^{13}$C).

To investigate the different contributions in the $g \sim 2$ region, fits were performed using the free Matlab package Easyspin (6.0.0) [3] (Fig. S1.2). Prior to the fitting, first integration was performed. The fitting is then made by simulating the first integral with Easyspin and minimizing the residual sum of squares. This procedure allows a more accurate estimation of the weight of the broad spectral features. Resulting fit parameters are given in Table S1.1.

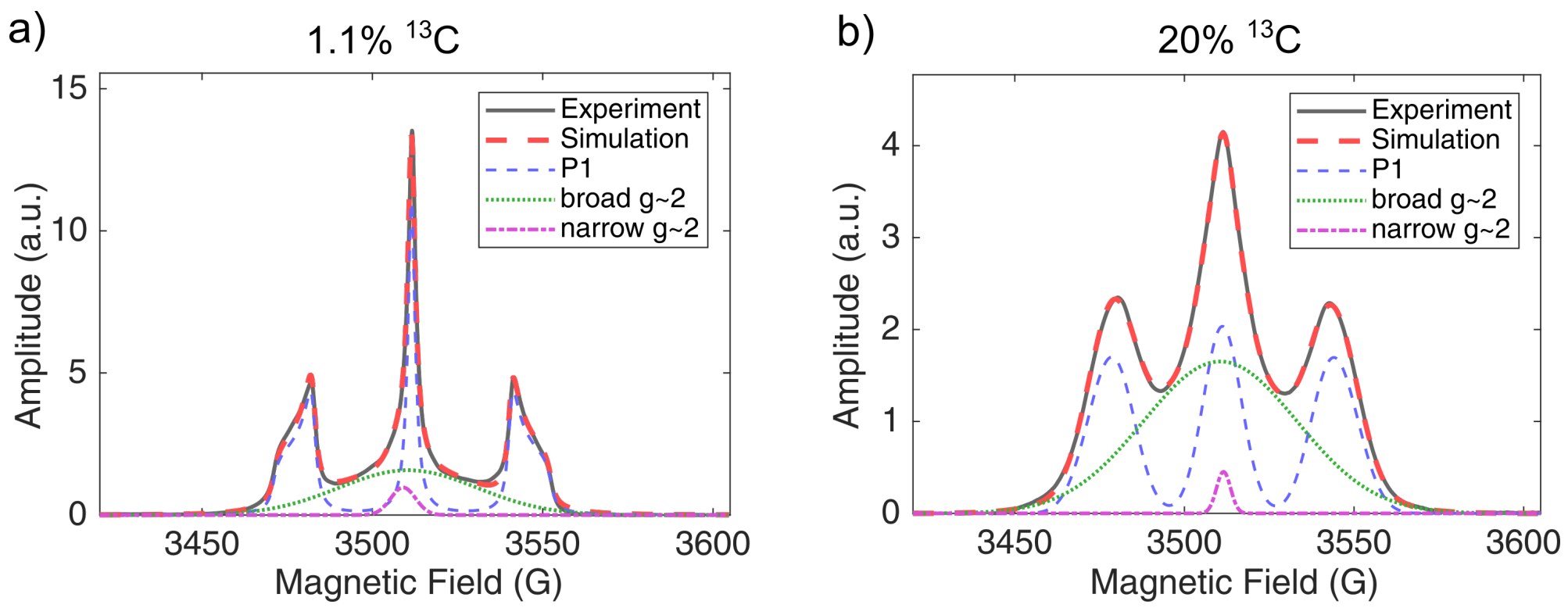


Fig. S1.2: **Fits of the first integral of the spectra in the $g \sim 2$ region** The fit includes three components: isolated P1 centers, and other dark spins divided into a broad and a narrow component. The fit was applied to the 1.1% $^{13}$C sample (a) and the 20% $^{13}$C sample (b), using Easyspin [3].

| sample | component | $g$-factor | $\Delta_{\text{Gauss}}$ (mT) | $\Delta_{\text{Lorentz}}$ (mT) | weight |
|---|---|---|---|---|---|
| 1.1% $^{13}$C sample | P1 (isolated) | 2.00240(2) | – | 0.250(7) | 0.60(1) |
| | broad | 2.0033(1) | 4.8(2) | – | 0.36(1) |
| | narrow | 2.0040(2) | 0.86(6) | – | 0.039(6) |
| 20% $^{13}$C sample | P1 (isolated) | 2.00240(2) | 1.320(6) | – | 0.472(4) |
| | broad | 2.00294(1) | 5.33(1) | – | 0.515(5) |
| | narrow | 2.00243(2) | 0.49(1) | – | 0.0130(3) |

Table S1.1: **Fitted parameters for the dark spins appearing in the $g \sim 2$ spectral region.** Isolated P1 centers are fitted with the Hamiltonian including the $^{14}$N coupling parameters, $A_{\parallel} = 114\,\text{MHz}$, $A_{\perp} = 81.3\,\text{MHz}$ [4]. The contribution from the remaining spins is divided into a 'broad' and a 'narrow' component, where the broad part could contain contribution from P1 present as clusters. $\Delta_{\text{Gauss}}$ and $\Delta_{\text{Lorentz}}$ are the linewidths corresponding to the full width at half maximum of the broadening function, taken as Gaussian or Lorentzian, depending on the spin species, to ensure the best fit. These parameters provide the fitted curve represented in Fig S1.2.

## S2 Additional datasets

### S2.1 NMR spectra, polarization buildup and $T_1$

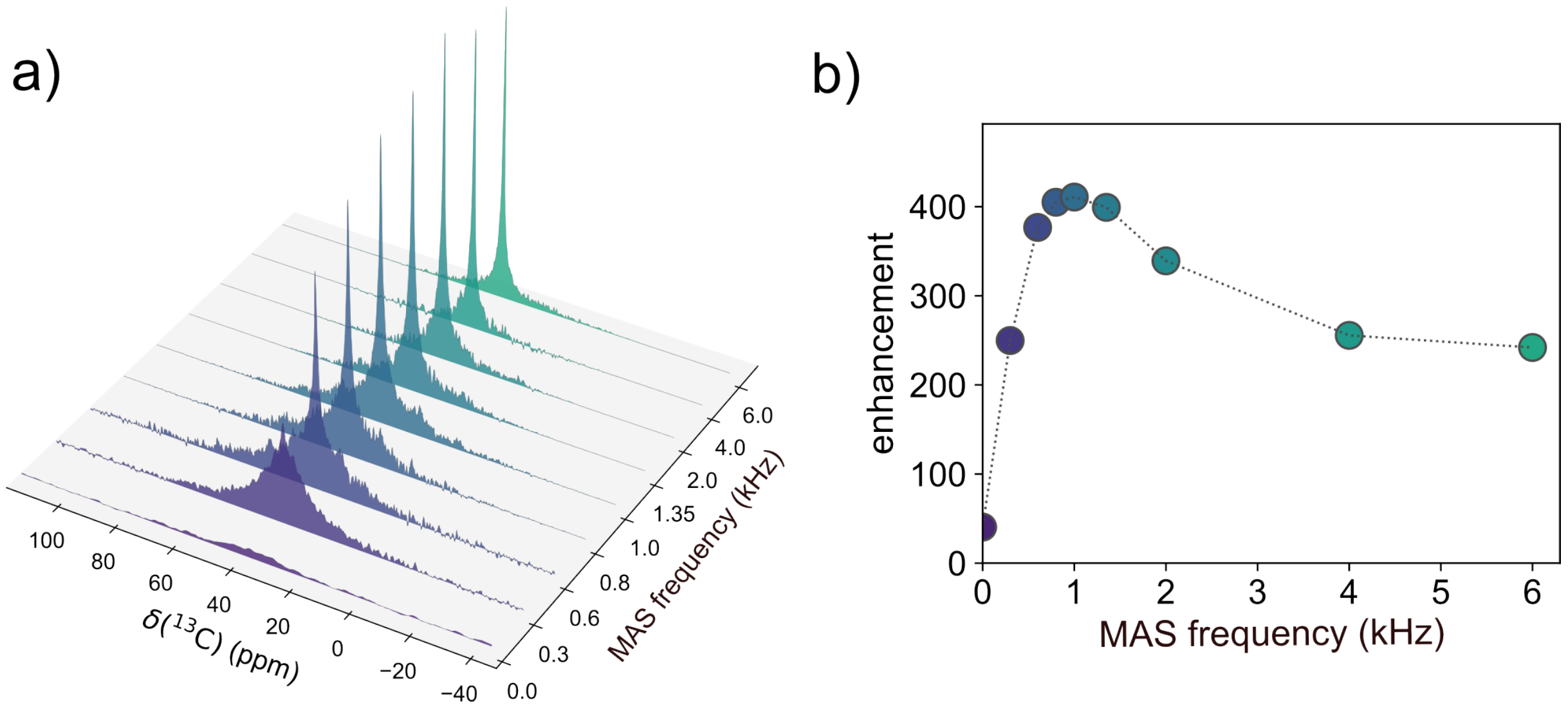


Fig. S2.1: **Light-induced hyperpolarized $^{13}$C signal at varying spinning speeds in the 1.1% $^{13}$C sample.** a) Light-induced signals varying the spinning speed between 0 and 6 kHz b) Enhancement factor as a function of the spinning speed. The data was acquired at 7.1 T magnetic field with $d4 = 120$ s of illumination time, at the wavelength 532 nm.

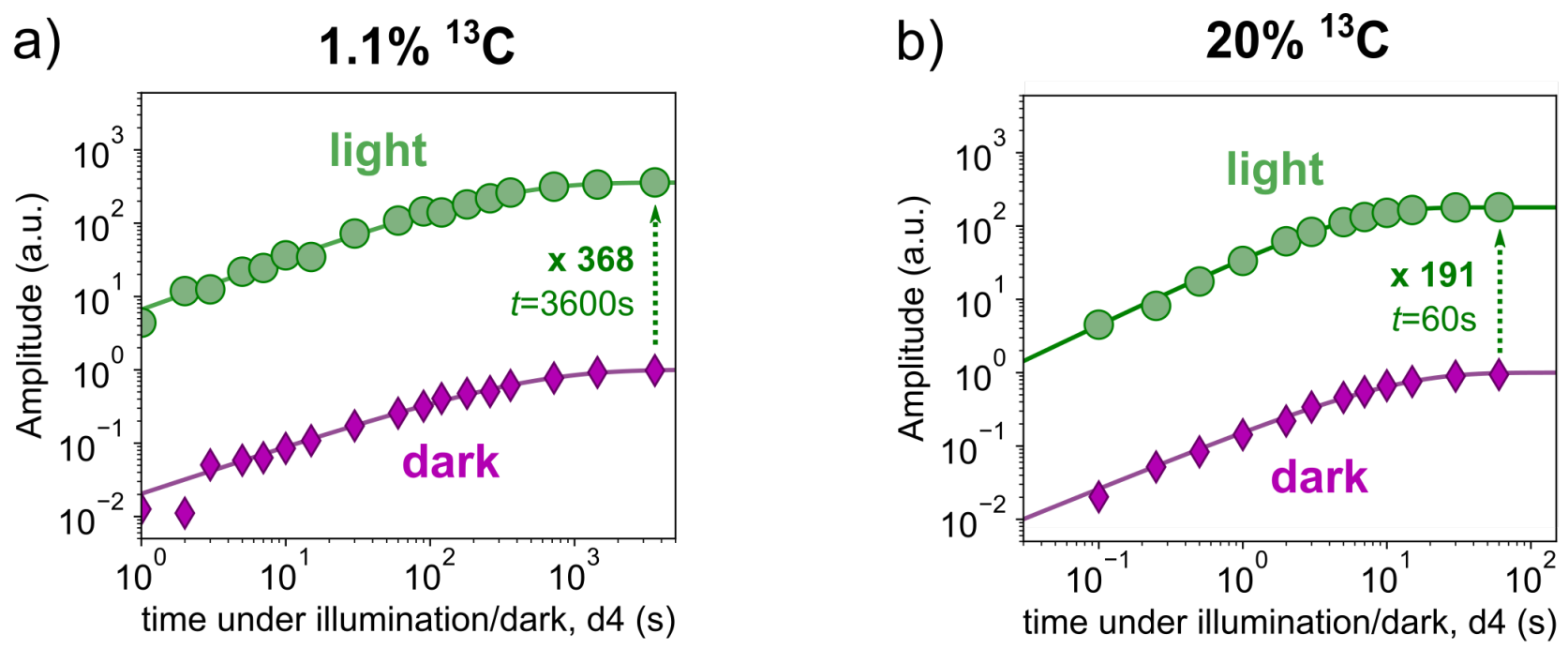


Fig. S2.2: **Comparison of polarization buildup curves under illumination and $T_1$ recovery curves in the dark, for determination of the absolute polarization under illumination in the steady-state, at 1 kHz spinning speed and 7.1 T field. (a) 1.1% sample and (b) 20% $^{13}$C sample.** The steady state enhancement is taken at $d4 = 3600$ s in the 1.1% $^{13}$C sample and $d4 = 60$ s in the 20% $^{13}$C sample, where both the curves of saturation-recovery in the dark and of buildup induced by light have reached saturation. Continuous lines are stretched exponential fits.

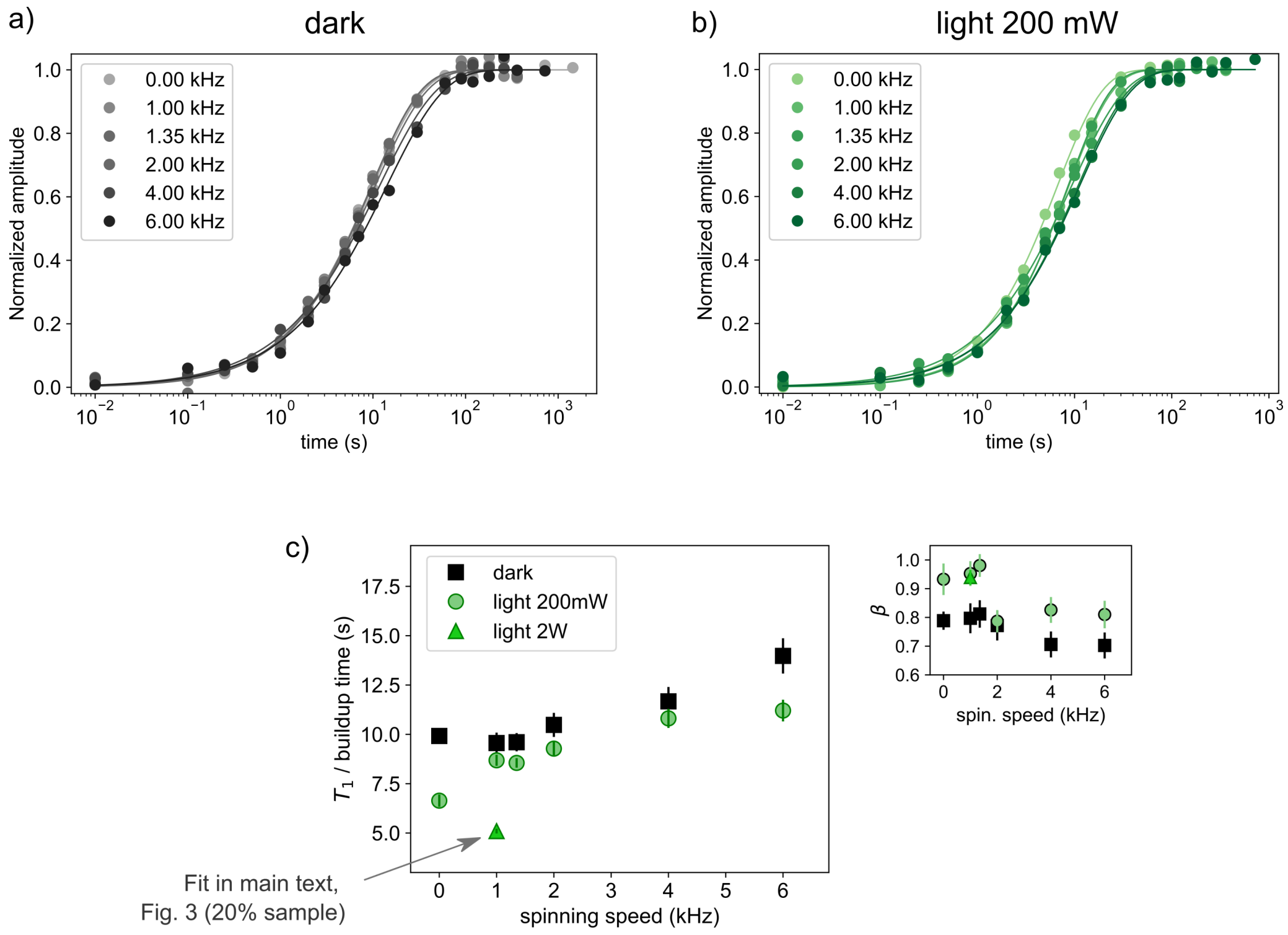


Fig. S2.3: **Comparison of $T_1$ and light-induced polarization buildup dynamics, in the 20% $^{13}$C sample at different spinning speeds ($7.1\,\mathrm{T}$ field).** a) Saturation-recovery $T_1$ measurements (dark) at different spinning speeds. b) Polarization buildup under green illumination at 200 mW laser power – lower than in the experiments discussed in main text. In a) and b), continuous lines are fits with a stretched exponential function $M(t) = M_\infty \cdot \left\{1 - \exp\left(-(t/T_{1,\mathrm{pol}})^\beta\right)\right\}$. c) Left panel: fitted values of $T_1$ and $T_{\mathrm{pol}}$ as a function of the spinning speed (squares: $T_1$, circles: $T_{\mathrm{pol}}$ under 200 mW illumination, triangle: $T_{\mathrm{pol}}$ under 2 W illumination as in main text, Fig. 4). Right panel: Stretch exponent $\beta$ extracted from the fits.

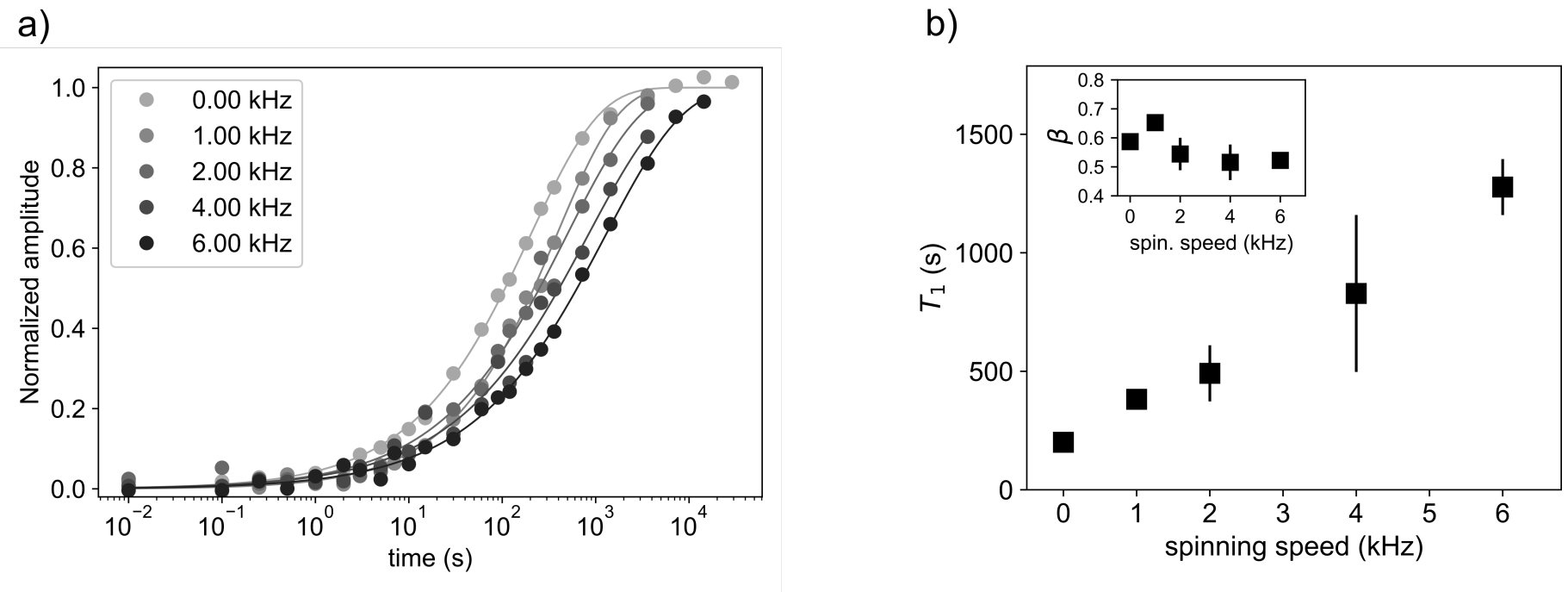


Fig. S2.4: $T_1$ **recovery curves in the 1.1% $^{13}$C sample at different spinning speeds** (7.1 T **field, experiments in the dark only), with fits.** a) Experimental data for $T_1$ by saturation-recovery. Continuous lines are stretched exponential fits. b) Corresponding fitted $T_1$ and stretch exponent $\beta$ (inset) as a function of the spinning speed.

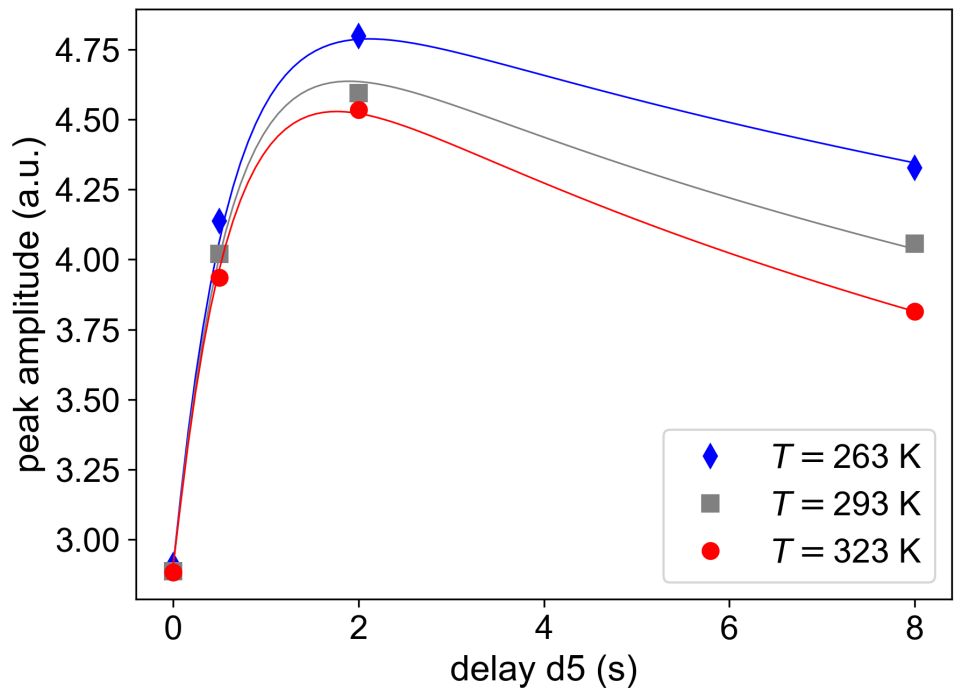


Fig. S2.5: **Peak amplitude of the NMR signal as a function of the delay** $d5$ **in the spin-diffusion experiment, at varying temperature (20% $^{13}$C sample).** To evidence the separation between the time constants describing the signal rise (spin diffusion) and decay ($T_1$), the same experiment protocol as in main text, Fig. 5 was repeated with the temperature regulated at $T = 263, 293, 323\,\mathrm{K}$. The experiment was performed at 9.4 T magnetic field, using 532 nm illumination at laser power 650 mW ($d4 = 0.5\,\mathrm{s}$). Continuous lines represent a fit with the phenomenological function $A \cdot \left(1 - \exp(-\frac{d4+d5}{T_D})\right) \cdot \exp\left\{-(\frac{d4+d5}{T_1^{\mathrm{pk}}})^\beta\right\}$. The fit was performed globally, with only the parameter $T_1^{\mathrm{pk}}$ allowed to change across datasets. Resulting parameters are $T_D = 0.63(2)\,\mathrm{s}, T_1^{\mathrm{pk}}[263\,\mathrm{K}] = 112(18)\,\mathrm{s}, T_1^{\mathrm{pk}}[293\,\mathrm{K}] = 67(10)\,\mathrm{s}, T_1^{\mathrm{pk}}[323\,\mathrm{K}] = 50(7)\,\mathrm{s}, \beta = 0.60(8)$. In these fits, as the peak amplitude is considered, the label $T_1^{\mathrm{pk}}$ is used, to differentiate from the $T_1$ extracted when considering the integral over the complete spectrum (in general, these two quantities are different).

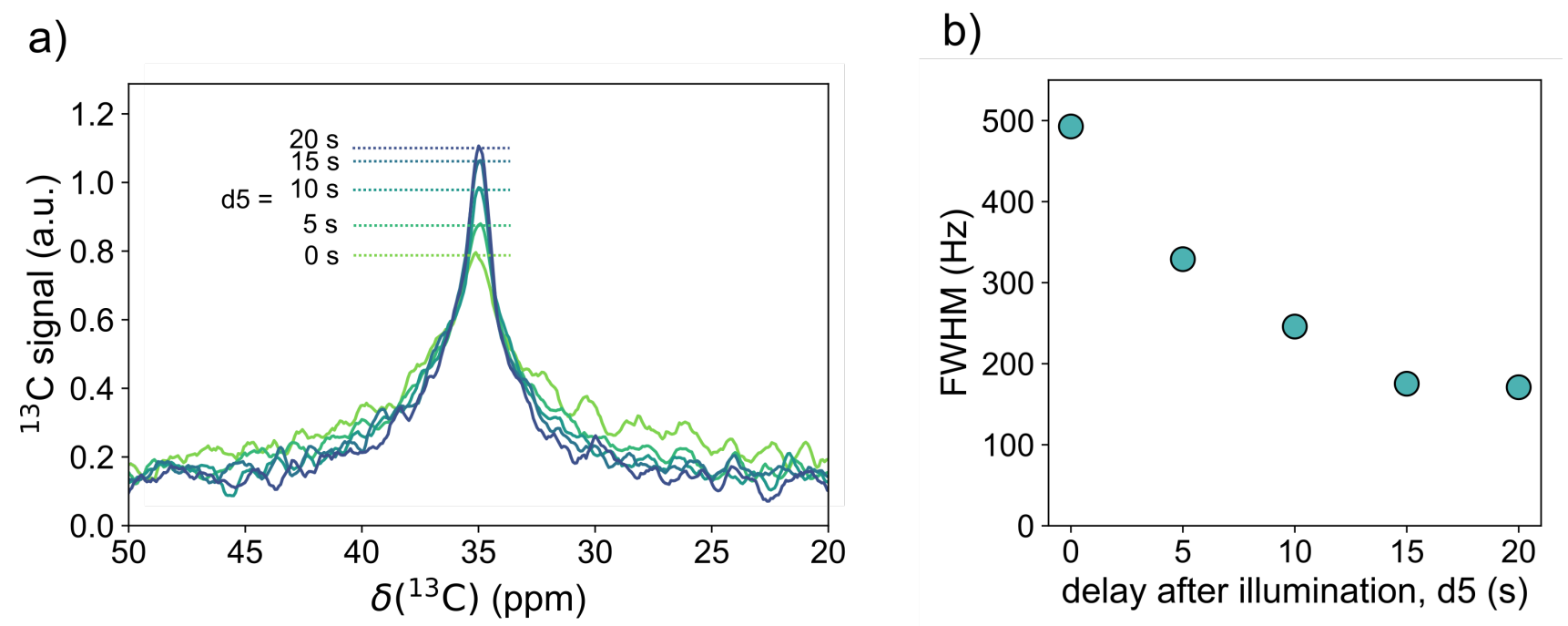


Fig. S2.6: **Spin-diffusion experiment at natural $^{13}$C abundance.** The protocol used in main text, Fig. 5 is repeated for Sample B with 1.1% $^{13}$C. a) Light-enhanced $^{13}$C NMR spectrum as a function of the post-illumination delay $d5$, which is varied between 0 and 20 s b) Corresponding FWHM of the NMR line. The experiment was performed at a MAS frequency of 1 kHz, using an illumination time $d4 = 10$ s.

## S2.2 NV zero-phonon line analysis for quantification of laser-induced heating

The coefficient of laser-induced heating in the experiments performed without temperature regulation (*i.e.*, at the magnetic field of 7.1 T) was estimated by analysis of the NV zero-phonon-line (ZPL) in the photoluminescence spectra acquired with Sample A in the NMR probe (Fig. S2.7). Fit of the background-subtracted data with a bigaussian function ($\propto e^{-(\lambda-\lambda_{\mathrm{ZPL}})^2/2w_1^2}$ $[\lambda < \lambda_{\mathrm{ZPL}}]$; $e^{-(\lambda-\lambda_{\mathrm{ZPL}})^2/2w_2^2}$ $[\lambda > \lambda_{\mathrm{ZPL}}]$) yields the maximum $\lambda_{\mathrm{ZPL}}$ at the laser powers of 1 and 2 W (Fig. S2.7b). The ZPL energy is known to depend on the sample temperature $T_{\mathrm{S}}$ according to a coefficient of 46 µeV K$^{-1}$ around room temperature [5], corresponding to $\frac{\partial\lambda_{\mathrm{ZPL}}}{\partial T_{\mathrm{S}}} = 0.015\,\mathrm{nm\,K^{-1}}$. Assuming a linear response of the change in $T_{\mathrm{S}}$ when the laser power $P_{\mathrm{las}}$ is increased, we obtain as heating coefficient:

$$\frac{\partial T_{\mathrm{S}}}{\partial P_{\mathrm{las}}} = \frac{[(638.430 \pm 0.036) - (638.366 \pm 0.066)]\,\mathrm{nm}}{(2\,\mathrm{W} - 1\,\mathrm{W}) \times (0.015\,\mathrm{nm\,K^{-1}})} = 4.2 \pm 5.0\,\mathrm{K\,W^{-1}} \tag{S1}$$

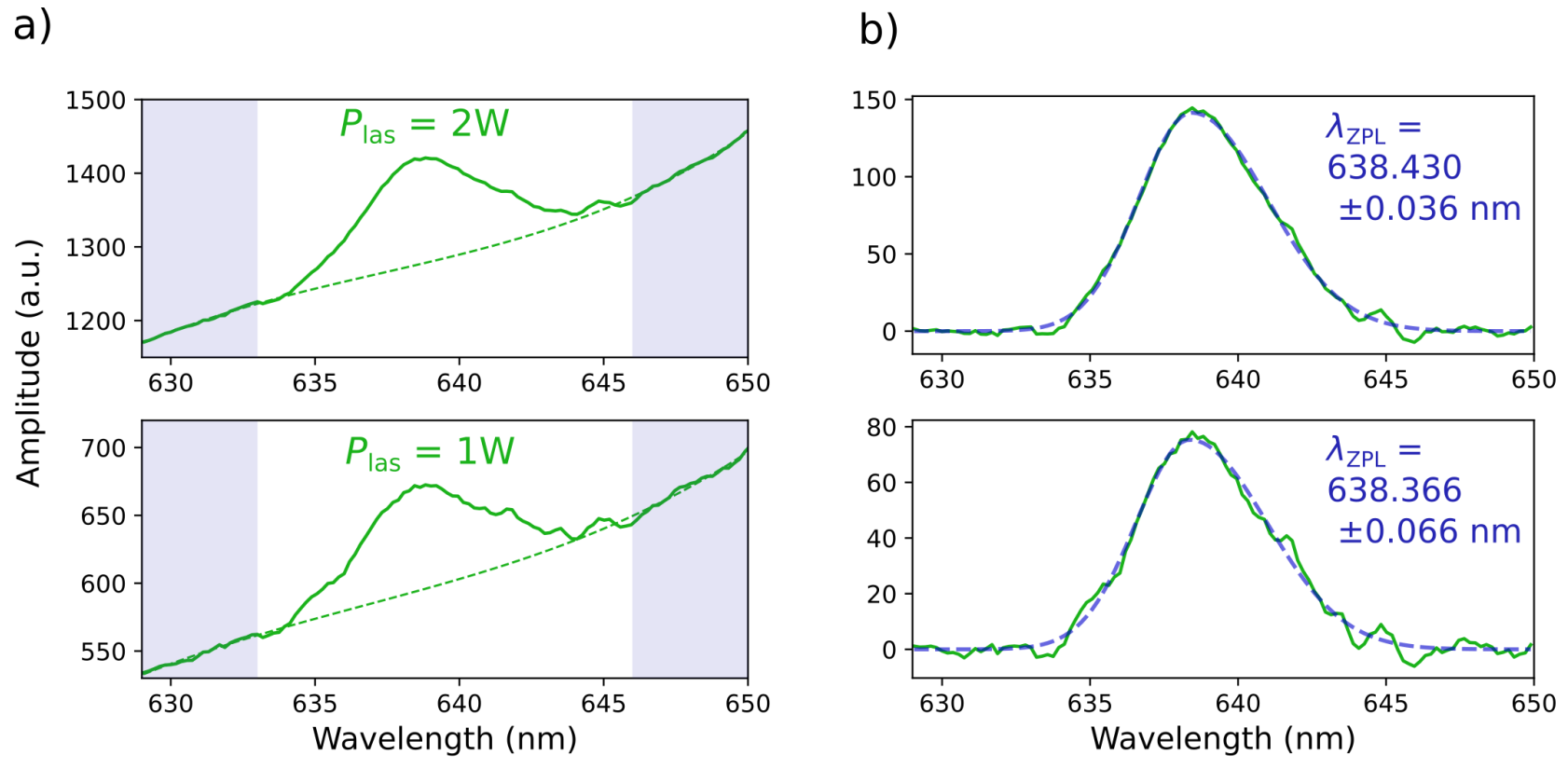


Fig. S2.7: **NV zero-phonon-line (ZPL) analysis for determination of the coefficient of laser-induced heating in the 7.1 T experiments.** a) Zoom in the ZPL region of the photoluminescence spectrum at the laser powers of 2 and 1 W. Dashed lines show polynomial (third-order) fits of the ZPL background using the shaded areas as fitting region b) Background-subtracted ZPLs and fits with a bigaussian function. The acquisitions were taken on Sample A (20% $^{13}$C), spinning at 2 kHz frequency, other conditions were identical as in the NMR experiments reported in the manuscript.

# S3 Model for the NV energies and optical pumping dynamics

## S3.1 Diagonalization of the NV Hamiltonian

The ground-state spin Hamiltonian of NV reads, neglecting the $^{14}$N hyperfine interaction [6]:

$$\hat{H}/h = |\bar{\gamma}_{\rm NV}|\mathbf{B}\cdot\hat{\mathbf{S}} + D\hat{S}_z^2, \tag{S2}$$

where $D = 2867\,\mathrm{MHz}$ is the zero-field splitting, $z$ is the defect axis, $\bar{\gamma}_{\rm NV} = \frac{\gamma_{\rm NV}}{2\pi} = -28.032\,\mathrm{GHz\,T^{-1}}$.

In the simulations represented in main text, exact numerical diagonalization of the full Hamiltonian of the NV-P1-$^{13}$C system, as discussed further in section S4.1, was performed.

However, to obtain approximate expressions of the NV transition frequencies (justifying that LACs in the NV-P1-$^{13}$C system occur for $\theta \approx \alpha_m$), one can derive analytical expressions for $B \gg D/|\bar{\gamma}_{\rm NV}|(\sim 0.1\,\mathrm{T})$, that is when the magnetic field is much above that of the ground state level anticrossing of NV. This condition is fulfilled in our experiments.

For this, we follow the approach detailed in ref. [7], *i.e.* we perform a transformation to an *effective Hamiltonian* in the laboratory frame.

**Diagonalization.** Defining $(\theta, \phi)$ as the spherical coordinates of $\mathbf{B}$ in the NV frame,

$$\mathbf{B} = (B\sin\theta\cos\phi, B\sin\theta\sin\phi, B\cos\theta)_{xyz[\rm NV]}, \tag{S3}$$

the effective Hamiltonian reads

$$\hat{H}_{\rm eff}/h = |\bar{\gamma}_{\rm NV}|B\hat{S}_z + D(\theta)\hat{S}_z^2 + \delta(\theta)\hat{S}_z + c(\theta), \tag{S4}$$

where

$$D(\theta) = \frac{1+3\cos(2\theta)}{4}D = \frac{3\cos^2\theta - 1}{2}D, \tag{S5}$$

$$c(\theta) = D\sin^2(\theta), \tag{S6}$$

$$\delta(\theta) = \frac{\bar{\gamma}_{\rm NV}B|G_1|^2}{(\bar{\gamma}_{\rm NV}B)^2 - [D(\theta)]^2} + \frac{|G_2|^2}{2|\bar{\gamma}_{\rm NV}|B}, \text{with} \tag{S7}$$

$$G_1 = \frac{D\sin\theta\cos\theta e^{i\phi}}{\sqrt{2}}; \quad G_2 = \frac{D(1-\cos 2\theta)e^{2i\phi}}{4}. \tag{S8}$$

Mind that in equation (S4), $z$ is the axis of the applied magnetic field. The effective Hamiltonian $\hat{H}_{\rm eff}$ is obtained by treating the zero-field splitting term ($DS_z^2$ in equation (S2)) at the second order of perturbation. To ensure consistency with the energies obtained numerically, we keep in the Hamiltonian the offset $c(\theta)$ (although this constant offset is subtracted in ref. [7]).

The eigenstates of $\hat{H}_{\rm eff}$ correspond to the eigenstates of $\hat{S}_z$, where $z$ is the axis of the applied magnetic field (since, as visible in equation (S2), $\hat{H}_{\rm eff}$ commutes with $\hat{S}_z$).

We label the eigenstates $|m_s\rangle = |-1\rangle, |0\rangle, |+1\rangle$. Note that $m_s$ corresponds to the spin-projection number along the magnetic field axis.

Taking $B = 7.1\,\mathrm{T}$, one can calculate $\delta(\theta) \lesssim 10\,\mathrm{MHz}$, which is small compared to $D(\theta)(\sim$ GHz). Therefore, in a good approximation, one can neglect the term $\delta(\theta)$ in the effective Hamiltonian and energies. The corresponding energies are:

$$\begin{cases} E_{-1} \approx -|\bar{\gamma}_{\rm NV}|B + D(\theta) = -|\bar{\gamma}_{\rm NV}|B + \frac{1}{2}(3\cos^2\theta - 1)D + \sin^2(\theta)D, \\ E_0 \approx \sin^2(\theta)D, \\ E_{+1} \approx +|\bar{\gamma}_{\rm NV}|B + D(\theta) = +|\bar{\gamma}_{\rm NV}|B + \frac{1}{2}(3\cos^2\theta - 1)D + \sin^2(\theta)D. \end{cases} \tag{S9}$$

The NV transition frequencies read:

$$\nu_{0,\pm 1} \approx |\bar{\gamma}_{\rm NV}|B \pm \frac{1}{2}(3\cos^2\theta - 1)D, \tag{S10}$$

for the $|0\rangle \leftrightarrow |+1\rangle$ and $|0\rangle \leftrightarrow |-1\rangle$ transitions, respectively.

**Eigenstates in the zero-field basis.** To describe the dynamics under optical pumping (in section S3.2), it is useful to decompose the on-field eigenstates in the basis of the zero-field eigenstates of the Hamiltonian (S2), that we label $|-1^0\rangle, |0^0\rangle, |+1^0\rangle$. The basis change requires applying rotations by the angles $\theta$ and $\phi$ along the axes $x$ and $z$ of the NV frame, respectively. One obtains:

$$\begin{cases} |-1\rangle = \frac{\sin\theta}{\sqrt{2}}|0^0\rangle - \frac{\cos\theta}{\sqrt{2}}|-\rangle_\phi + \frac{1}{\sqrt{2}}|+\rangle_\phi, \\ |0\rangle = \cos\theta|0^0\rangle + \sin\theta|+\rangle_\phi, \\ |+1\rangle = \frac{\sin\theta}{\sqrt{2}}|0^0\rangle + \frac{\cos\theta}{\sqrt{2}}|-\rangle_\phi + \frac{1}{\sqrt{2}}|+\rangle_\phi, \end{cases} \tag{S11}$$

where we defined: $|\pm\rangle_\phi = \frac{1}{\sqrt{2}}\left(|+1^0\rangle e^{i\phi} \pm |-1^0\rangle e^{-i\phi}\right)$.

## S3.2 NV dynamics under optical pumping

The seven-level model is a general model for describing the dynamics of NV under optical pumping, which includes the dynamics of optical excitation to the $^3E$ excited state, relaxation within the ground spin state, radiative and non-radiative decay through inter-system crossing (Fig. S3.1) [8, 9]:

For our simulations,

- The model is adapted to the case of a magnetic field misaligned from the NV axis, as formulated in Tetienne *et al.* [9].
- The resulting rate equation system is reduced to the ground spin state levels (three levels) by adiabatic elimination of the fast-relaxing states. This allows reducing the computational space, in order to enable fast simulations when the NV-P1-$^{13}$C system is considered.

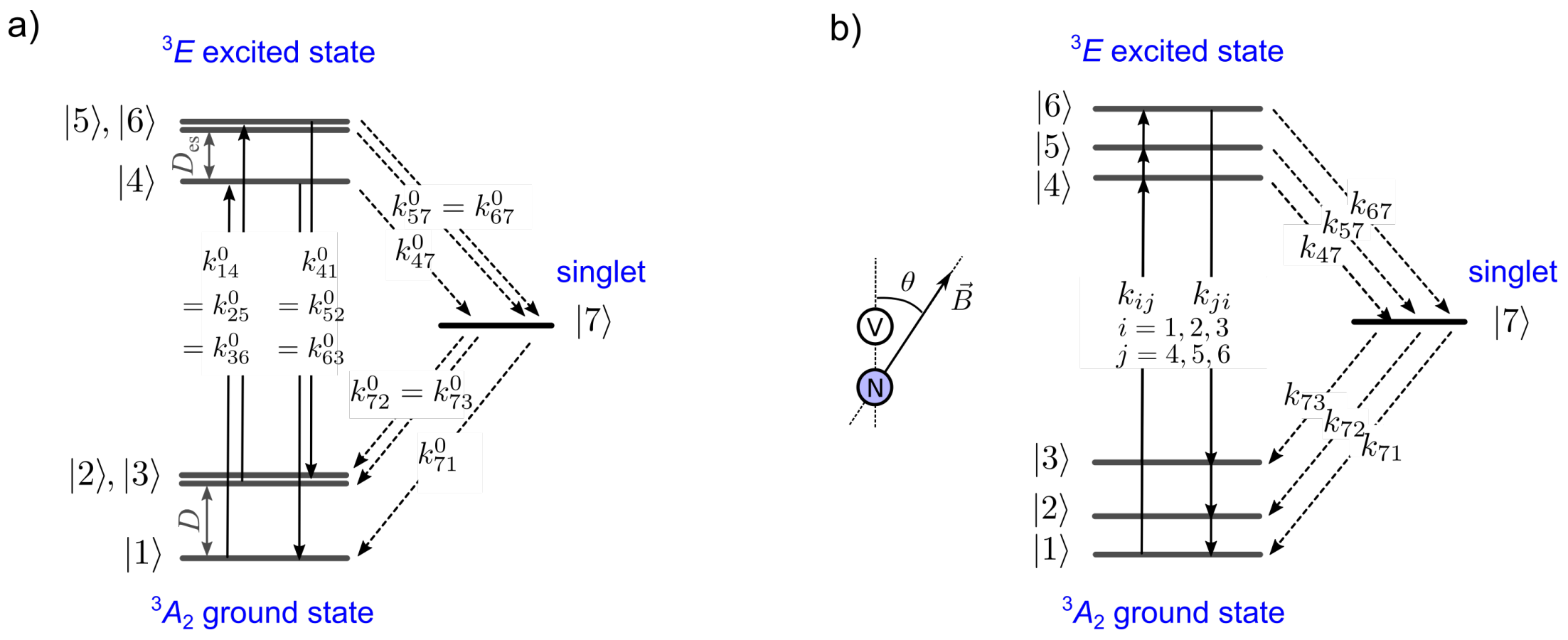


Fig. S3.1: **Description of the seven-level model.** a) At zero magnetic field and b) under a magnetic field. For clarity, only the transitions related to optical excitation, radiative and non-radiative decay through inter-system crossing are shown. Relaxation within the ground spin state is not represented here, but is nevertheless also included in the model.

## Summary of the seven-level model

We write the population dynamics of the seven-level model under a given magnetic field $\boldsymbol{B}$ as

$$\dot{n}_i = \sum_{j \neq i} k^{\boldsymbol{B}}_{ij} n_j - \sum_{j \neq i} k^{\boldsymbol{B}}_{ji} n_i, \tag{S12}$$

where $k_{ij}^{\boldsymbol{B}}$ denotes the transition rate from state $j$ to state $i$, where states are numbered as in Fig. S3.1b. These rates are deduced from the *zero-field* rates $k_{pq}^0$ using statistical averaging [9]:

$$k_{ij}^{\boldsymbol{B}} = \sum_{p=1}^{7} \sum_{q=1}^{7} |\langle i|p^0\rangle|^2 |\langle j|q^0\rangle|^2 k_{pq}^0, \tag{S13}$$

where $|i^0\rangle$ (resp. $|i\rangle$) denotes the zero-field (resp. on-field) eigenstate for $i = 1, 2, ...7$.

Equivalently to eq. (S12), one can write:

$$\dot{\mathbf{n}} = K\mathbf{n}, \tag{S14}$$

with elements of rate matrix $K$ being

$$K_{ij} = \begin{cases} k_{ij}^{\boldsymbol{B}}, & i \neq j, \\ -\sum_{\ell \neq i} k_{\ell i}^{\boldsymbol{B}}, & i = j. \end{cases} \tag{S15}$$

The diagonal terms therefore represent the total rate out of each state. With this convention, each column of $K$ sums to zero,

$$\sum_i K_{ij} = 0, \tag{S16}$$

which ensures conservation of the total population,

$$\frac{d}{dt} \sum_i n_i = 0. \tag{S17}$$

Following the references [8, 9], we note that $K$ incorporates optical excitation, radiative decay, non-radiative decay, spin mixing due to finite transverse magnetic field in NV basis, and thermal contribution through spin lattice relaxation. The rate matrix is constructed in the energy eigenbasis, so the resulting populations are also expressed in that basis.

In the present work, we use the rates of radiative and non-radiative decay from Robledo *et al.* [8], which have been used to describe NV ensemble experiments in Patel *et al.* [10] (Table S3.1).

| Parameter | Value |
|---|---|
| $k_{41,52,63}^0$ | 62.7 MHz |
| $k_{47}^0$ | 12.97 MHz |
| $k_{57,67}^0$ | 80 MHz |
| $k_{71}^0$ | 3.45 MHz |
| $k_{72,73}^0$ | 1.08 MHz |

Table S3.1: Rates of radiative and non-radiative decay through inter-system crossing (Fig. S3.1a) as used in ref. [8, 10].

The rates of optical excitation at a given laser intensity $I$ are obtained using the formula [10]:

$$k_{14,25,36}^0 = \frac{\sigma}{4h\nu} \times I, \tag{S18}$$

where $\sigma = 9.3 \times 10^{-17}\,\mathrm{cm}^2$ is an absorption cross-section under 532 nm excitation and $\nu$ is the laser frequency. Considering the intensity $I_{\max} = 30.5\,\mathrm{W\,cm}^{-2}$ (see section S4.2 for experimental parameters), we get $k_{14,25,36}^0 = 1.90\,\mathrm{kHz}$.

Relaxation within the ground spin state is taken into account by considering both the single quantum (SQ) and double quantum (DQ) processes [11]. First, we define the rates for the zero-field eigenstates $k_{12,21,31,13}^0 = (3T_{1,\mathrm{NV}}^{\mathrm{SQ}})^{-1}$, $k_{23,32}^0 = (2T_{1,\mathrm{NV}}^{\mathrm{DQ}})^{-1}$. Here, we assume $T_{1,\mathrm{NV}}^{\mathrm{SQ}} = T_{1,\mathrm{NV}}^{\mathrm{DQ}} = 2\,\mathrm{ms}$ (see also section S4.2 for a summary of simulation parameters). Second, we calculate the on-field rates from eq. (S13). Last, we perform a finite temperature correction to account for the Boltzmann distribution

of NV polarization generated by thermal excitation of the environment. In thermal equilibrium, the spin-lattice relaxation rates satisfy the detailed balance condition

$$k_{ji}^{\mathrm{th}}\, n_i^{\mathrm{th}} = k_{ij}^{\mathrm{th}}\, n_j^{\mathrm{th}}, \tag{S19}$$

where $n_i^{\mathrm{th}}$ is the thermal equilibrium population of state $i$. For a transition between states $i$ and $j$ with energy difference

$$\Delta E_{ij} = E_i - E_j > 0, \tag{S20}$$

the corresponding Boltzmann factor is

$$r_{ij} = \exp\left(-\frac{\Delta E_{ij}}{k_B T}\right). \tag{S21}$$

The corrected upward transition rate is

$$k_{ij}^{\mathrm{th}} = \frac{2r_{ij}}{1+r_{ij}} k_{ij}, \tag{S22}$$

where $k_{ij}$ is the on-field rate obtained from eq. (S13) (for clarity, we omit the $\boldsymbol{B}$ superscript). The corrected downward transition rate reads

$$k_{ji}^{\mathrm{th}} = \frac{2}{1+r_{ij}} k_{ji}, \tag{S23}$$

For the calculation of $r_{ij}$, a temperature $T = 320\,\mathrm{K}$ is assumed (see also section S4.2).

### Constructing an effective ground-state model using adiabatic elimination

We separate the seven-level population vector into ground-state and fast-state components,

$$\mathbf{n} = \begin{pmatrix} \mathbf{n}_g \\ \mathbf{n}_f \end{pmatrix}, \qquad \mathbf{n}_g = \begin{pmatrix} n_1 \\ n_2 \\ n_3 \end{pmatrix}, \qquad \mathbf{n}_f = \begin{pmatrix} n_4 \\ n_5 \\ n_6 \\ n_7 \end{pmatrix}. \tag{S24}$$

Correspondingly, the rate matrix is written in block form as

$$K = \begin{pmatrix} K_{gg} & K_{gf} \\ K_{fg} & K_{ff} \end{pmatrix}. \tag{S25}$$

The block rate equations are then

$$\dot{\mathbf{n}}_g = K_{gg}\mathbf{n}_g + K_{gf}\mathbf{n}_f, \tag{S26}$$

$$\dot{\mathbf{n}}_f = K_{fg}\mathbf{n}_g + K_{ff}\mathbf{n}_f. \tag{S27}$$

Here, $K_{gf}$ maps fast-state populations into the ground-state dynamics, whereas $K_{fg}$ maps ground-state populations into the fast-state dynamics.

The adiabatic elimination assumes that the excited and singlet populations relax much faster than the ground-state populations. The fast populations are therefore approximated as being in quasi-steady state, meaning that they remain close to the steady-state values determined by the instantaneous ground-state populations. We thus set

$$\dot{\mathbf{n}}_f \approx 0. \tag{S28}$$

Using Eq. (S27), this gives

$$\mathbf{n}_f \approx -K_{ff}^{-1} K_{fg} \mathbf{n}_g. \tag{S29}$$

Substitution into Eq. (S26) gives

$$\dot{\mathbf{n}}_g \approx \left(K_{gg} - K_{gf} K_{ff}^{-1} K_{fg}\right) \mathbf{n}_g. \tag{S30}$$

Thus the effective three-level rate equation is

$$\dot{\mathbf{n}}_g = K_{\text{eff}} \mathbf{n}_g, \qquad K_{\text{eff}} = K_{gg} - K_{gf} K_{ff}^{-1} K_{fg}. \tag{S31}$$

Figure S3.2 compares the NV spin-population dynamics under optical pumping obtained from the full seven-level model and the effective three-level model. A small but nonzero population remains in states outside the ground-state triplet manifold, leading to a slight discrepancy between the ground-state populations predicted by the seven-level and effective three-level models. When the ground-state populations of the seven-level model are renormalized to sum to unity, the predictions of the two models agree closely under the simulation conditions considered here.

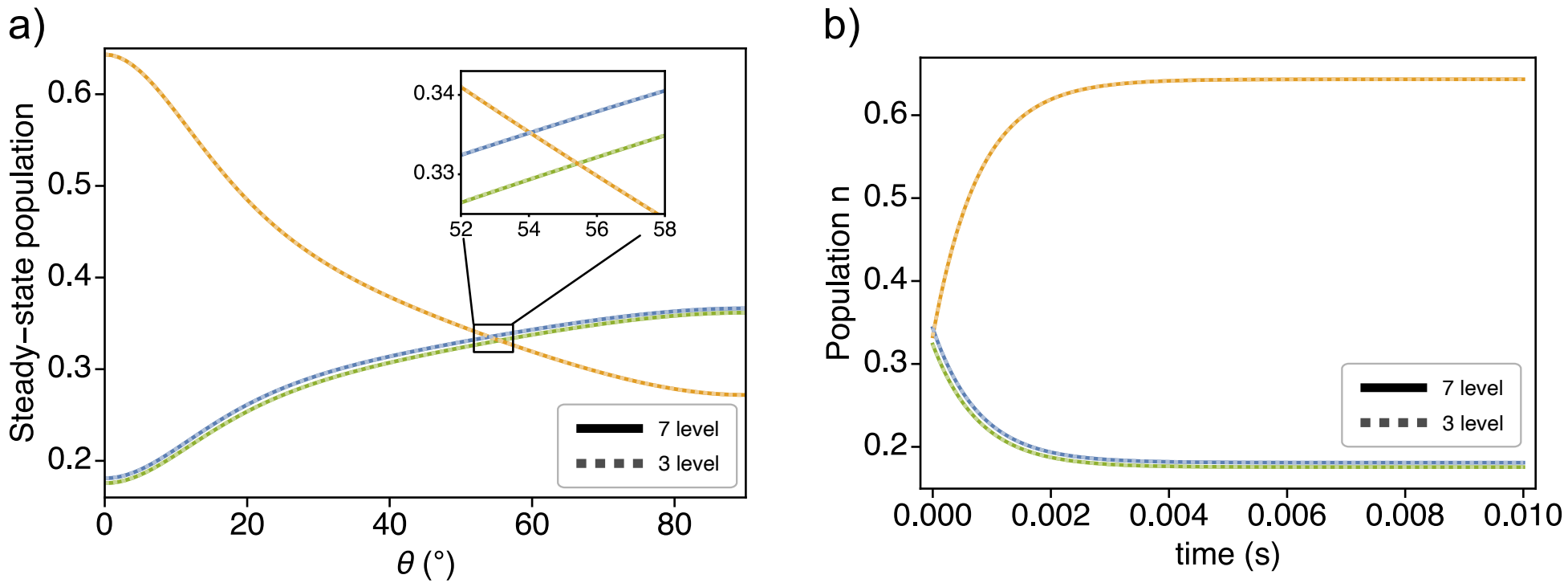


Fig. S3.2: **Comparison of NV spin-population dynamics under optical pumping calculated using the full seven-level model and the effective three-level model**. Blue, yellow, and green denote the populations of the eigenstates $|m_s\rangle = |0, -1, +1\rangle$, corresponding to states labeled $|1, 2, 3\rangle$ in Fig. S3.1b, respectively. (a) Steady-state spin populations obtained from the two models. (b) Time evolution of the spin populations for the two models, starting from the thermal population at $\theta = 0$. A laser intensity $I = 30.5\,\mathrm{W\,cm^{-2}}$ is assumed.

# S4 Description of the NV-P1-$^{13}$C system

## S4.1 Hamiltonian

To describe the NV-P1-$^{13}$C system, the NV Hamiltonian in equation (S2) is complemented with terms related to P1, $^{13}$C and the spin-spin interactions:

$$\hat{H}/h = \underbrace{|\bar{\gamma}_{\mathrm{NV}}|\mathbf{B}\cdot\hat{\mathbf{S_1}} + D(\hat{S}_{1,z})^2}_{\mathrm{NV}} + \underbrace{|\bar{\gamma}_{\mathrm{P1}}|\mathbf{B}\cdot\hat{\mathbf{S_2}}}_{\mathrm{P1}} - \underbrace{\bar{\gamma}_I\mathbf{B}\cdot\hat{\mathbf{I}}}_{^{13}\mathrm{C}} + \underbrace{\hat{\mathbf{S_1}}\cdot A\cdot\hat{\mathbf{I}}}_{\mathrm{NV}-^{13}\mathrm{C}} + \underbrace{a_{\mathrm{ee}}\,\hat{\mathbf{S_1}}\cdot\hat{\mathbf{S_2}}}_{\mathrm{NV}-\mathrm{P1}}, \tag{S32}$$

where the operators $\hat{\mathbf{S_1}}, \hat{\mathbf{S_2}}, \hat{\mathbf{I}}$ correspond to NV, P1, $^{13}$C respectively. The factors $\bar{\gamma}_{\mathrm{NV,P1},I}$ are the gyromagnetic ratios in ordinary frequency units ($\bar{\gamma} = \gamma/2\pi$), with values $\bar{\gamma}_{\mathrm{P1}} = -28.024\,\mathrm{GHz\,T^{-1}}$, $\bar{\gamma}_{\mathrm{NV}} = -28.032\,\mathrm{GHz\,T^{-1}}$, $\bar{\gamma}_I = 10.705\,\mathrm{MHz\,T^{-1}}$ .

The Hamiltonian in Eq. (S32) represents a simplified model where the interaction of the NV and P1 electron spins to their respective $^{14}$N nuclei is *not* included. Including these hyperfine interaction terms would lead to energy levels shifts and additional polarization transfer pathways. Nevertheless, LACs involving the NV, P1 electron spins and the $^{13}$C nuclear spin are as well expected and should remain effective in terms of polarization transfer to $^{13}$C. We now detail the different interaction terms remaining in Eq. (S32).

*Hyperfine interaction to* $^{13}$C. The $^{13}$C nuclear spin is assumed to be coupled to the NV, as represented by the term $\hat{\mathbf{S_1}}\cdot A\cdot\hat{\mathbf{I}}$, with $A$ the hyperfine coupling tensor. The principal values of $A$ for the different considered cases of $^{13}$C hyperfine coupling are given in main text, Table 2. The orientation of the principal axes is given by the angles $(\alpha_{\mathrm{h}}, \varphi_{\mathrm{h}})$ as defined in Figure S4.1. For the first-shell $^{13}$C and the $^{13}$C with 'Moderate' coupling, it is known that $\varphi_{\mathrm{h}} = 0$, while $\alpha_{\mathrm{h}} = 70.52°$ and $70.6°$, respectively [12].[1] For the 'Weak 1' and 'Weak 2' cases, we assume the same principal axes orientation as in the 'Moderate' case.

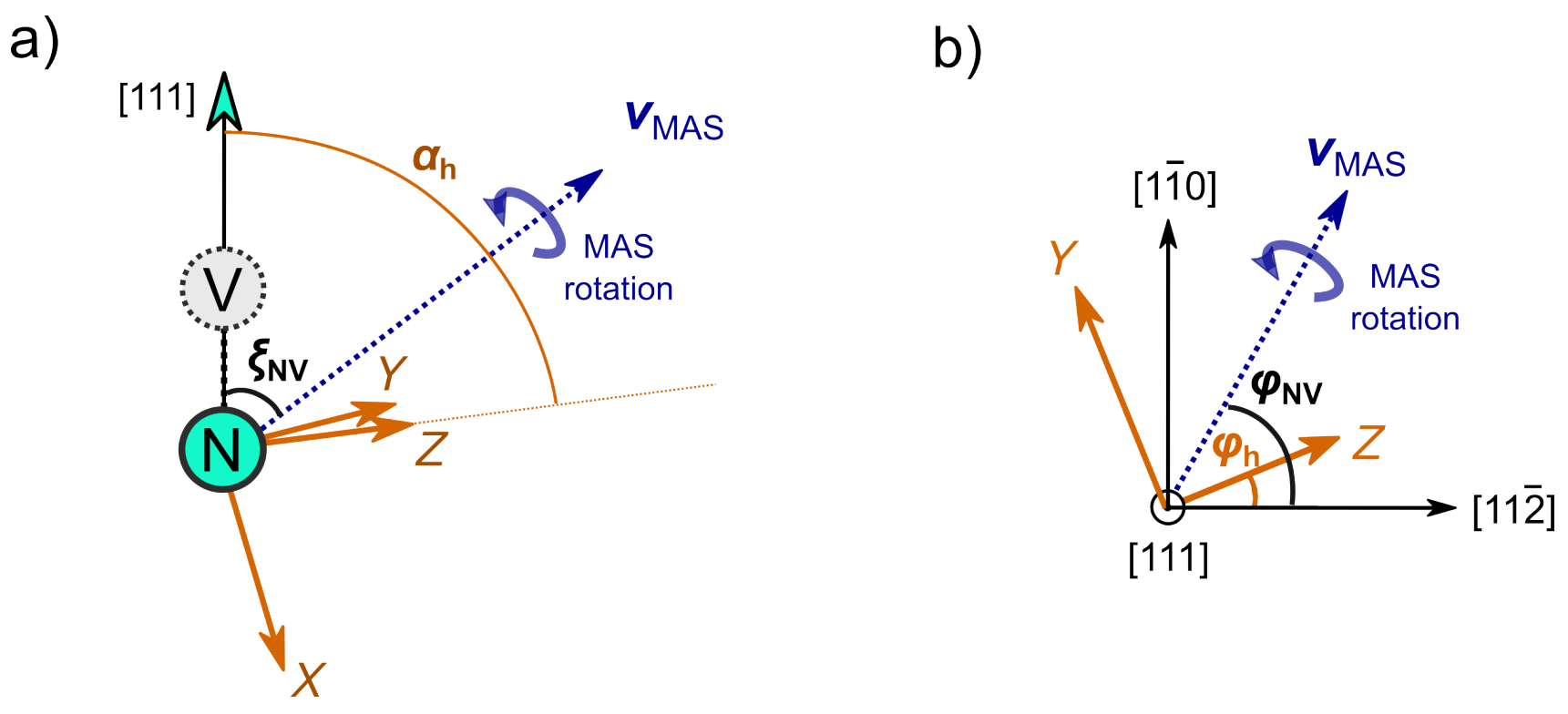


Fig. S4.1: **Coordinates system definitions.** a) Representation of the principal axes ($XYZ$) of the hyperfine tensor, and the MAS rotation axis ($\mathbf{v}_{\mathrm{MAS}}$) b) Top-view, using the crystal-related frame consisting of the axes [111], [11$\bar{1}$], [1$\bar{1}$0] (where [111] is the direction of the considered NV). The angles $(\xi_{\mathrm{NV}}, \varphi_{\mathrm{NV}})$ therefore define the orientation of the MAS rotation axis, while $(\alpha_{\mathrm{h}}, \varphi_{\mathrm{h}})$ define the orientation of the hyperfine coupling tensor to a given $^{13}$C.

*Electron-electron interaction.* In equation (S32), the NV-P1 interaction is assumed to be scalar, as a simplification of the general dipolar Hamiltonian. Indeed, the general dipolar Hamiltonian can be decomposed into three parts: (1) $zz$ interaction, (2) flip-flop and (3) non-secular terms while only terms (1) and (2) are present in the scalar interaction:

[1] as the angles in ref. [12] are taken from the [100] crystal direction, a subtraction of $\alpha_m \approx 54.74°$ is needed to match the definition in Fig. S4.1 where the [111] direction is used.

$$a_{\rm ee}\,\hat{\mathbf{S_1}}\cdot\hat{\mathbf{S_2}} = a_{\rm ee}\,\underbrace{\left(\hat{S}_{1,z}\hat{S}_{2,z}\right)}_{zz} + \frac{a_{\rm ee}}{2}\,\underbrace{\left(\hat{S}_{1,+}\hat{S}_{2,-}+\hat{S}_{1,-}\hat{S}_{2,+}\right)}_{\text{flip}-\text{flop}}, \tag{S33}$$

where we have used the ladder operators ($\hat{S}_{i,\pm}$). The first consequence of assuming a scalar interaction is, therefore, that it removes the non-secular terms (of the form $\hat{S}_{i,\pm}\hat{S}_{j,z}$, $\hat{S}_{i,+}\hat{S}_{j,+}$, and $\hat{S}_{i,-}\hat{S}_{j,-}$). These terms are however not expected to impact the dynamics as the system follows the Zeeman order at these high magnetic fields. The second consequence is that the strength of the flip-flop interaction is directly given by $\frac{a_{\rm ee}}{2}$, independently on the crystal frame orientation of the $B-$field at a given time. The value of $a_{\rm ee}$ can be deduced from the electron-electron distance by considering the corresponding term in the dipolar Hamiltonian, as we now detail. After removing the non-secular terms mentioned above, the dipolar Hamiltonian reads:

$$\hat{H}_{\rm dip}/h = \frac{1}{2\pi}\frac{\mu_0\hbar}{4\pi r^3}\gamma_{\rm NV}\gamma_{\rm P1}(3\cos^2(\theta_{\rm ee})-1)\left\{\hat{S}_{1,z}\hat{S}_{2,z}-\frac{1}{4}\{\hat{S}_{1,+}\hat{S}_{2,-}+\hat{S}_{1,-}\hat{S}_{2,+})\right\}, \tag{S34}$$

where $(r,\theta_{\rm ee})$ are spherical coordinates describing the relative position of the two electron spins. Using again the compact notation $\bar{\gamma}_{\rm NV,P1}=\frac{\gamma_{\rm NV,P1}}{2\pi}$ and equating the flip-flop terms in equations (S33) and (S34), we get:

$$a_{\rm ee} = -\frac{\mu_0\hbar}{4r^3}\bar{\gamma}_{\rm NV}\bar{\gamma}_{\rm P1}(3\cos^2(\theta_{\rm ee})-1). \tag{S35}$$

A typical value corresponding to the distance $r$ is obtained as

$$a_{\rm ee} \approx \frac{\mu_0\hbar}{4r^3}\bar{\gamma}_{\rm NV}\bar{\gamma}_{\rm P1}. \tag{S36}$$

### S4.2 Simulation parameters

In Table S4.1 are given parameters used for the simulations of sections S5-S6. The experimental parameters correspond to the measurements taken at a magnetic field of 7.1 T under 532 nm illumination.

At a laser power of 2 W and considering the peak intensity of a gaussian beam out of a narrow fiber with NA=0.2, from a distance of 5 mm, we get as the laser intensity $I_{\rm max} = 30.5\,\mathrm{W\,cm^{-2}}$.

A temperature $T = 320\,\mathrm{K}$ is defined to take into account moderate heating induced by illumination and spinning.

| Parameter | Value | Description |
|---|---|---|
| $B$ | 7.1 T | Magnetic field |
| $I_{\rm max}$ | $30.5\,\mathrm{W\,cm^{-2}}$ | Laser intensity |
| $T$ | 320 K | Temperature |
| $a_{\rm ee}$ | 5 MHz | NV-P1 coupling |
| $\bar{\gamma}_{\rm NV}$ | $-28.032\,\mathrm{GHz\,T^{-1}}$ | gyromagnetic ratio of NV |
| $\bar{\gamma}_{\rm P1}$ | $-28.024\,\mathrm{GHz\,T^{-1}}$ | " P1 |
| $\bar{\gamma}_I$ | $10.705\,\mathrm{MHz\,T^{-1}}$ | " $^{13}$C |
| $T_{1,\rm NV}$ | 2 ms | NV relaxation time (both single and double-quantum) |
| $T_{1,\rm P1}$ | 1 ms | P1 relaxation time |
| $T_{1,\rm 13C}$ | 5 s | $^{13}$C relaxation time |

Table S4.1: **Parameters used in the simulations**. These parameters were taken for the numerical implementations in sections S5-S6, unless specified otherwise.

### S4.3 NV-$^{13}$C hyperfine coupling distribution

Polarization transfer at LACs is driven by the $A_{zx}, A_{zy}$ terms of the hyperfine coupling in the laboratory frame [13]. In the context of our experiments, it is therefore sensible to define the hyperfine coupling strength as

$$a_{\mathrm{h}} = \sqrt{A_{zx}^2 + A_{zy}^2}. \tag{S37}$$

Now considering a powder ensemble, for a carbon with defined hyperfine elements in the hyperfine tensor frame $A_{XX}, A_{YY}, A_{ZZ}$, one can define the strength as the RMS of eq. (S37). Frame transformation and powder averaging yields

$$\begin{aligned} a_{\mathrm{h}} &= \sqrt{\langle A_{zx}^2 + A_{zy}^2 \rangle_{\theta,\varphi}} \\ &= \sqrt{\frac{2}{15\pi}} \sqrt{(A_{YY} - A_{XX})^2 + (A_{ZZ} - A_{XX})^2 + (A_{ZZ} - A_{YY})^2}. \end{aligned} \tag{S38}$$

For each of the carbons considered in our simulations, one can find the number $N_{\mathrm{h}}$ of lattice sites that would provide a more important coupling strength. For a given $^{13}$C content ($n_{13\mathrm{C}}$), the probability that one $^{13}$C occupies one site with hyperfine strength stronger than $a_{\mathrm{h}}$ is given by

$$p_{\mathrm{h}+}(n_{13\mathrm{C}}) = 1 - (1 - n_{13\mathrm{C}})^{N_{\mathrm{h}+}}. \tag{S39}$$

For the first-shell $^{13}$C and the $^{13}$C with moderate coupling, one has $N_{\mathrm{h}+} = 0$ and $N_{\mathrm{h}+} = 3$, respectively, owing to the three equivalent positions in the first shell. For $^{13}$C with weaker hyperfine couplings, the value of $N_{\mathrm{h}+}^{+}$ can be determined by comparison to the tables from Takács et al. [14], who determined hyperfine tensor between NV and surrounding carbons numerically up to 3 nm distance.

| $^{13}$C coupling | $a_h$ (MHz) | $N_{\mathrm{h}+}$ | $p_{\mathrm{h}+}$(1.1%) | $p_{\mathrm{h}+}$(20%) |
|---|---|---|---|---|
| First-shell | 22.55 | 0 | 0 | 0 |
| Moderate | 1.53 | 3 | 0.032 | 0.488 |
| Weak 1 | 0.44 | 34 | 0.306 | 0.999 |
| Weak 2 | 0.22 | 61 | 0.481 | 1.000 |

Table S4.2: Hyperfine coupling strength driving polarization transfer ($a_{\mathrm{h}}$) and probability $p_{\mathrm{h}+}$ of finding a $^{13}$C with stronger hyperfine strength, for 1.1% and 20% $^{13}$C content respectively.

# S5 Polarization transfer to $^{13}$C: non-spinning case

## S5.1 Constructing the jump operators for the coupled spin dynamics using Lindblad master equation

The open-system dynamics of the coupled NV-P1-$^{13}$C system with Hamiltonian given in Eq. S32 are modeled using the Lindblad master equation

$$\frac{d\rho}{dt} = -\frac{i}{\hbar}[H,\rho] + \sum_j \left( L_j \rho L_j^\dagger - \frac{1}{2}\left\{ L_j^\dagger L_j, \rho \right\} \right), \tag{S40}$$

where the operators $L_j$ are the jump (collapse) operators describing incoherent relaxation processes.

### Effective three-Level dissipation of NV

The NV-center dissipative dynamics are modeled within the ground-state spin triplet manifold as mentioned above

$$|1\rangle, |2\rangle, |3\rangle.$$

The rates obtained from the seven-level model in section S3.2 are used to construct the quantum jump operators for NV dynamics in the ground state.

We denote the effective transition rate from level $|j\rangle$ to level $|i\rangle$ by $\Gamma_{ij}^{(\mathrm{NV})}$ which are the off-diagonal matrix elements of $K_{\mathrm{eff}}$. The associated jump operators are constructed as

$$L_{ij}^{(\mathrm{NV})} = \sqrt{\Gamma_{ij}^{(\mathrm{NV})}}\, |i\rangle\langle j|, \qquad i \neq j, \tag{S41}$$

where $i, j \in \{1,2,3\}$.

Explicitly, the six allowed incoherent transitions are

$$\begin{aligned}
L_{1,2}^{(\mathrm{NV})} &= \sqrt{\Gamma_{1,2}}\, |1\rangle\langle 2|, \\
L_{2,1}^{(\mathrm{NV})} &= \sqrt{\Gamma_{2,1}}\, |2\rangle\langle 1|, \\
L_{1,3}^{(\mathrm{NV})} &= \sqrt{\Gamma_{1,3}}\, |1\rangle\langle 3|, \\
L_{3,1}^{(\mathrm{NV})} &= \sqrt{\Gamma_{3,1}}\, |3\rangle\langle 1|, \\
L_{3,2}^{(\mathrm{NV})} &= \sqrt{\Gamma_{3,2}}\, |3\rangle\langle 2|, \\
L_{2,3}^{(\mathrm{NV})} &= \sqrt{\Gamma_{2,3}}\, |2\rangle\langle 3|.
\end{aligned}$$

These effective Lindblad operators capture both optical pumping and spin relaxation, under the presence of spin mixing induced by the magnetic-field misalignment.

### Thermal relaxation model for P1 and $^{13}$C

We follow the same approach as used for modeling the thermal relaxation of NV (section S3.2), that is, the upward and downward transition rates are chosen to satisfy detailed balance at temperature $T$.

Given an energy splitting

$$\Delta E = E_1 - E_0,$$

the Boltzmann factor is

$$r = \exp\left(-\frac{\Delta E}{k_B T}\right),$$

where $k_B$ is Boltzmann's constant, and $T$ is the temperature.

The downward relaxation rate is

$$\gamma_\downarrow = \frac{1}{T_1}\frac{1}{1+r},$$

while the upward thermal excitation rate is

$$\gamma_\uparrow = \frac{1}{T_1}\frac{r}{1+r}.$$

These rates satisfy $\frac{\gamma_\uparrow}{\gamma_\downarrow} = e^{-\Delta E/(k_B T)}$ ensuring thermal equilibrium in the absence of driving.

In the simulations, $T_{1,\mathrm{P1}} = 1\,\mathrm{ms}$, $T_{1,\mathrm{13C}} = 5\,\mathrm{s}$ and $T = 320\,\mathrm{K}$ are taken (see also section S4.2 for simulation parameters).

### $^{13}$C nuclear spin jump operators

For the nuclear spin, with eigenstates $|\alpha\rangle, |\beta\rangle$, we define the lowering and raising operators

$$\sigma_-^{(\mathrm{C})} = |\alpha\rangle\langle\beta|, \quad \sigma_+^{(\mathrm{C})} = |\beta\rangle\langle\alpha|$$

The corresponding Lindblad jump operators are

$$L_\downarrow^{(\mathrm{C})} = \sqrt{\gamma_\downarrow^{(\mathrm{C})}}\,\sigma_-^{(\mathrm{C})},$$
$$L_\uparrow^{(\mathrm{C})} = \sqrt{\gamma_\uparrow^{(\mathrm{C})}}\,\sigma_+^{(\mathrm{C})}.$$

### P1 electron spin jump operators

Similarly for the P1 electron spin, with eigenstates $|\downarrow\rangle, |\uparrow\rangle$,

$$\sigma_-^{(\mathrm{P1})} = |\downarrow\rangle\langle\uparrow|, \quad \sigma_+^{(\mathrm{P1})} = |\uparrow\rangle\langle\downarrow|,$$

with jump operators

$$L_\downarrow^{(\mathrm{P1})} = \sqrt{\gamma_\downarrow^{(\mathrm{P1})}}\,\sigma_-^{(\mathrm{P1})},$$
$$L_\uparrow^{(\mathrm{P1})} = \sqrt{\gamma_\uparrow^{(\mathrm{P1})}}\,\sigma_+^{(\mathrm{P1})}.$$

The full set of collapse operators used in the simulation is therefore

$$\mathcal{C} = \left\{L_\uparrow^{(\mathrm{C})}, L_\downarrow^{(\mathrm{C})}, L_\uparrow^{(\mathrm{P1})}, L_\downarrow^{(\mathrm{P1})}, L_{ij}^{(\mathrm{NV})}\right\}. \tag{S42}$$

### Implementation of dephasing

To model inhomogeneous dephasing, an additional static magnetic-field offset is included in the Hamiltonian of each species. This is done by incorporating the following term in the Hamiltonian:

$$\frac{\hat{H}_{\mathrm{noise}}}{h} = \bar{\gamma}\sum_i b_i \sigma_z$$

where $\bar{\gamma}$ is the gyromagnetic ratio of the corresponding spin species, and $b_i$ is the magnetic field noise that the $i$th spin of that species experiences, and $\sigma_z$ is the spin-$z$ operator in the lab basis.

For each realization, the field noise is independently sampled from a zero-mean Gaussian distribution with standard deviation for the three spin species:

$$\sigma_B = \frac{1}{\sqrt{2}\pi|\bar{\gamma}|T_2^*},$$

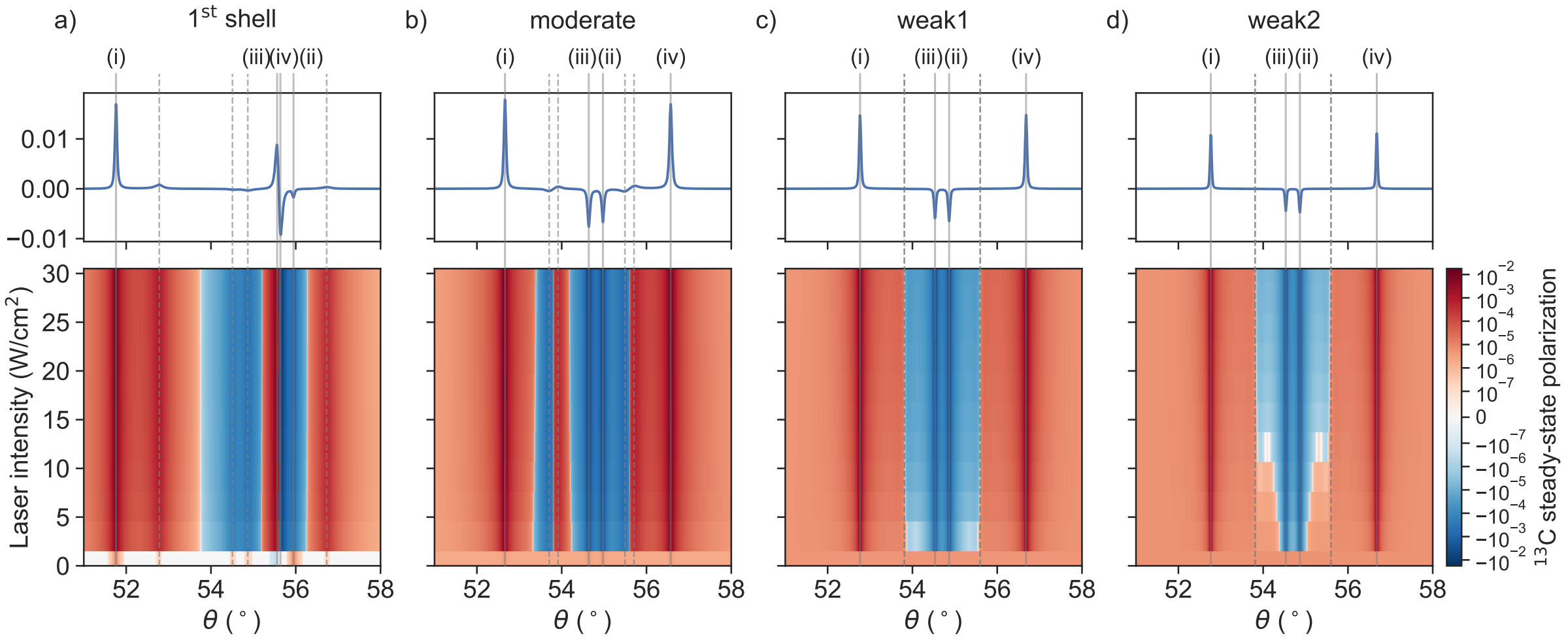


Fig. S5.1: **Steady-state polarization of $^{13}$C in the vicinity of $\theta = \alpha_m$ in the non-spinning case.** While the data being represented is the same as in main text Fig. 6, logarithmic scaling has been used to enhance fine polarization features. The values with $|P| < 1e^{-7}$ were set to zero to reduce color variation near zero. The dotted lines indicate the electron-electron LAC, while the solid lines indicate the nuclear-electron LACs.

where $T_2^*$ is its inhomogeneous dephasing time. The field fluctuations acting on different spins are assumed to be uncorrelated. We also verified that introducing correlated fluctuations does not qualitatively change the results. The steady-state $^{13}$C polarization is calculated independently for $N = 500$ field noise realizations, and the resulting observables are ensemble averaged. The $T_2^*$ values used for the simulations in the non-spinning case are listed in Table S5.1.

| $T^*_{2,\mathrm{NV}}$ | $T^*_{2,\mathrm{P1}}$ | $T^*_{2,\mathrm{13C}}$ |
|---|---|---|
| 300 ns | 300 ns | 5 ms |

Table S5.1: Default $T_2^*$ used throughout the non-spinning simulation.

## S5.2 Numerical implementation

The Lindblad master equation is numerically solved for the steady-state. Besides the $T_2^*$ values listed in Table S5.1, simulation parameters are that given in section S4.2.

As seen in Figure S5.1, increasing the laser intensity used for optical polarization of NV spins, polarization buildup of $^{13}$C is observed.

Figure S5.2 shows the spherical average of the steady-state polarization, defined as $\langle P(\theta)\rangle = 2\langle I_z(\theta)\rangle$. The powder-averaged polarization is calculated as

$$\langle P\rangle_{\mathrm{sph}} = \int_0^{\pi/2} \langle P(\theta)\rangle \sin\theta \, \mathrm{d}\theta. \tag{S43}$$

For reducing computational load of spherical averaging simulations, we calculate polarisation of nuclear spins in rough steps of 1° from $\theta$= 0° to 51°, and 58° to 90°. For $\theta$=51° to 58°, we take a finer scan of angle and measure the polarisation in steps of 0.05°.

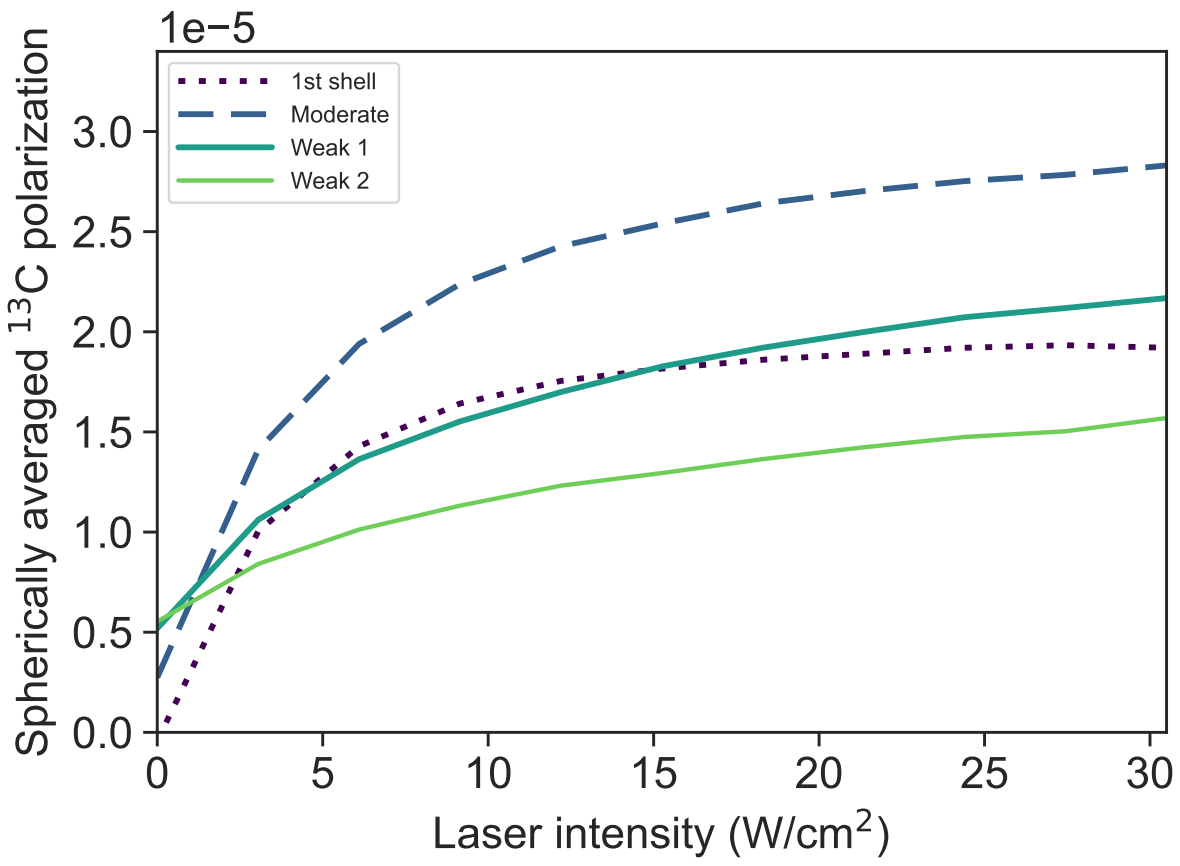


Fig. S5.2: **Steady-state $^{13}$C polarization after spherical averaging for powder samples (non-spinning case simulation).**

## S6 Polarization transfer to $^{13}$C: spinning case

The evolution of the NV-P1-$^{13}$C system under magic angle spinning and illumination is described as follows:

- Aside the LACs, the evolution is described by a system of rate equations on the populations.
- At the LACs discrete jumps occur, with probabilities determined by the Landau-Zener formula.

Following a common approximation in the description of MAS-DNP from randomly oriented systems [13], it is considered that coherences do not impact the dynamics.

The respective models as well as the numerical implementation are discussed in the following sections.

### S6.1 Rate equations: evolution between LACs

In between LACs eigenstates of the NV-P1-$^{13}$C system are taken as tensor products $\{|m_{s,\mathrm{NV}}\rangle \otimes |m_{s,\mathrm{P1}}\rangle \otimes |s_{13\mathrm{C}}\rangle\}$ where:

- $|m_{s,\mathrm{NV}}\rangle, |m_{s,\mathrm{P1}}\rangle$ run over the various eigenstates of NV and P1.
- $|s_{13\mathrm{C}}\rangle$ corresponds to eigenstates of $^{13}$C taking into account the hyperfine interaction to NV.

This allows building a global rate equation system describing the dynamics of relaxation, and optical pumping:

$$\frac{\mathrm{d}\tilde{\mathbf{n}}}{\mathrm{d}t} = \mathbf{K}_{\mathrm{NV}} \cdot \tilde{\mathbf{n}} + \mathbf{K}_{\mathrm{rel}} \cdot (\tilde{\mathbf{n}} - \tilde{\mathbf{n}}^{\mathrm{th}}), \tag{S44}$$

where $\tilde{\mathbf{n}}$ is the population vector for the NV-P1-$^{13}$C system ($12 \times 1$) and $\tilde{\mathbf{n}}^{\mathrm{th}}$ is the vector containing thermal (Boltzmann) populations. The first term in eq. (S44) represents the NV optical pumping and relaxation dynamics. The rate of population transfer between states belonging to $m_{s,\mathrm{NV}} = i$ and $m'_{s,\mathrm{NV}} = j$ manifolds, $|s\rangle = |m_{s,\mathrm{NV}}\rangle \otimes |m_{s,\mathrm{P1}}\rangle \otimes |s_{13\mathrm{C}}\rangle$ and $|s'\rangle = |m'_{s,\mathrm{NV}}\rangle \otimes |m'_{s,\mathrm{P1}}\rangle \otimes |s'_{13\mathrm{C}}\rangle$ is given by the matrix element:

$$\mathbf{K}_{\mathrm{NV},|s\rangle\to|s'\rangle} = (K_{\mathrm{eff}})_{i,j} \times \left|\langle m'_{s,\mathrm{P1}} \,|\, m_{s,\mathrm{P1}}\rangle\right|^2 \times \left|\langle s'_{13\mathrm{C}}|s_{13\mathrm{C}}\rangle\right|^2 \tag{S45}$$

where $\mathbf{K}_{\mathrm{NV}}$ is obtained in section S3.2. The definition in equation (S45) takes into account the fact that the quantization axis of a $^{13}$C coupled to NV can vary across the different $m_s-$manifold of NV, and that correspondingly, flip of the $^{13}$C nuclear spin can occur during the optical pumping / relaxation dynamics of NV.

The second term in eq. (S44) represents the relaxation of P1 and $^{13}$C. It is taken as

$$\mathbf{K}_{\mathrm{rel}} = \mathbb{1}_{\mathrm{NV}} \otimes K_{\mathrm{rel,P1}} \otimes \mathbb{1}_{13\mathrm{C}} + \mathbb{1}_{\mathrm{NV}} \otimes \mathbb{1}_{\mathrm{P1}} \otimes K_{\mathrm{rel,13C}}, \tag{S46}$$

where $\mathbb{1}_{\mathrm{NV}}, \mathbb{1}_{\mathrm{P1}}, \mathbb{1}_{13\mathrm{C}}$ are the identity matrices for the NV, P1 and $^{13}$C populations, respectively, and :

$$K_{\mathrm{rel,P1}} = \frac{1}{2}\begin{pmatrix} -\frac{1}{T_{1,\mathrm{P1}}} & \frac{1}{T_{1,\mathrm{P1}}} \\ \frac{1}{T_{1,\mathrm{P1}}} & -\frac{1}{T_{1,\mathrm{P1}}} \end{pmatrix}, \qquad K_{\mathrm{rel,13C}} = \frac{1}{2}\begin{pmatrix} -\frac{1}{T_{1,13\mathrm{C}}} & \frac{1}{T_{1,13\mathrm{C}}} \\ \frac{1}{T_{1,13\mathrm{C}}} & -\frac{1}{T_{1,13\mathrm{C}}} \end{pmatrix}. \tag{S47}$$

In the simulations, $T_{1,\mathrm{P1}} = 1\,\mathrm{ms}$ and $T_{1,13\mathrm{C}} = 5\,\mathrm{s}$ are taken (see also section S4.2 for full simulation parameters).

## S6.2 Landau-Zener formula and dynamics at LACs

### The Landau-Zener formula

The Landau-Zener formula is used to describe the probability $P_{\text{LZ}}$ of a non-adiabatic jump at the individual LACs. If $E_i$ and $E_j$ are the time-dependent energies of the diabatic (crossing) states, the LZ formula reads:

$$\begin{aligned} P_{\text{LZ}} &= e^{-2\pi\Gamma} \\ \Gamma &= \frac{\Delta^2/4\hbar}{\left|\frac{\partial}{\partial t}(E_j - E_i)\right|}, \end{aligned} \tag{S48}$$

where $\Delta$ is the full energy gap at the anticrossing. Correspondingly the probability of an adiabatic process is $1 - P_{\text{LZ}}$ .

### Validity regime and powder averaging

*Validity regime of LZ.* Whenever the LZ formula predicts $P_{\text{LZ}} > 0$, for the predicted probability to be accurate, the LAC region must be crossed over a time shorter than $T_2$, where $T_2$ is the coherence time of a superposition of the two states involved in the anticrossing. Defining the span (in energy) of the LAC region as $3\Delta$, this condition reads

$$\left|\frac{\partial}{\partial t}(E_j - E_i)\right| > \frac{3\Delta}{T_2}, \tag{S49}$$

where $T_2$ is the coherence time of a superposition of the states of energies $E_i$ and $E_j$. $T_2$ is mostly determined by the paramagnetic spin bath.

Here, $T_2 = 3\,\mu\text{s}$ is taken based on coherence times (single quantum) observed in EPR for samples of similar paramagnetic impurity composition. [2]

For one NV trajectory (one value of $\xi_{\text{NV}}$), the simulation is considered to be valid is (S49) is verified for all LACs along the trajectory.

*Powder averaging.* When simulating a powder average, all values of $\xi_{\text{NV}}$ in the range $0 - 90°$ are considered. In the simulations, it is assumed that the powder average is valid when equation (S49) is verified along the trajectories for values of $\xi_{\text{NV}}$ covering $>$ **95**% of the full solid angle.

With this criterion, it is found for each set of $^{13}$C hyperfine coupling parameters a minimum MAS rate at which the simulation is valid. For $a_{ee} = 5\,\text{MHz}$, the limit evaluates to 200 17, 4 and 2 Hz for the following hyperfine coupling configurations: 1$^{\text{st}}$-shell, Moderate, 'Weak 1' and 'Weak 2' $^{13}$C, respectively.

## S6.3 Numerical implementation

The following steps are implemented in Python using packages `qtip` and scipy's `odeint`:

- **Hamiltonian diagonalization.** Using `qutip`, the Hamiltonian is numerically diagonalized. For a given NV trajectory, we diagonalize it for a set of points describing one rotor period. For each value of the rotor coordinate $\varphi_m$, qutip returns a set of eigenenergies $E_i(\varphi_m)$ in ascending order.

- **Finding LAC parameters.** LACs are found based on local minima of the energy difference $E_{i+1}(\varphi_m) - E_i(\varphi_m)$ (to avoid false detection, minima are selected when the energy difference is below a threshold $T$ defined so that $\sqrt{2}a_{\text{ee}} < T < \nu_{13\text{C}}$). Once all the LACs are found, the states are reindexed to map into defined states of the NV-P1-$^{13}$C system (on which the matrices $\mathbf{K}_{\text{NV}}$ and $\mathbf{K}_{\text{rel}}$ used in eq. (S44) are acting). Simultaneously, the energy gap and the slope $(\frac{\partial}{\partial\varphi_m}\{E_{i+1}(\varphi_m) - E_i(\varphi_m)\})$ of the LACs is extracted from a local quadratic fit.

- **Calculating LZ probabilities.** For a given MAS frequency, $P_{\rm LZ}$ is calculated for each LAC according to the LZ formula (S48).

- **Solving evolution: rate equations + LZ.** To solve the evolution, we first set the population as thermal: $\tilde{\mathbf{n}}(t=0) = \tilde{\mathbf{n}}^{\rm th}$. For the dynamics in between LACs, SciPy `odeint` is used to solve the rate equation system. At each LAC, populations of the involved energy levels are redistributed following the probabilities predicted by the LZ formula. The new populations are taken as initial condition for the next round of solving with `odeint`, etc. This allows simulating the dynamics over a small number (typically $\sim 10$) rotor periods.

- **Solving for $n \gg 1$ rotor periods.** To predict the dynamics for a higher number (e.g. $k = 100000$) of rotor periods, the transfer matrix for the populations corresponding to exactly one rotor period, $\mathcal{P}$, is first computed. Then the population vector at the beginning of the $k-$th rotor period is obtained as:

$$\tilde{\mathbf{n}}(t = kT_R) = \mathcal{P}^k \tilde{\mathbf{n}}(t=0). \tag{S50}$$

  This 'stroboscopic' approach is used to compute the polarization buildup curves represented in main text as well as the steady-state polarization.

- **Powder averaging.** The powder-averaged population vector $\langle \tilde{\mathbf{n}} \rangle_{\rm Sph}$ is computed based on the populations obtained for different geometrical parameters $(\xi_{\rm NV}, \varphi_{\rm NV})$ described in Fig. S4.1:

$$\langle \tilde{\mathbf{n}} \rangle_{\rm Sph} = \frac{\sum\limits_{\varphi_{\rm NV}=0,90^\circ} \sum\limits_{\xi_{\rm NV}=0,5,10,\ldots 90^\circ} \sin(\xi_{\rm NV}) \tilde{\mathbf{n}}_{(\xi_{\rm NV},\varphi_{\rm NV})}}{\sum\limits_{\varphi_{\rm NV}=0,90^\circ} \sum\limits_{\xi_{\rm NV}=0,5,10,\ldots 90^\circ} \sin(\xi_{\rm NV})}, \tag{S51}$$

  where the $\sin(\xi_{\rm NV})$ factor is needed to take into account the contribution to the solid angle from NV in the range $[\xi_{\rm NV}, \xi_{\rm NV} + d\xi_{\rm NV}]$. The sum considers a $5^\circ$ step on $\xi_{\rm NV}$. As the parameter $\varphi_{\rm NV}$ was found to affect only weakly the populations (by a few %), average is only performed over two values, $\varphi_{\rm NV} = 0, 90^\circ$.

  From the powder averaged populations $(\langle \tilde{\mathbf{n}} \rangle_{\rm Sph}$, the $^{13}$C polarization is computed as

$$P_{\rm 13C} = 2 \langle I_z \rangle$$

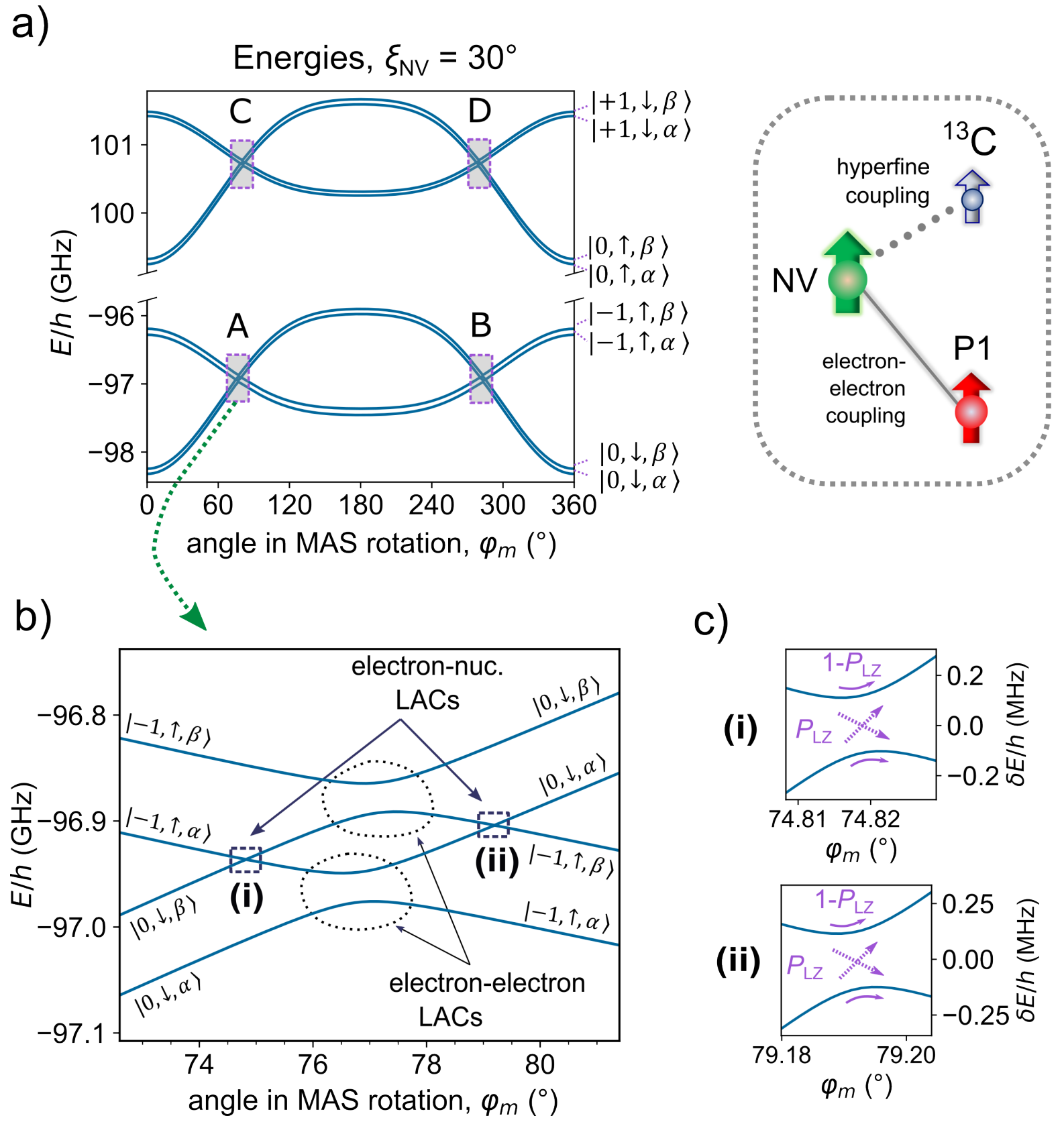


Fig. S6.1: **Description of the level crossings in the NV-P1-$^{13}$C system under MAS.** a) Energies in the NV-P1-$^{13}$C system as a function of the rotor angular coordinate $\varphi_m$. Energies in the manifold $m_{s,\text{NV}} + m_{s,\text{P1}} = \pm 1/2$ are represented (LACs do not occur within the remaining manifolds, $m_{s,\text{NV}} + m_{s,\text{P1}} = \pm 3/2$). b) Magnification of the energies in one of the four anticrossing regions, showing two electron-nuclear LACs (i,ii) and two electron-electron LACs. c) Zoom on the electron-nuclear LACs and illustration of the LZ probabilities. The simulation was performed at $B = 7.1\,\text{T}$ magnetic field, with the angle $\xi_{\text{NV}} = 30°$, the electron-electron coupling $a_{\text{ee}} = 15\,\text{MHz}$ and considering the 'Moderate' NV-$^{13}$C hyperfine coupling case.

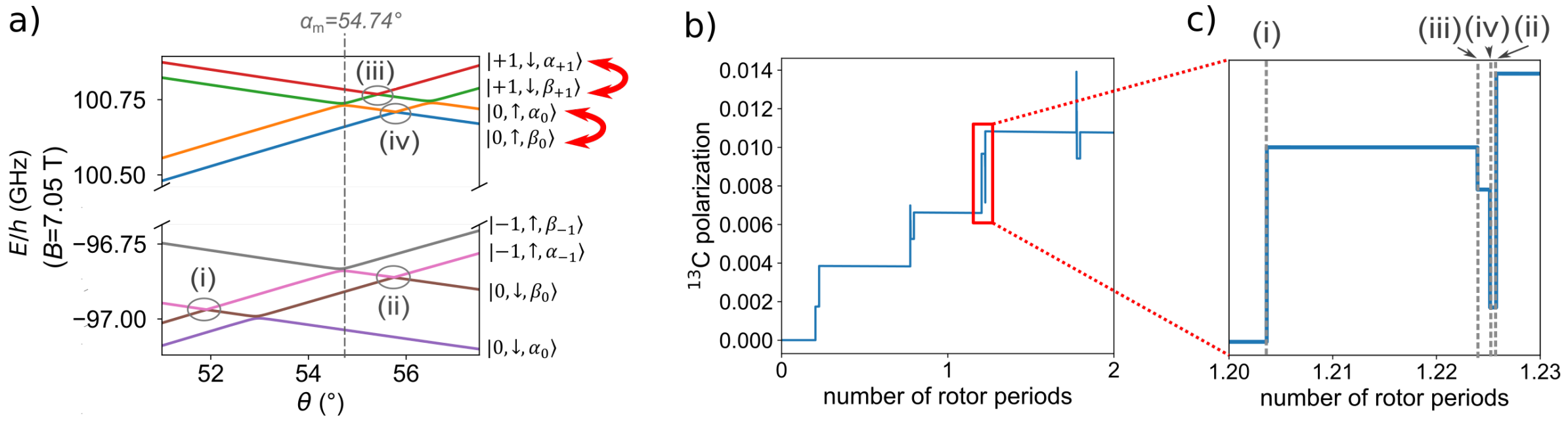


Fig. S6.2: **Polarization dynamics of the first-shell $^{13}$C under MAS.** a) Representation of the energies in the NV-P1-$^{13}$C system as a function of $\theta$ (angle of NV to the $B-$field) in the vicinity of LACs, at a magnetic field of 7.1 T and for $a_{\text{ee}} = 5\,\text{MHz}$. The arrows describe the inversion of the $\alpha$ and $\beta$ nuclear states in the upper energy manifold ($m_{s,\text{NV}} + m_{s,\text{P1}} = 1/2$) occurring as a consequence of the strong hyperfine coupling. b) Simulated polarization buildup for 1 kHz MAS frequency, taking $\xi_{\text{NV}} = 30°$. c) Zoom on a time window where electron-nuclear LACs are met, showing that LACs in the upper energy manifold (iii,iv) lead to negative polarization transfer and thereby lower the total transferred polarization.

# S7 Parameters relevant to the direct polarization of external $^1$H spins

Among the near-surface paramagnetic defects present in nanodiamonds, a fraction typically shows ∼MHz hyperfine coupling to surface $^1$H spins, as revealed by electron-nuclear double resonance [15]. The polarization of $^1$H part of NV-X-$^1$H cluster –where 'X' is a near-surface paramagnetic spin– can be simulated with the same model used for the simulations shown in main text.

This scheme would require the NV depth to not exceed a certain threshold. Sushkov *et al.* [16] found a NV at a depth of 3.3(2) nm in a diamond with typical (carboxyl/hydroxyl) surface termination, whereas the observation of NV into monodisperse detonation nanodiamond particles with ∼ 5 nm diameter [17] suggests that NV can even survive at lower depths.

Assuming the defect 'X' is located exactly at the surface, above NV, depths of 2 and 3 nm correspond –using equation (S36)– to an electron-electron (NV-X) coupling strength $a_{ee} = 3.2$ MHz and 0.96 MHz respectively.

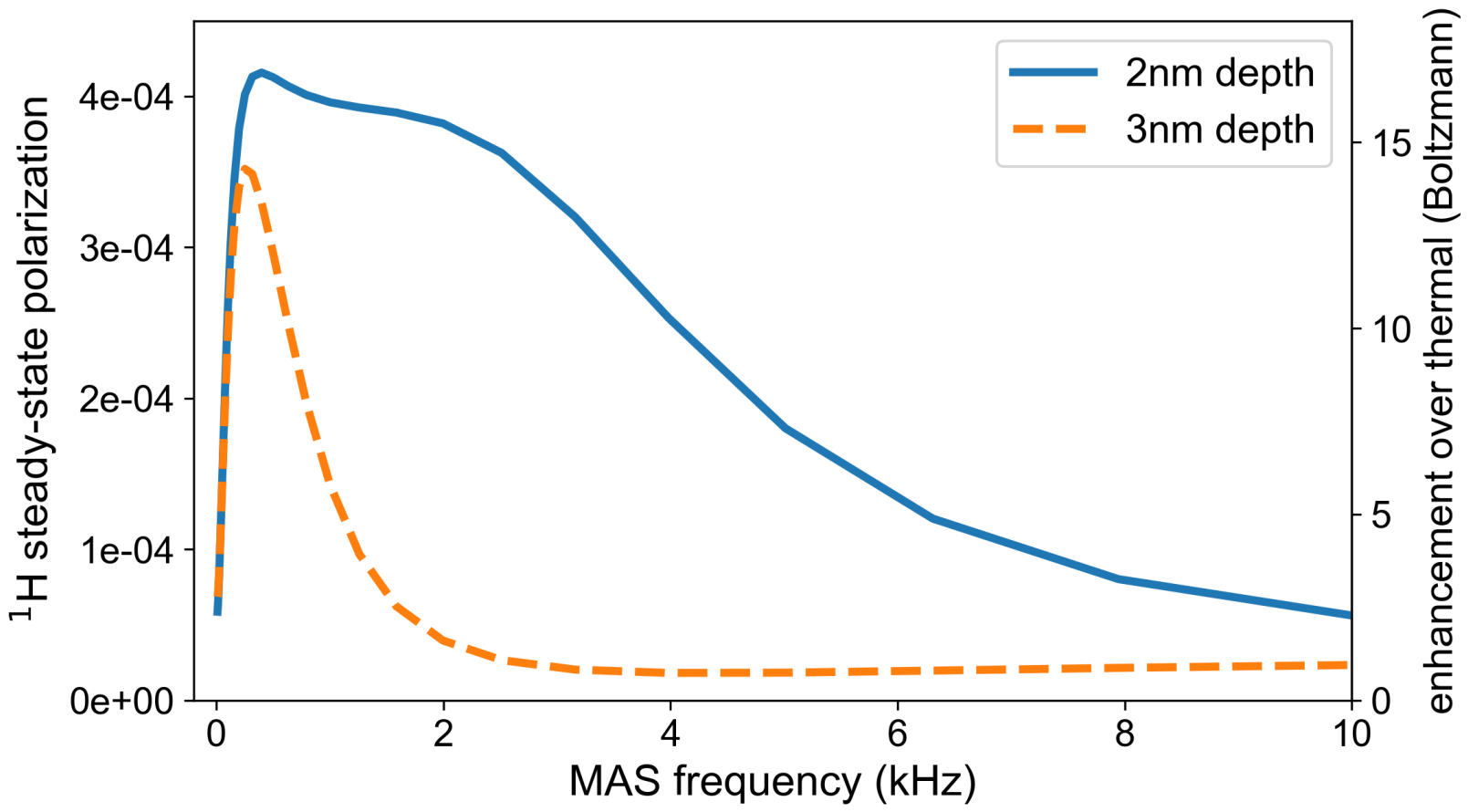


Fig. S7.1: **Simulated steady-state polarization of a surface $^1$H coupled to a subsurface defect.** In this simulation, a $^1$H spin part of a NV-X-$^1$H cluster is considered, where 'X' is a $S = 1/2$ defect ($g = 2.0023$) at the diamond surface. The simulations shown as continuous and dashed line consider NV depths of 2 and 3 nm, respectively, corresponding to NV-X couplings $a_{ee} = 3.2$ and 0.96 MHz (for simplification, the defect 'X' is assumed to be located at the surface, directly above NV). An angle between NV and the MAS axis $\xi_{NV} = 65°$ is assumed. Relaxation time is taken as $T_{1,1H} = 100$ ms for $^1$H and $T_{1,X} = 1$ ms for X. Other parameters are same as those listed in section S4.2.

In Figure S7.1 are represented simulation of the steady state polarization of a $^1$H coupled to the subsurface defect with hyperfine couplings parameters $A_{XX} = A_{YY} = -0.5$ MHz, $A_{ZZ} = 1$ MHz. As we see, positively enhanced $^1$H polarization can be generated. Nevertheless, at the NV depth of 3 nm, the process is only efficient at MAS frequencies below ∼ 1 kHz. This is consistent with the onset of non-adiabaticity of electron-electron crossings (discussed in main text) occurring at lower MAS frequencies in this case of weak NV-X coupling, compared to the 2 nm depth case.



\end{document}